\documentclass{emulateapj}
\usepackage{graphicx}
\usepackage{mathrsfs}
\usepackage{amsmath}
\usepackage{subfigure}
\usepackage{color}

\shorttitle{SLACS. XII.}
\shortauthors{Shu et al. 2015}
 
\begin{document}
 
\title{The Sloan Lens ACS Survey. XII. Extending Strong Lensing to Lower Masses$^{\dag}$}

\altaffiltext{$^{\dag}$}{Based on observations made with the NASA/ESA Hubble Space Telescope, obtained at the Space Telescope Science Institute, which is operated by AURA, Inc., under NASA contract NAS 5-26555. These observations are associated with programs \#10174, \#10494, \#10587, \#10798, \#10886, and \#12210.}

\author{\mbox{Yiping Shu\altaffilmark{1}}}
\author{\mbox{Adam S. Bolton\altaffilmark{1}}}
\author{\mbox{Joel R. Brownstein\altaffilmark{1}}}
\author{\mbox{Antonio D. Montero-Dorta\altaffilmark{1}}}
\altaffiltext{1}{Department of Physics and Astronomy, University of Utah,
115 South 1400 East, Salt Lake City, UT 84112, USA ({\tt yiping.shu@utah.edu, bolton@astro.utah.edu})}

\author{\mbox{L\'eon V. E. Koopmans\altaffilmark{2}}}
\altaffiltext{2}{
Kapteyn Astronomical Institute, University of Groningen, PO Box 800, NL-9700 AV Groningen, the Netherlands
}

\author{\mbox{Tommaso Treu\altaffilmark{3, 4}$^{\ddag}$}}
\altaffiltext{3}{
Department of Physics, University of California, Santa Barbara, CA 93106, USA
}
\altaffiltext{4}{
Department of Physics and Astronomy, University of California, Los Angeles, CA 90095-1547, USA
}
\altaffiltext{$^{\ddag}$}{Packard Fellow}

\author{\mbox{Rapha\"el Gavazzi\altaffilmark{5}}}
\altaffiltext{5}{Institut d'Astrophysique de Paris, CNRS, UMR 7095, Universit«e Pierre et Marie Curie, 98bis Bd Arago, 75014 Paris, France}

\author{\mbox{Matthew W. Auger\altaffilmark{6}}}
\altaffiltext{6}{
Institute of Astronomy, University of Cambridge, Madingley Road, Cambridge CB3 0HA, UK
}

\author{\mbox{Oliver Czoske\altaffilmark{7}}}
\altaffiltext{7}{
Institut f\"{u}r Astronomie, der Universit\"{a}t Wien, T\"{u}rkenschanzstra{\ss}e 17, 1180 Wien, Austria
}

\author{\mbox{Philip J. Marshall\altaffilmark{8}}}
\altaffiltext{8}{
Kavli Institute for Particle Astrophysics and Cosmology, Stanford University, 452 Lomita Mall, Stanford, CA 94305, USA
}

\author{\mbox{Leonidas A. Moustakas\altaffilmark{9}}}
\altaffiltext{9}{
Jet Propulsion Laboratory, California Institute of Technology, MS 169-506, 4800 Oak Grove Drive, Pasadena, CA 91109, USA
}

\begin{abstract}

We present observational results from a new Hubble Space Telescope (\textsl{HST}) Snapshot program to extend the methods of the Sloan Lens ACS (SLACS) Survey to lower lens-galaxy masses.  We discover 40 new galaxy-scale strong lenses, which we supplement with 58 previously discovered SLACS lenses. In addition, we determine the posterior PDFs of the Einstein radius for 33 galaxies (18 new and 15 from legacy SLACS data) based on single lensed images. We find a less-than-unity slope of $0.64\pm0.06$ for the $\log_{10} {\sigma}_*$-$\log_{10} {\sigma}_{\rm SIE}$ relation, which corresponds to a 6-$\sigma$ evidence that the total mass-density profile of early-type galaxies varies systematically in the sense of being shallower at higher lens-galaxy velocity dispersions. 
The trend is only significant when single-image systems are considered, highlighting the importance of including both ``lenses'' and ``non-lenses'' for an unbiased treatment of the lens population when extending to lower mass ranges. By scaling simple stellar population models to the \textsl{HST} I-band data, we identify a strong trend of increasing dark-matter fraction at higher velocity dispersions, which can be alternatively interpreted as a trend in the stellar initial mass function (IMF) normalization.  
Consistent with previous findings and the suggestion of a non-universal IMF, we find that a Salpeter IMF is ruled out for galaxies with velocity dispersion less than $180$ km/s. Considered together, our mass-profile and dark-matter-fraction trends with increasing galaxy mass could both be explained by an increasing relative contribution on kiloparsec scales from a dark-matter halo with a spatial profile more extended than that of the stellar component.
\end{abstract}


\slugcomment{Accepted for Publication in the ApJ}

\maketitle

\section{Introduction}

Early-type galaxies (ETGs), classified by their morphology, compose one of the two main categories of galaxies \citep{Hubble26, Hubble36}. Although considered to be relatively \emph{``dead''} and \emph{``featureless''} as a consequence of their little star formation activities and smooth light distributions, ETGs play a crucial role in studying the evolution of galaxies, the nature of dark matter, and cosmology. Believed to be the endproducts of hierarchical merging scenario \citep{TT72, White91, Kauffmann93, Cole2000}, their structures, properties, and formation histories can be used as a compelling test of the $\rm \Lambda$ Cold Dark Matter ($\rm \Lambda CDM$) paradigm. Additionally, ETGs can be extremely luminous and therefore can be used as powerful cosmological tracers of the large-scale structure \citep{Eisenstein05, Percival07, Anderson12}. 

However, the formation and evolution of ETGs are still puzzling and further investigations are highly demanded. Concerning the mass-density profile of ETGs, N-body DM-only numerical simulations have revealed a somewhat ``universal'' density profile with a $r^{-1}$ inner profile and a $r^{-3}$ drop-off at large radii, independent of the halo mass \citep{NFW96, NFW97}. Later on, various observations of DM-dominated galaxies yield inconsistent inner density slopes with numerical simulations \citep{Moore99, Graham06, Navarro10}, the tension of which can be loosened by taking baryonic physics into account. Gas cooling permits baryons to condense in the central regions of galaxies, and therefore it is believed to make the mass distribution more centrally concentrated \citep[e.g.][]{Gnedin04, Gustafsson06, Abadi10, Velliscig14}. Heating due to dynamical friction and supernovae (SN)/Active Galactic Nucleus (AGN) feedback, in contrast, can soften the central density concentration \citep[e.g.][]{Nipoti04, Romano-Diaz08, Governato10, Duffy10, Martizzi12, Dubois13, Velliscig14}. The strength of these competing effects differs from galaxy to galaxy and hence studying the dependences of the shape of the mass-density profile in the central region on galaxy mass, redshift, and other structural quantities unravels the formation and evolution of ETGs. 

The stellar initial mass function (IMF) is an empirical relation quantifying the relative fraction of stars as a function of the stellar mass at the time when the whole population formed. \citet{Salpeter55} first quantified the IMF as a simple power-law function using main-sequence stars in the solar neighborhood. Later on, various modifications have been considered at the low-mass end and the most commonly used forms are the Kroupa IMF \citep{Kroupa01} and the Chabrier IMF \citep{Chabrier03}. The IMF of a galaxy could depend on the environmental properties of the molecular cloud it originated from such as metallicity, temperature, and density, and could therefore be non-universal. Having knowledge of the form and the variation of the IMF provides deep insights in understanding the role of the environment during star formation and galaxy evolution processes. Recently, several pieces of evidence suggest that the IMF indeed varies \citep[e.g.][]{Treu10, Auger10, vanDokkum10, Strader11, Cappellari12, Sonnenfeld12, Spiniello12, Ferreras13, LaBarbera13, Conroy13, Tortora13, Brewer14, Spiniello14}. 

Strong gravitational lensing (GL) has its unique power among the many techniques for the study of ETGs. As a pure gravity-dependent effect, GL provides highly accurate measurements of total mass that are robust against different models and assumptions about galaxy properties. Therefore, it provides the best estimation of the total projected mass within the so-called \emph{Einstein radius} enclosed by the lensed images of the background object. 

One of the known issues in estimating the enclosed mass by GL is the mass-sheet degeneracy (MSD), in which a proper mass-density transformation preserves the dimensionless lensing observables such as image positions, shapes, flux ratios, etc. \citep[e.g,][]{Falco85, Gorenstein88, Narayan96, Schneider13}. Given lensing data alone, breaking this degeneracy requires some additional physical assumptions. However, following \citet{SLACSVIII}, a qualitative estimations of the effect of MSD on our final results are presented in Section 4. Further information including stellar kinematics can fully break the MSD degeneracy. 

Various lensing surveys have been conducted in the past decade and led to numerous important results. The Lenses Structure and Dynamics (LSD) Survey aimed to measure the stellar kinematics of a small sample of E/S0 galaxy lenses and combine it with GL to constrain the central mass distribution \citep{KT02, KT03, TK02a, TK04}. The SLACS survey \citep{SLACSI, SLACSII, SLACSIII, SLACSIV, SLACSV, SLACSVI, SLACSVII, SLACSVIII, SLACSIX, SLACSX, SLACSXI} is by far the most productive survey for galaxy-scale strong lenses with known lens and source redshifts, with a discovery of over 90 spectroscopically-selected lenses confirmed by high-resolution Hubble Space Telescope (\textsl{HST}) follow-up.  SLACS observes relatively low-redshift ($z_L \lesssim 0.4$) ETG lens candidates selected from the Luminous Red Galaxy \citep[LRG,][]{Eisenstein2001} and MAIN \citep{Strauss2002} galaxy samples of the Sloan Digital Sky Survey \citep[SDSS, ][]{SDSSI}. The SLACS survey has yielded multiple novel results on the structure and dynamics of ETGs, which are detailed in the previous papers of this series.  Recently, the technique of spectroscopic lens selection has been extended to earlier cosmic time (higher redshift) by the BOSS Emission-Line Lens Survey \citep[BELLS,][]{Brownstein2012}, which has confirmed 25 strong lenses ($0.4 \lesssim z_L \lesssim 0.7$) using data from the Baryon Oscillation Spectroscopic Survey \citep{Dawson2013} of the SDSS-III \citep{SDSSIII}\@. 

The SLACS sample is a unique resource for studies of the structure of ETGs. As studied by \citet{Arneson12}, no significant bias in the mass axis ratio relative to the parent population is detected in spectroscopically selected lens samples (See also \citet{Sonnenfeld14}). And the selection bias in the mass-density profile slope is rather small, estimated by \citet{BELLSII} to be on the order of $\sim 0.01$ for a combined sample of SLACS and BELLS lenses. However, it is known that the SLACS selection function favors the high-mass end due to several related factors. First, strong lensing cross section (an approximation of the lensing possibility in general) increases with the lens galaxy mass, so high-mass ETGs are more likely to act as strong lenses. Second, even if a low-mass galaxy acts as a strong lens, the characteristic angular separation of the lensed images will be small and hard to resolve even at space-based imaging resolution. Third, low-mass galaxies can be intrinsically too faint to be selected for SDSS spectroscopy. Fourth, for the preceding reasons, high-mass SLACS candidates have been prioritized for \textsl{HST} follow-up during previous SLACS programs, in order to maximize the survey success rate.

In order to extend the power of strong lensing to low-mass galaxies, an extension of the SLACS survey known as ``SLACS for the Masses'' (hereafter S4TM, \textsl{HST} Snapshot Program 12210) was initiated in April 2012 with a focus on lens candidates with lower masses and smaller predicted Einstein radii as compared to SLACS lenses.  While the lensing confirmation rate of S4TM is lower than that of previous SLACS \textsl{HST} programs, it importantly achieves a wider lens-mass baseline in combination with previous SLACS lenses.  
Note that the Sloan WFC Edge-on Late- type Lens Survey \citep[SWELLS,][]{SWELLSI} also probes low-mass lenses albeit with a focus on spiral galaxies.
We refer the readers to the S4TM catalog paper by Brownstein et al.\ 2015 (in preparation) for a full description of the S4TM program details. In this paper, we present the first scientific results on the total mass-density profile and dark-matter content of an extended ETG sample combining the S4TM lenses and previous SLACS lenses.  We use a hierarchical Bayesian method to infer the mass-profile scaling relation of the combined lens sample, and estimate stellar masses through single stellar-population (SSP) model scalings to the observed \textsl{HST} photometry.

This paper is organized as follows. In Section 2, we briefly describe our lens identification technique using the SDSS spectroscopy and high-resolution imaging data observed by the \textsl{HST}\@. Section 3 describes our parametric lens-modeling method. We then derive the main findings in Sections 4 and 5 with regards to the study of the mass-density profile and dark-matter fraction of the ETGs. Discussion and conclusions are presented in Section 6. Throughout the paper, we assume a standard cosmology with $\rm \Omega_m = 0.274$, $\rm \Omega_{\Lambda} = 0.726$ and $\rm H_0 = 70\,km\,s^{-1}\,Mpc^{-1}$ \citep[WMAP7, ][]{WMAP7}. 

\section{Lens Candidate Identification}

The S4TM survey is a snapshot program designed to extend strong gravitational lensing observations toward lower masses and relatively smaller Einstein radii as compared to previous SLACS programs. Using the same lens searching technique as SLACS, 137 lens candidates were identified from the seventh data release (DR7) of the SDSS \citep{DR7} and awarded as \textsl{HST} snapshot targets in Observing Cycle 18. The details about the lens selection technique can be found in the papers by \citet{SLACSI, SLACSV} and  Brownstein et al.\ 2015 (in preparation).  The basic approach is to search for high-redshift emission lines such as [O\textsc{ii}] doublets, H$\beta$, and [O\textsc{iii}] superimposed on the spectra of SDSS target galaxies at lower redshifts. Such emission lines, associated with star-forming galaxies more distant along the same line of sight, indicate the presence of a candidate lensing system, and also allow us to simultaneously determine the redshifts of the background objects. 

Between 2010 September and 2012 June, 118 out of 137 candidates  were successfully observed with an exposure time of 420\,s each with a single exposure through the F814W filter of the Wide Field Channel (WFC) of the Advanced Camera for Surveys (ACS).  The images were visually inspected using the \emph{ACSPROC} software, a GUI tool implemented by \citet{Brownstein2012}. By searching for lensed features in the b-spline-subtracted residual images \citep{SLACSI, SLACSV}, we have confirmed 40 strong gravitational lenses with clear and definite multiple lensed images or even complete Einstein rings (classified as ``grade-A''), 8 systems with strong evidence of multiple imaging but insufficient SNR for definite conclusion and/or modeling (classified as ``grade-B''), as well as 18 systems showing clear images of the background objects but no clear counter-images (classified as ``grade-C''). 
We exclude one grade-C lens (\emph{$\rm SDSSJ1310+0220$}) from now on as it turned out to be a face-on late-type galaxy with strong emission lines after an examination of its SDSS spectrum. 

\section{Lens Modeling and Sample Definition}

For the foreground lens galaxies, we consider a singular isothermal ellipsoid (SIE) lens model \citep{Kassiola93, Kormann94, Keeton98, SLACSV} that is generalized from a singular isothermal sphere (SIS) model in which the 2D surface mass density falls off as $R^{-1}$, but consists of elliptical iso-density contours specified by  position angle $P.A.$ and minor-to-major axis ratio $q_{\rm SIE}$. 
We do not include external shear in our lens model as it has been shown by \citet{SLACSIII}, \citet{SLACSVIII}, and confirmed again using our grade-A lens sample to be a minor effect. Discussions on the effect of external shear on the final results are presented in Section~\ref{sect:systematics}.
The SIE lens model is characterized by the lensing strength $b_{\rm SIE}$ (specified according to an ``intermediate axis'' convention for elliptical models). 
\begin{equation}
b_{\rm SIE} = 4 \pi \frac{\sigma_{\rm SIE}^2}{c^2} \frac{d_{LS}}{d_S}, 
\label{eq:b_SIE}
\end{equation}
where $d_{LS}$ and $d_S$ are the angular diameter distances from the lens and the observer to the source, respectively. 

The relationship between the $b_{\rm SIE}$ Einstein-radius parameter and the ``lens-model velocity dispersion'' $\sigma_{\rm SIE}$ given in Equation~\ref{eq:b_SIE} is, strictly speaking, only valid for the case of circular symmetry.  For the elliptical case, we take Equation~\ref{eq:b_SIE} to be our implicit definition of $\sigma_{\rm SIE}$.  By this convention, $b_{\rm SIE}$ is the intermediate axis (geometric mean of semi-major and semi-minor axes) of the isodenisty contour within which the average convergence $\kappa$ is equal to unity. As discussed by \citet{Huterer05}, an alternative convention is to adopt the radius of a circle (denoted $\theta_{\rm E}$ in that work) within which the average $\kappa$ is unity. We adopt the $b_{\rm SIE}$ convention in this work to enable us to use the empirical relation of \citet[][]{SLACSX} (which adopts the same convention) between mass-density profile and the ratio of stellar to lensing velocity dispersions. We explore the sensitivity of our results to this choice in Section~\ref{sect:systematics} using the ellipicity-dependent $b_{\rm SIE}$-$\theta_{\rm E}$ relation from \citet{Huterer05}, which is accurate to $< 1\%$ on average within the axis-ratio ranges of our lenses. 

The surface brightness distribution of sources are represented by either one or multiple S\'{e}rsic components with the form 
\begin{equation}
I(x, y) \propto {\rm exp}[-\frac{1}{2}(\frac{qx^2+y^2/q}{\sigma^2})^{n/2}], 
\end{equation}
with the axis ratio $q$, width $\sigma$, and exponent $n$ as free parameters. 

For a particular SIE lens model and specific composition of the source, one can generate the predicted lensed images via the ray-tracing technique according to the analytical expressions of the lens equation \citep{Kormann94}. We apply different fitting strategies to grade-A and grade-C lenses as explained in the following subsections. Note that for all the lens imaging data, we rescale the corresponding errors such that the average error of the background matches the standard deviation of the background to correct for possible correlations in the errors caused by image resampling.

\subsection{Lens Modeling: Grade-A Lenses}

\begin{figure}[hbtp]
\centering
\includegraphics[width=0.49\textwidth]{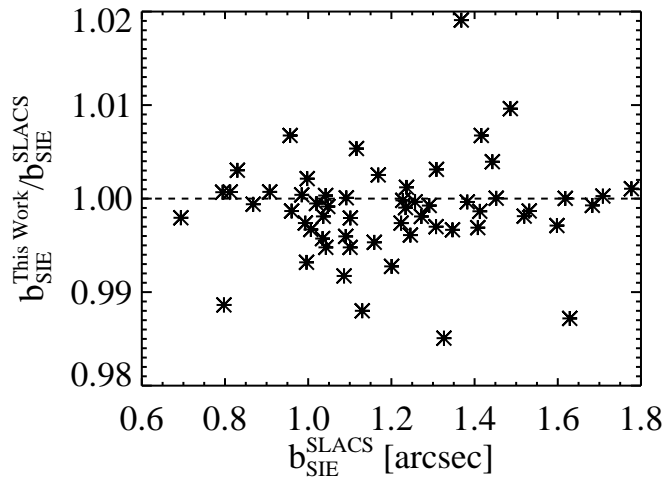}
\caption{\label{fig:theta_comp}
A comparison of the derived Einstein radii between this work and the published SLACS values for 58 grade-A lenses confirmed by the SLACS survey. The ratios are consistent with 1 with a $\rm rms=0.005$ which suggests that our code is robust.}
\end{figure}

For each of the 40 newly discovered grade-A lenses in the S4TM survey, the fitting strategy is relatively straightforward. A Levenberg-Marquardt non-linear least-squares fit \citep[MPFIT,][]{More78, MPFIT} to the observed lensed images is performed to obtain the best-fit parameters for both the lens and the source. Distinct multiple lensed images/rings ensure accurate and unambiguous model fits. We use the position angle $P.A._*$ and axis ratio $q_*$ extracted from the b-spline fit to the light distribution as initial guesses. The starting value for $b_{\rm SIE}$ is determined from the separation between a lensed image and its counter-image. Depending on the configuration of the lensing features, one or multiple source components are considered to ensure a reasonably good fit. 

We also apply the same fitting strategy to 58 grade-A lenses found in the SLACS survey, and show the results in Figure~\ref{fig:theta_comp}. It is clear that the derived Einstein radii of this work are consistent with the published values in \citet{SLACSV} with an average scatter of $0.5\%$. In subsequent analyses, we adopt this scatter as the fractional uncertainty in the measured Einstein radii for grade-A lenses since the statistical errors significantly underestimate the uncertainties as discussed by \citet{SLACSV}.

\begin{figure}[htbp]
\includegraphics[width=0.49\textwidth]{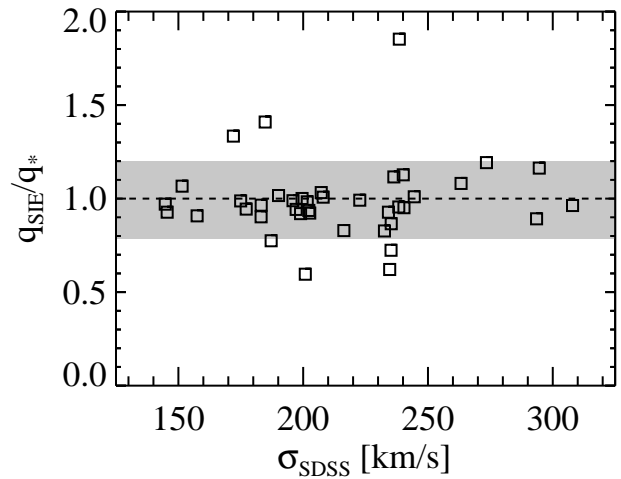}
\caption{\label{fig:light_mass_1}
Ratio between the minor-to-major axis ratio determined from SIE model fitting $q_{\rm SIE}$ and that measured from light distribution $q_*$ as a function of velocity dispersion for 40 S4TM grade-A lenses. The shaded gray region indicates the rms spread (see text for details). }
\end{figure}

\begin{figure}[hbtp]
\includegraphics[width=0.49\textwidth]{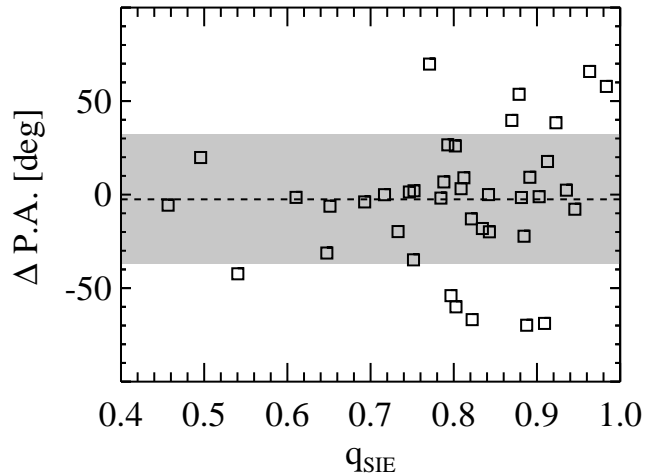}
\caption{\label{fig:light_mass_2}
Discrepancy in the position angle determined from SIE model fitting $P.A.$ and that measured from light distribution $P.A._*$ as a function of $q_{\rm SIE}$ for 40 S4TM grade-A lenses. The shaded gray region indicates the rms spread (see text for details). }
\end{figure}

Figures~\ref{fig:light_mass_1} and \ref{fig:light_mass_2} compare the axis ratios and position angles of the light distribution to those of the mass distribution for the grade-A subsample. Figure~\ref{fig:light_mass_1} displays the ratios of the minor-to-major axis ratio as a function of the SDSS stellar velocity dispersion $\sigma_{\rm SDSS}$ which is consistent with $1.0$ with a rms scatter of $0.2$. No correlation with the velocity dispersion (an approximation of the total mass) is observed. Figure~\ref{fig:light_mass_2} visualizes the difference in the position angle $\Delta P.A. = P.A. - P.A._*$ of lenses with respect to the axis ratios. $\langle \Delta P.A. \rangle = -2^\circ$ with a rms spread of $34^{\circ}$. Clearly, as either $q_{\rm SIE}$ or $q_{*}$ goes to $1$, the position angle becomes ill-determined and the scatter increases significantly. Note that $| \Delta P.A. |$ are smaller than $31^{\circ}$ for all but one S4TM grade-A lens with both $q_{*}$ and $q_{\rm SIE}$ less than $0.8$. Therefore, in general, the hypothesis of light tracing mass is valid in terms of both the match of isophotal and isodensity contours and the position-angle alignment and indicates little external perturbing potential.  
In general, our SIE lens models with multiple parameterized sources can successfully recover the overall lensing features as well as small details. 

\subsection{Lens Modeling: Grade-C Lenses}
\label{sect:grade-C}

In comparison with ``grade-A'' lenses, the systems we refer to as ``grade-C lenses'' (i.e., without counter-images) are less informative about the mass structure of the foreground galaxy (in addition to being less visually striking). 
However, while these grade-C lenses do not provide accurate lens mass measurements, they do provide accurate lens-mass \textit{probability distributions}. 
Furthermore, as will be seen later, grade-C systems are relatively less massive, and hence, the lens-mass probability distributions that they provide are an essential element of our program to extend strong-lensing science to lower lens masses. 
Indeed, the inclusion of such single-image systems to a lens ensemble analysis makes the selection function less sharply dependent on lens galaxy mass as compared to the grade-A lens sample alone.

Nevertheless, in most previous gravitational lensing studies, grade-C lens systems have been ignored because of the difficulty or impossibility of obtaining definite lens mass models.  (A significant counterexample is the use of lensing "flexion" to constrain mass models in systems that allow this technique: \citet{Goldberg07}.)  This difficulty is particularly pronounced within pure imaging surveys for lenses, because background galaxies with no clear multiple imaging can easily be confused with satellite galaxies of the foreground lens.  For the case of the S4TM program, however, we have a strong prior for the identification of singly imaged background galaxies due to the original spectroscopic detection of a second redshift along the line of sight.

In this work, we explicitly incorporate grade-C lenses into our analysis, by first determining their posterior probability density functions (PDFs) of the Einstein-radius, and subsequently incorporating these PDFs into our ensemble analysis of the scaling relations of the lens population via a hierarchical Bayesian method. In order to obtain the posterior PDFs of $b_{\rm SIE}$, we utilize the \textsc{MultiNest} package \citep{Feroz08, Feroz09, Feroz13, Buchner14} which is a Bayesian inference tool designed to efficiently compute the evidence of models in high dimensions through nested sampling. For highly complex multimodal distributions with pronounced degeneracies, such as the grade-C lens modelings in our case, it has been demonstrated that \textsc{MultiNest} significantly outperforms traditional Markov Chain Monte Carlo (MCMC) techniques. We assume uniform priors with appropriate bounds for all the lens and source parameters. In particular, we require the total mass profiles of grade-C lenses to be not flatter than the light distributions by explicitly setting $q_{\rm SIE} \geq q_*$. This special requirement stems from the fact that the fitting code can converge to unphysical lens models with very small axis ratios and small lensing strength for grade-C systems. As shown previously by \citet{SLACSIII, SLACSVII, Barnabe11} and confirmed again by the grade-A subsample in the S4TM survey (Figure~\ref{fig:light_mass}), the hypothesis of a mass quadrupole following the light quadrupole is generally valid, and the requirement of mass being not flatter than light is well-motivated. Figures ~\ref{fig:PPDF_12210_1} and \ref{fig:PPDF_slacs_1} in the Appendix Section show the foreground-subtracted images and posterior PDFs of $b_{\rm SIE}$ for S4TM and SLACS grade-C lenses, respectively.

\subsection{Combined Sample}

We also run \textsc{MultiNest} on grade-A lenses and find their posterior PDFs of $b_{\rm SIE}$ can be well described by extremely narrow Gaussian functions with means consistent with the best-fit $b_{\rm SIE}$ values from MPFIT. Therefore we approximate the posterior PDFs of $b_{\rm SIE}$ for grade-A lenses as delta functions. We then combine both posterior PDFs to perform unbiased analyses of the mass structure of ETGs across a wider range of galaxy masses. In the following sections, we combine these measurements with stellar velocity dispersions and broadband photometry to constrain the mass-density profile and dark-matter fraction of ETGs as a function of galaxy mass.

\begin{figure}[hbtp]
\centering
\includegraphics[width=0.49\textwidth]{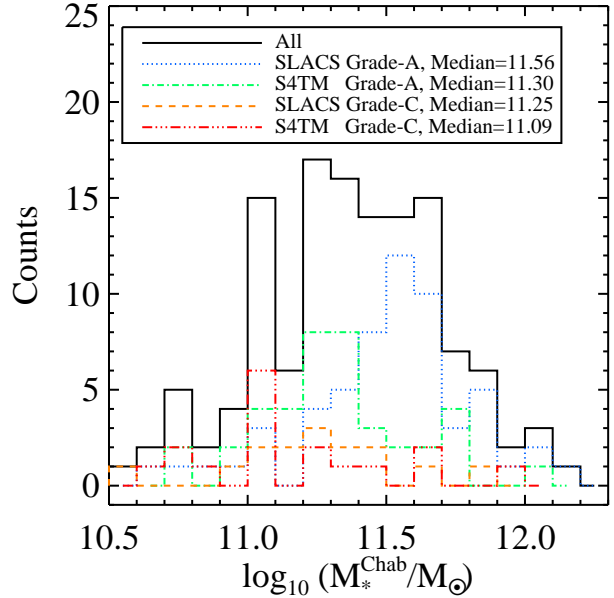}
\caption{\label{fig:hist_SM}
Distributions of the stellar mass $M_*^{\rm Chab}$ for the whole grade-A+C sample (black), SLACS grade-A subsample (blue), S4TM grade-A subsample (green), SLACS grade-C subsample (chocolate), and S4TM grade-C subsample (red) and the median values for each subsample. The stellar mass is derived from the \textsl{HST} I-band photometry by the SPS analysis assuming a Chabrier IMF and other parameters (please refer to Section 5 for details).}
\end{figure}

\begin{figure}[htbp]
\centering
\includegraphics[width=0.49\textwidth]{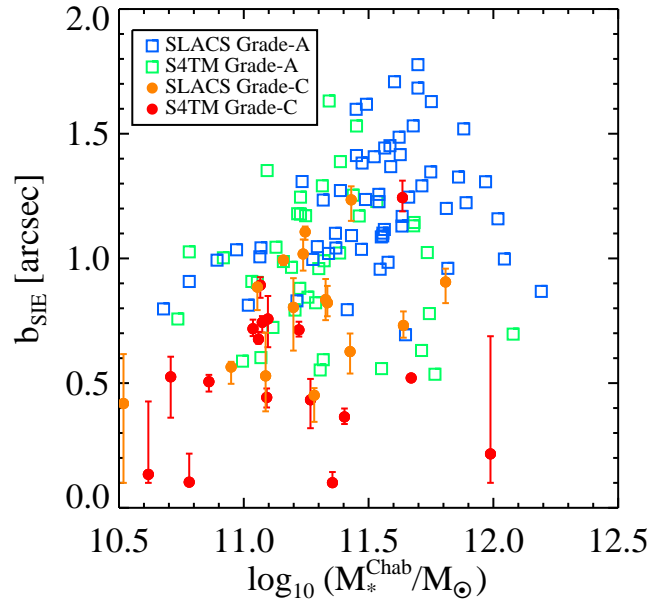}
\caption{\label{fig:distribution}
Distribution of the predicted Einstein radii $b_{\rm SIE}$ and the stellar mass $M_*^{\rm Chab}$ for all the modeled lenses. Open squares represents grade-A lenses, with 40 from the S4TM survey (green) and 58 from the SLACS survey (blue). We do not show uncertainties for grade-A lenses as they have been quantified to be on the level of $0.5\%$ (Figure~\ref{fig:theta_comp}) which are much smaller than the symbol size. Grade-C lenses are filled circles with error bars indicating the 1$\sigma$ uncertainties inferred from the posterior PDFs. 
}
\end{figure}

Combining the S4TM survey and the SLACS survey generates a data set including 98 (40+58) grade-A and 33 (18+15) grade-C ETG lens systems. The mean redshift for the foreground lenses is $\langle z_L \rangle = 0.18$ and $\langle z_S \rangle = 0.58$ for the background sources. The distributions and median values of the stellar masses (derived from the \textsl{HST} F814W photometry assuming a Chabrier IMF and the fiducial stellar-population model of Section 5 below) for various subsamples are also plotted in Figure~\ref{fig:hist_SM}. Note that the S4TM lenses are generally less massive compared to the SLACS lenses and grade-C lenses are generally less massive than grade-A lenses. Figure~\ref{fig:distribution} shows the distribution of the Einstein radius $b_{\rm SIE}$ and stellar mass for all the 131 lenses. For grade-A lenses, only the measurements of $b_{\rm SIE}$ are plotted as the uncertainties have been shown to be on the level of $0.5\%$ which are much smaller than the symbol size. For grade-C lenses, the most likely $b_{\rm SIE}$ values and the 1$\sigma$ uncertainties inferred from the posterior PDFs of $b_{\rm SIE}$ are displayed.

\section{Mass Structure Analysis}

In this section, we combine the posterior PDFs of $b_{\rm SIE}$ for both grade-A and grade-C lenses with lens-galaxy stellar velocity dispersions measured from SDSS spectroscopy in order to constrain the mass-density profile of ETG lenses as a function of lens-galaxy mass. The inclusion of new S4TM systems and grade-C lenses allows us to explore a broader range of galaxy masses than previous studies.

\subsection{Velocity-Dispersion Proxy}

To investigate the degree of central concentration of ETG mass profiles, we employ an observational proxy defined by the ratio of stellar velocity dispersion to the SIE model velocity dispersion $\sigma_{\rm SIE}$, defined in relation to the observable parameter $b_{\rm SIE}$ through Equation~(\ref{eq:b_SIE}). It was suggested by \citet{Kochanek94} that this ratio, later denoted as $f_{\rm SIE} = \sigma_{*} / \sigma_{\rm SIE}$, should be approximately unity for isothermal mass models, greater than unity for models more centrally concentrated than isothermal, and less than unity for models less centrally concentrated.  Successive studies have confirmed this hypothesis and showed that $f_{\rm SIE}$ can be used as an empirical indicator of the mass-density slope \citep{Kochanek2000, TK02a, TK04, SLACSII, SLACSIII, SLACSVII, SLACSVIII, SLACSX}.  We take this approach in the current work in order to investigate physical trends and their significance in the simplest possible manner, and defer a more detailed and self-consistent joint analysis of gravitational lensing and stellar kinematics to future papers.

As mentioned above, the posterior PDFs of $\sigma_{\rm SIE}$ are converted directly from the posterior PDFs of $b_{\rm SIE}$. The other ingredient for our present analysis is the stellar velocity dispersion $\sigma_*$, which is the 
standard deviation of velocities of stars 
within a galaxy. This quantity is determined spectroscopically by measuring the broadening of the galaxy spectrum due to the luminosity-weighted superposition of Doppler-shifted absorption lines from individual stars. Instead of adopting the preexisting SDSS stellar velocity-dispersion values calculated using a set of stellar spectra templates derived from ELODIE stellar spectrum library \citep{ELODIE} with wavelength coverage 4100\AA-6800\AA, we generate a new set of stellar templates from the Indo-US library \citep{Indo-US} patched and extended by selected synthetic spectra in POLLUX database \citep{POLLUX}, covering the full wavelength range of observed spectra. More details about this procedure can be found in \citet{Bolton12, BELLSII}. The resulting velocity dispersions $\sigma_{\rm SDSS}$ are then corrected to values within one half of the half-light radius $\rm R_{half}/2$ (a quantity that is comparable to the lens-galaxy Einstein radius as shown in \citet{SLACSV} and confirmed again by our lens sample) following a compromise prescription between \citet{Jorgensen95, Mehlert03} and \citet{Cappellari06} as
\begin{equation}
\sigma_* = \sigma_{\rm SDSS} \times (\frac{1.5^{\prime \prime}}{\rm R_{half}/2})^{0.05},
\end{equation} 
where $1.5^{\prime \prime}$ is the angular radius of the SDSS fiber. The half-light radius $R_{\rm half}$ is derived from a core-S\'{e}rsic fit \citep{Graham03} to the surface brightness distribution of each lens galaxy as explained in Brownstein et al.\ 2014 (in preparation). 

\subsection{Hierarchical Bayesian Analysis}

Our primary interest here is in the physical scaling relation between $\sigma_*$ and $\sigma_{\rm SIE}$ that encodes the variation of the mass-density profile within the ETG lens population.  In the formalism of statistics, we treat this as a \textit{conditional} probability density function (PDF) of $\sigma_*$ given $\sigma_{\rm SIE}$.

We parameterize the mean scaling relation between $\log_{10}{\sigma}_*$ and $\log_{10}{\sigma}_{\rm SIE}$ as 
\begin{equation}
(\log_{10}{\sigma}_* -2.35) = a \times (\log_{10}{\sigma}_{\rm SIE} - 2.35) + b, 
\label{eq:relation}
\end{equation}
with a slope $a$ and intercept $b$. The constant shift of $2.35$ is introduced to reduce the correlation between $a$ and $b$. And the value itself is chosen to match the average value of $\log_{10}{\sigma}_{*}$. To constrain the parameters of this relation, we use the hierarchical Bayesian method \citep[see e.g.,][]{Shu12, BELLSII, Brewer13, Sonnenfeld13, Sonnenfeld14}.  This method makes full use of all the observed information, deconvolves the observational uncertainties, and offers unbiased estimations of population ``hyperparameters'' that we are interested in. The hierarchical Bayesian approach also allows for straightforward inclusion of the posterior PDFs of $b_{\rm SIE}$ in the analysis.

\begin{table}[tbp]
\centering
\caption{Key ingredients for the hierarchical Bayesian analysis.}
\label{tb:ingredients}
\begin{tabular}{ c l }
\hline
\hline
Symbol & Definition \\
\hline
$x$ & Parameter 1: $\log_{10} \sigma_{\rm SIE}$ \\
$y$ & Parameter 2: $\log_{10} \sigma_{*}$ \\
$I_i$ & Data 1: HST imaging data for lens $i$\\
$y_i$ & Data 2: measured value of $\log_{10} \sigma_{*}$ for lens $i$ \\
$a$ & Hyperpar. 1: slope of $\log_{10}{\sigma}_*$-$\log_{10}{\sigma}_{\rm SIE}$ \\ 
$b$ & Hyperpar. 2: intercept of $\log_{10}{\sigma}_*$-$\log_{10}{\sigma}_{\rm SIE}$ \\ 
$\delta$ & Hyperpar. 3: intrinsic scatter in $\log_{10} \sigma_{*}$ \\ 
$\vec{\theta}$ & Vector of hyperparameters \\
$\mathrm{Pr}(y_i | y)$ & Likelihood function of $y$ given $y_i$ \\
$\mathrm{Pr}(y, x | \vec{\theta}, \mathscr{H})$ & Joint PDF of parameters given $\vec{\theta}$ \\
$\mathrm{Pr}(y | x, a, b, \delta, \mathscr{H})$ & PDF of parameter $y$ given $x$, $a$, $b$, and $\delta$ \\
$\mathrm{Pr}(b_{\rm SIE} | I_i, q, \phi)$ & Posterior PDF of $b_{\rm SIE}$ given $I_i$, $q$, and $\phi$ \\
\hline
\hline
\end{tabular}
\end{table}

Here we describe all the key ingredients in the hierarchical Bayesian analysis. $x$, $y$ represent the two physical parameters $\log_{10} \sigma_{\rm SIE}$, $\log_{10} \sigma_{*}$, and $I_i$, $y_i$ are the corresponding observed data for the \emph{$i^{th}$} lens galaxy. The three hyperparameters (compacted as $\vec{\theta}$) we want to determine are i) $a$: the slope of $\log_{10}{\sigma}_*$-$\log_{10}{\sigma}_{\rm SIE}$ relation; ii) $b$: the intercept; iii) $\delta$: the intrinsic scatter in $\log_{10}{\sigma}_*$. 

Following the same strategy used by \citet{Shu12}, the likelihood function of hyperparameters $\vec{\theta}$ given the observed data $\{\vec{d}\}$ and a hypothesis $\mathscr{H}$ is defined as

\begin{align}
\label{eq:log-likelihood}
\mathscr{L}(\vec{\theta} | \{\vec{d}\} &, \mathscr{H}) \equiv \mathrm{Pr} (\{\vec{d}\} | \vec{\theta}, \mathscr{H}) = \prod_{i=0}^{N} \, \mathrm{Pr}(y_i, I_i | \vec{\theta}, \mathscr{H}) \nonumber \\
& = \prod_{i=0}^{N} \iint \mathrm{Pr}(y_i, I_i | y, x) \mathrm{Pr}(y, x | \vec{\theta}, \mathscr{H}) \mathrm{d}x \mathrm{d}y. \nonumber
\end{align}
In general, $y$ and $x$ are independent variables and $\mathrm{Pr}(y_i, x_i | y, x)$ can be split as
\begin{align}
\mathscr{L}(\vec{\theta} | \{\vec{d}\}, \mathscr{H}) & = \prod_{i=0}^{N} \iint \mathrm{Pr}(y_i | y) \mathrm{Pr}(I_i | x) \mathrm{Pr}(y, x | \vec{\theta}, \mathscr{H}) \mathrm{d}x \mathrm{d}y \nonumber \\
&= \prod_{i=0}^{N} \iint \mathrm{Pr}(y_i | y) \mathrm{Pr}(y | x, a, b, \delta, \mathscr{H}) \mathrm{Pr}(I_i | x) \nonumber \\
& \hspace{1.66cm} \mathrm{Pr}(x) \mathrm{d}x \mathrm{d}y.
\end{align}
The last two terms of the above equation is related to the posterior PDF of $b_{\rm SIE}$ as
\begin{equation}
\mathrm{Pr}(I_i | x) \mathrm{Pr}(x) \mathrm{d}x = \mathrm{Pr}(b_{\rm SIE} | I_i) \mathrm{Pr} (b_{\rm SIE}) \mathrm{d} b_{\rm SIE}.
\end{equation}
Then the likelihood function becomes
\begin{align}
\mathscr{L} (\vec{\theta} | \{\vec{d}\}, \mathscr{H}) = \prod_{i=0}^{N} \iiiint \mathrm{d}y \mathrm{d}q \mathrm{d}{\phi} \mathrm{d}b_{\rm SIE} \, \mathrm{Pr}(y_i | y) \nonumber \\ 
\mathrm{Pr}(y | x, a, b, \delta, \mathscr{H}) \mathrm{Pr} (b_{\rm SIE} | I_i, q, \phi) \mathrm{Pr}_i (q) \mathrm{Pr}_i (\phi).
\end{align}

Looking at each term, $\mathrm{Pr}(y | y_i)$ is the likelihood function of $y$ given $y_i$, which can be written as
\begin{equation}
\label{eq:pdf_y}
\mathrm{Pr}(y_i | y) = \frac{1}{\sqrt{2\pi}\delta_i} \mathrm{exp}[-\frac{(y-y_i)^2}{2\delta_i^2}].
\end{equation}
$\mathrm{Pr}(y | x, a, b, \delta, \mathscr{H})$ is the conditional PDF of $y$ given $x$ and three hyperparameters. Following the parameterized model stated by Equation~(\ref{eq:relation}), $\mathrm{Pr}(y | x, a, b, \delta, \mathscr{H})$ can be expressed as
\begin{align}
\label{eq:pdf_yx}
& \mathrm{Pr}(y | x, a, b, \delta, \mathscr{H}) = \nonumber \\
& \,\,\, \frac{1}{\sqrt{2\pi}\delta} \mathrm{exp}\{-\frac{[a(x-2.35)+b-(y-2.35)]^2}{2\delta^2}\},
\end{align}
with an assumed intrinsic scatter $\delta$ in $y$. $\mathrm{Pr}(y | x, a, b, \delta, \mathscr{H})$ is of physical interest as it quantifies the relation between $y$ and $x$ and can be adopted in related studies. $\mathrm{Pr}(b_{\rm SIE} | I_i, q, \phi)$ is the posterior PDF of $b_{\rm SIE}$ as returned by \textsc{MultiNest} given the imaging data $I_i$, axis ratio $q$, and position angle $\phi$. $\mathrm{Pr}_i (q)$ and $\mathrm{Pr}_i (\phi)$ are the prior PDFs of $q$ and $\phi$ for the $i^{th}$ lens galaxy, which are assumed to be uniform distributions. 

The integration over $y$ can be easily done and yields
\begin{align}
\mathscr{L}
= \prod_{i=0}^{N} \, \iiint \mathrm{d}q \mathrm{d}{\phi} \mathrm{d}b_{\rm SIE} \, & \frac{\exp[-(\Delta-y_i)^2/(2(\sigma_i^2+\delta^2))]}{\sqrt{2\pi (\sigma_i^2+\delta^2)}} \nonumber \\
& \mathrm{Pr} (b_{\rm SIE} | I_i, q, \phi) \mathrm{Pr}_i(q) \mathrm{Pr}_i({\phi}),
\end{align}
where $\Delta = a \times [x(b_{\rm SIE})-2.35]+b+2.35$. 

For grade-A lenses, we can treat $\mathrm{Pr} (b_{\rm SIE} | I_i, q, \phi)$, $\mathrm{Pr}_i(q)$, and $\mathrm{Pr}_i({\phi})$ as delta functions. The likelihood contribution from grade-A lenses is 
\begin{equation}
\ln \mathscr{L}_A = - \sum_{i=0}^{N_A} \{\frac{(\Delta (b_{\rm SIE}^{i})-y_i)^2}{2(\sigma_i^2+\delta^2)} +0.5 \times \ln [2\pi (\sigma_i^2+\delta^2)] \}.
\end{equation}
For the $i^{th}$ grade-C lens, \textsc{MultiNest} generates samples of $b_{\rm SIE}^{ij}$ with derived posteriors (or weights) $\mathrm{Pr} (b_{\rm SIE}^{ij} | I_i, q, \phi)$. The likelihood contribution is therefore approximated as 
\begin{align}
\ln \mathscr{L}_C = \sum_{i=0}^{N_C} \ln \{ \sum_{j=0}^{N} & \frac{\exp[-(\Delta(b_{\rm SIE}^{ij})-y_i)^2/(2(\sigma_i^2+\delta^2))]}{\sqrt{2\pi (\sigma_i^2+\delta^2)}} \times \nonumber \\
& \mathrm{Pr} (b_{\rm SIE}^{ij} | I_i, q, \phi) \}.
\end{align}
The posterior PDF of the hyperparameters that we are interested in is simply related to the likelihood via the \emph{Bayes' rule} as
\begin{align}
\mathrm{Pr}(\vec{\theta} | \{\vec{d}\}, \mathscr{H}) &=& \frac{\mathrm{Pr}(\{\vec{d}\} | \vec{\theta}, \mathscr{H}) \, \mathrm{Pr}(\vec{\theta} | \mathscr{H})}{\mathrm{Pr}(\{\vec{d}\} | \mathscr{H})} \nonumber \\
&=& \frac{\mathscr{L}(\vec{\theta} | \{\vec{d}\}, \mathscr{H}) \, \mathrm{Pr}(\vec{\theta} | \mathscr{H})}{\mathrm{Pr}(\{\vec{d}\} | \mathscr{H})},
\end{align}
from which we can infer the relation between $\log_{10} \sigma_{*}$ and $\log_{10} \sigma_{\rm SIE}$. 

Known as $f_{\rm SIE}$, the ratio of $\sigma_*$ to $\sigma_{\rm SIE}$ has been used as an empirical estimator of the logarithmic mass-density slope $\gamma^\prime$ \citep{SLACSVIII, SLACSX}. Using the $\sigma_{*}$-$\sigma_{\rm SIE}$ relation found above, by definition, we have
\begin{align}
f_{\rm SIE} = \frac{\sigma_*}{\sigma_{\rm SIE}}  = 10^{y-x} & = 10^{[({a}-1) (y-2.35) + {b}]/{a}} \nonumber \\
& = 10^{(a-1)(x-2.35) + b}. 
\label{eq:f_SIE}
\end{align}
The physical interpretation of hyperparameters $a$ and $b$ then become straightforward. If $a$ is exactly unity, $f_{\rm SIE}$ (or equivalently the logarithmic mass-density slope $\gamma^\prime$) is independent of $\sigma_{*}$, an indicator of the mass of lens galaxies. A less-than-unity $a$ indicates an anti-correlation between $\gamma^\prime$ and $\sigma_*$, namely galaxies with smaller/larger $\sigma_*$ are more/less centrally concentrated. $b$ is related to the $f_{\rm SIE}$ value at the mean $\sigma_{\rm SIE}$ value ($\sigma_{\rm SIE} \approx 220$ km/s). The mass-density profile is (approximately speaking) isothermal if $b=0$, sub-isothermal if $b < 0$, and super-isothermal if $b > 0$. 

\begin{figure}[htbp]
  \includegraphics[width=0.49\textwidth]{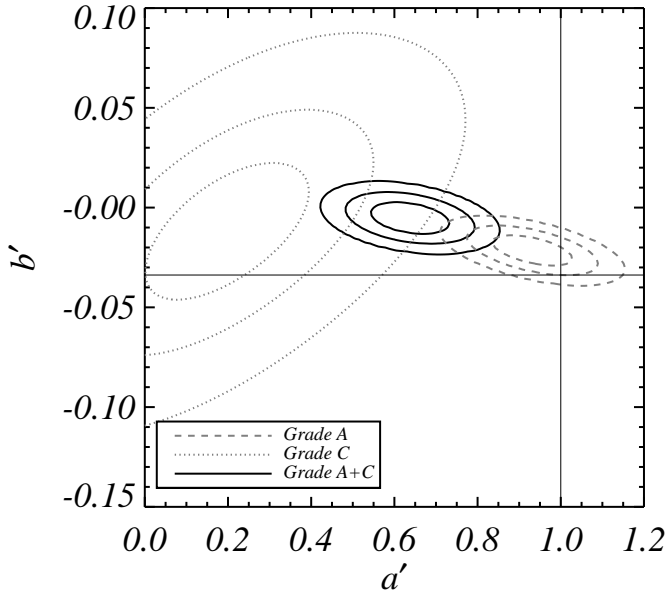}
\caption{\label{fig:prob2d_ab}
The posterior PDFs of $(a, b)$ for the combined sample (solid black contours), grade-A subsample (dashed gray contours), and grade-C subsample (dotted gray contours). $68\%$, $95\%$, and $99.7\%$ confidence levels are plotted accordingly. The horizontal line ($b=-0.0338$ as suggested by the empirical relation in \citet{SLACSX}) corresponds to an isothermal mass-density profile at $\sigma_{\rm SIE} \approx 220$ km/s, and the vertical line ($a=1$) corresponds to a velocity-dispersion-independent mass-density profile.} 
\end{figure}

\begin{figure}[htbp]
\centering
\includegraphics[width=0.49\textwidth]{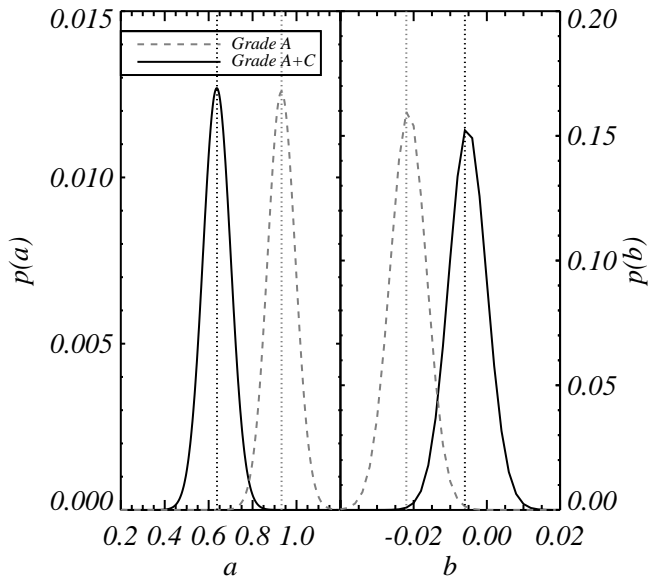}
\caption{\label{fig:prob_ab}
The marginal PDFs of $a$ and $b$ for the combined sample (grade-A+C lenses, solid black curves) and grade-A lens subsample (dashed gray curves). Dotted lines corresponds to the peak positions in the marginal PDFs.}
\end{figure}

Marginalizing over the $\delta$ dimension, the 2D posterior PDF of $a$ and $b$ is shown in Figure~\ref{fig:prob2d_ab}. There are three sets of contours for grade-A subsample (dashed gray), grade-C subsample (dotted gray), and the combined sample (solid black). The marginal PDFs of $a$ and $b$ are obtained by marginalizing over either $b$ or $a$ and shown in Figure~\ref{fig:prob_ab}. The best-estimated values for $a$ and $b$ and the uncertainties are extracted from simple Gaussian fits to the corresponding marginal PDFs, and summarized in Table~\ref{tb:results}. 

\begin{table}[htbp]
\centering
\caption{The best-estimated values of the two hyperparameters $a$ and $b$ derived from the corresponding marginal PDFs for grade-A subsample (2nd row) and grade-A+C sample (3rd row). By including grade-C lenses, the slopes become significant shallower.}
\label{tb:results}
\begin{tabular}{ c | c | c }
\hline
\hline
Lens Sample & $a_{\rm best}$ & $b_{\rm best}$ \\
\hline
Grade A & $0.93 \pm 0.06$ & $-0.021 \pm 0.005$ \\
\hline
Grade A+C & $0.64 \pm 0.06$ & $-0.005 \pm 0.005$ \\
\hline
\hline
\end{tabular}
\end{table}

We see that for the grade-A lens subsample, the overall mass-density profile is essentially consistent with a mass-independent model at about 1.1$\sigma$ ($P(a < 1.0) = 85.49\%$), as found in previous SLACS studies \citep[e.g.,][]{SLACSVII,SLACSX}. Grade-C lenses alone can not provide meaningful constraints on the hyperparameters given such broad posterior PDF distribution. However, the combination of grade-A and grade-C systems in the analysis significantly shifts the grade-A results and provides evidence for a mass-dependent density profile at about 6$\sigma$. The sense of this trend is for lower mass (i.e., lower velocity-dispersion) ETGs to have steeper mass-density profiles and higher mass ETGs to have shallower profiles. 

Note that our current results for grade-A systems are consistent with previous SLACS results of a velocity-dispersion-independent density profile. Similarly, the parameters from our current analysis of grade-A+C systems are substantially consistent with the results of our current grade-A-only analysis. Hence our detection of a velocity-dispersion-dependent density profile in this work is fundamentally due to the inclusion of new data for new lenses covering a wider baseline in mass, and is not an artifact of the new methods we have applied in the analysis of these data.

Consulting the empirical relation between the mass-density profile slope $\gamma^\prime$ and the observable $f_{\rm SIE}$ from \citet{SLACSX}, we find that a value of $b=-0.0338$ corresponds on average to an isothermal profile ($\gamma^\prime = 2$). Considering the best-fit value for $b$ for the grade-A subsample, we find that it is somewhat inconsistent with isothermal at about 2.3$\sigma$.  Including grade-C systems as well, the best $b$ value is more strongly inconsistent with isothermal at $> 5\sigma$.  In both cases, the offset is in the sense of having a slightly super-isothermal profile at the central lensing velocity-dispersion of the samples, consistent with previous SLACS findings \citep{Koopmans09}.

\subsection{Systematic Effects on Mass Structure}
\label{sect:systematics}

As previously mentioned, several issues in the lens modeling procedure have not been fully addressed yet, including the environmental contributions, discrepancy between $b_{\rm SIE}$ and $\theta_{\rm E}$, and the intrinsic nonisothermality of the lens galaxies. In this section, we investigate the individual effects of the above issues on our mass-structure results. 

\subsubsection{Environmental effects}

\begin{figure}[htbp]
\centering
\includegraphics[width=0.49\textwidth]{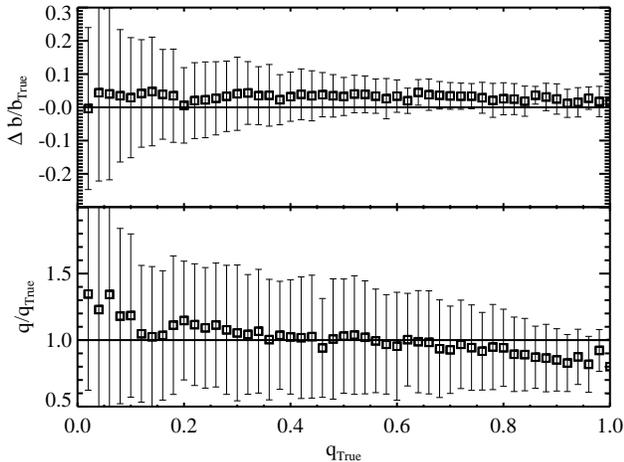}
\caption{\label{fig:shear_verify}
Fractional discrepancy of the Einstein radius (\textit{top panel}) and ratio of axis ratio (\textit{bottom panel}) as a function of the ``true'' axis ratio when fitting mock images generated from an SIE+external shear model by an SIE model (please refer to the text for details).}
\end{figure}

Here we discuss the environmental effects on our mass-structure estimation caused by neglecting possible extra contributions from the environment, i.e., MSD and external shear. Following the recipe in \citet{SLACSVIII}, we write the ``true'' total convergence as 
\begin{equation}
\kappa (x, y) = \frac{\sqrt{q}}{2} \frac{R_{\rm SIE}^0}{\sqrt{x^2 + q^2 y^2}} + \kappa_{\rm ext}, 
\end{equation}
where the first term on the right-hand-side is the contribution from the SIE lens, and the second term is used to quantify the contribution from external convergence (i.e. mass sheet). $R_{\rm SIE}^0$ is the ``true'' Einstein radius for the lens galaxy assuming an SIE model. 

The total projected mass within an ellipse of axis ratio $q$ and semi-minor axis $R^\prime$ is
\begin{align}
M_{\rm total} &= \Sigma_{\rm crit} \iint \kappa (x, y) \mathrm{d}x \mathrm{d}y \nonumber \\
  &= \pi \Sigma_{\rm crit} ({R_{\rm SIE}^0 R^\prime} + {{R^\prime}^2} \kappa_{\rm ext}), 
\end{align}
where $R^\prime = R/\sqrt{q}$ (geometric mean of the semi-major and semi-minor axises), and $\Sigma_{\rm crit}$ is the critical density. Note that here, for first-order approximation, an uniform external convergence $\kappa_{\rm ext}$ is assumed. In the ``intermediate axis'' convention we have been working with, the Einstein radius $R_{\rm SIE}$ we would infer by neglecting the external convergence is given by 
\begin{equation}
R_{\rm SIE}^2 = \frac{M_{\rm total}}{\pi \Sigma_{\rm crit}} = R_{\rm SIE}^0 R_{\rm SIE} + R_{\rm SIE}^2 \kappa_{\rm ext}. 
\end{equation}
Therefore, the ``true'' Einstein radius $R_{\rm SIE}^0$ differs from the value inferred from our SIE model $R_{\rm SIE}$ by a factor of $(1-\kappa_{\rm ext})$
\begin{equation}
\label{eq:external}
R_{\rm SIE}^0 = R_{\rm SIE}(1-\kappa_{\rm ext}). 
\end{equation}

From the scaling law between total projected mass and radius, \citet{SLACSVIII} estimated the external convergence $\kappa_{\rm ext}$ for SLACS lenses is no more than a few percent, and no significant bias was detected. Therefore we expect the effect of neglecting external convergence in our analysis is subdominant to measurement errors. Also as suggested by \citet{BELLSII}, the external convergences are fluctuations about the mean densities, and only introduce random noise rather than bias, which will be smoothed out within our analysis. 

The external shear $\gamma$ that we choose to dismiss turns out to be also a minor effect. In order to quantify it, we fit an SIE+external shear model with fixed axis ratio and position angle to the S4TM grade-A lenses and compare the resulted Einstein radii to values obtained by an SIE model. The overall fractional discrepancy is on the level of $2\%$, which suggest that the effect is again subdominant. To explore any degeneracy between external shear and axis ratio, we generate a set of mock lensed images from an SIE+external shear model with uniformly distributed axis ratios for the SIE profile. The lens galaxy and source are placed at typical lens and source redshifts, respectively. Typical noise level is also added to the mock images. The amplitude of the external shear is set to the mean shear found by \citet{Holder03} at $0.11$ (also consistent with results in \citet{Wong11}), and the position angle of the external shear is marginalized over. We then fit the mock images using an SIE model, and compare the so-obtained Einstein radii to the fiducial value. The results are shown in Figure~\ref{fig:shear_verify}. It is clear that we are able to recover the Einstein radii within an accuracy of about $4\%$ for the entire axis-ratio range. More importantly, we do not observe any significant correlation between the discrepancy in the Einstein radius and the axis ratio. 

\subsubsection{$b_{\rm SIE}$ convention VS. $\theta_{\rm E}$ convention}

As previously mentioned, the $b_{\rm SIE}$ convention for an elliptical case is generally not identical to the $\theta_{\rm E}$ convention for a circular case. The conversion between $b_{\rm SIE}$ and $\theta_{\rm E}$ has been derived in \citet{Huterer05} as a function of ellipticity (or axis ratio $q$ in our model). By definition, the deviation from $\theta_{\rm E}$ converges to 0 as $q$ approaches 1. Here we use all the 98 grade-A lenses to explore the sensitivity of our final results to the different conventions. 

We first convert $b_{\rm SIE}$ obtained by the SIE modeling to $\theta_{\rm E}$ following \citet{Huterer05}, and then incorporate $\sigma_{\rm SIE}$ predicted by $\theta_{\rm E}$ into the Bayesian analysis. Given the facts that the mean axis ratio of the grade-A sample is $0.789$ and there is no significant correlation between $\sigma_{*}$ and $q$, converting $b_{\rm SIE}$ to $\theta_{\rm E}$ will only introduce an almost constant offset of $-0.002$ on average in $\log_{10} \sigma_{\rm SIE}$, which consequently gets propagated into an offset in $b$, but leaves the slope $a$ untouched. This expectation has been verified as we find that the posterior PDF of $a$ and $b$ is almost identical except for a tiny positive shift ($\sim 0.001$) in the marginal PDF of $b$. 

We think the grade-A subsample is representative of the entire sample in terms of the axis-ratio distribution. Therefore, we conclude that the choice of different conventions in the analysis is only a minor effect on the final mass structure results. 

\subsubsection{Intrinsic nonisothermality}

It has been suggested in this paper that the intrinsic mass profile is a varying function of galaxy mass, being steeper than isothermal in lower-mass galaxies. This deviation from isothermality confirms recent findings by \citet{Sonnenfeld13, Dutton14}, and \citet{Sonnenfeld14}. The effect of neglecting the intrinsic nonisothermality in the SIE modeling needs to be investigated. In order to do that, we fit 40 S4TM grade-A lenses by a singular power-law ellipsoid (SPLE) model for 5 discrete logarithmic mass-density slopes $\gamma^{\prime}$ from $1.5$ to $2.5$, and compare the so-obtained lensing strength $b_{\rm SPLE}$, which we think is a better approximation for the lensing velocity dispersion, to $b_{\rm SIE}$. We find that the majority of them suggests a correlation between the deviation from the ``true'' value defined as $b_{\rm SIE}-b_{\rm SPLE}$ and logarithmic slope $\gamma^{\prime}$. Following the mass-structure trend observed in this paper, it indicates that we tend to overestimate $\sigma_{\rm SIE}$ in low-mass galaxies ($\gamma^{\prime} > 2$) and underestimate $\sigma_{\rm SIE}$ in high-mass galaxies ($\gamma^{\prime} < 2$). That would in turn yield an even more pronounced mass-structure trend than we have seen. We defer further quantitative studies to future papers. Nevertheless, we confirm that neglecting the intrinsic nonisothermality can not alleviate the detected mass dependence of the total mass-density profile in the central region. 

\section{Stellar Masses and Initial Mass Function}

In addition to allowing measurements of the shape of the lens mass-density profile, strong lensing data can be combined with photometry and stellar-population diagnostics to constrain the dark-matter fraction and/or stellar IMF in the lens \citep[e.g.][]{Treu10, Ferreras10, Spiniello11, Grillo11, Spiniello12, Sonnenfeld12, Brewer14}. For this purpose, we estimate stellar masses of lens galaxies based on scaling of SSP models to \textsl{HST} I-band photometry under a range of stellar-population assumptions, and adopting either a Chabrier or Salpeter IMF.  This simplicity is motivated by uniformity, since although multi-band \textsl{HST} photometry is available for many of the SLACS lenses, all the new S4TM systems currently have I-band data alone.  High-resolution \textsl{HST} imaging is essential to masking the contribution from lensed background sources when performing lens-galaxy photometric modeling (Brownstein et al.\ 2014 in preparation).  For this reason, we disregard multi-band SDSS photometric magnitudes in our analysis.

\begin{figure*}[htbp]
\centering
\includegraphics[width=0.9\textwidth]{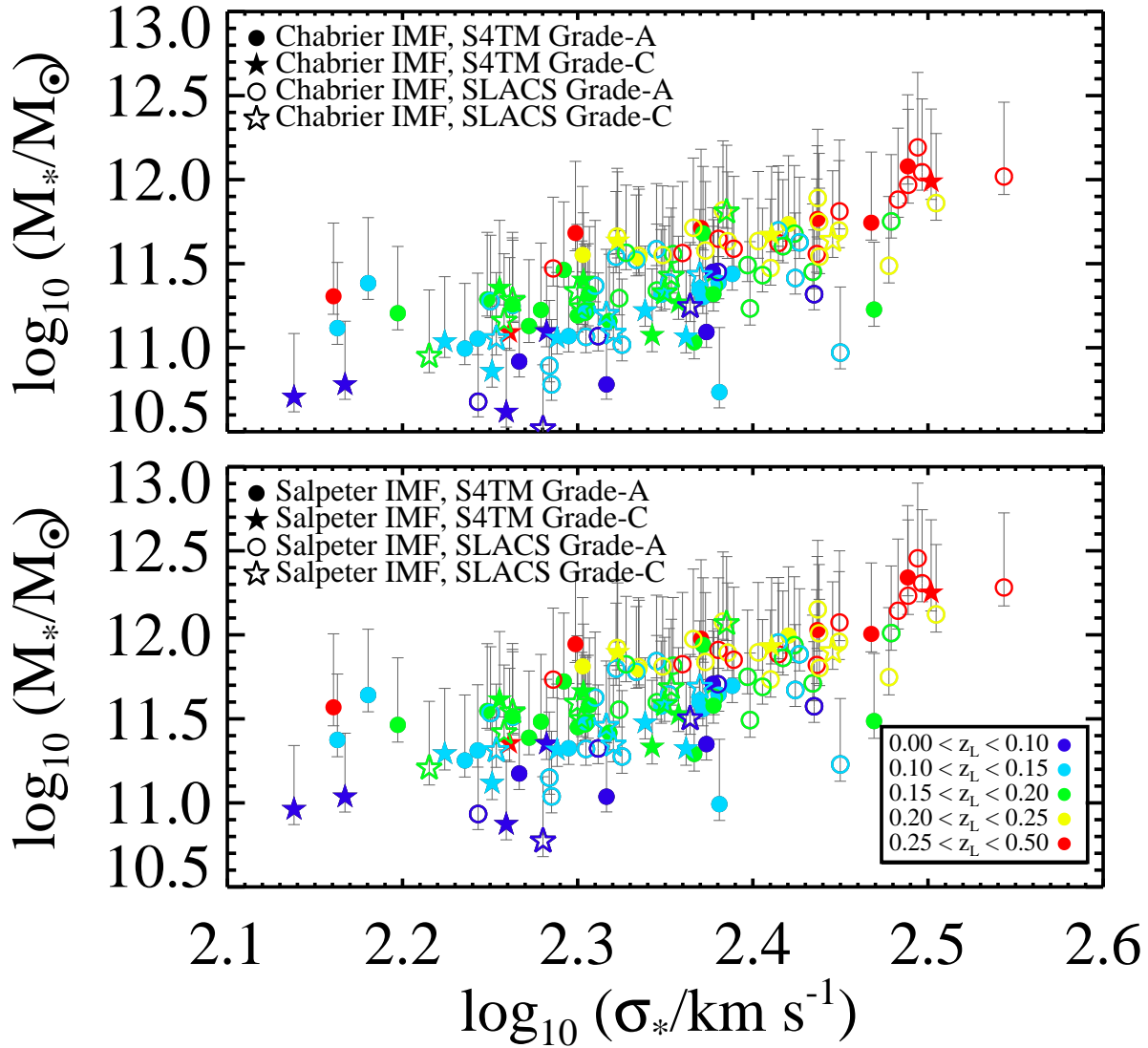}
\caption{\label{fig:M_*}
Stellar masses $M_{*}$ of the 130 lenses inferred from the SPS analysis for the two IMFs as a function of the stellar velocity dispersion. Filled symbols are for the S4TM lenses and open symbols for the SLACS lenses. Grade-C lenses are shown by stars and grade-A lenses are circles. Colors represent the redshifts of the lens galaxies. Gray error bars represent the systematic variations (see text for details on how to determine the error bars). The correlation of scatter with redshift is primarily driven by Malmquist bias in the parent samples.}
\end{figure*}

\begin{figure*}[htbp]
\centering
\subfigure[]{\includegraphics[width=0.49\textwidth]{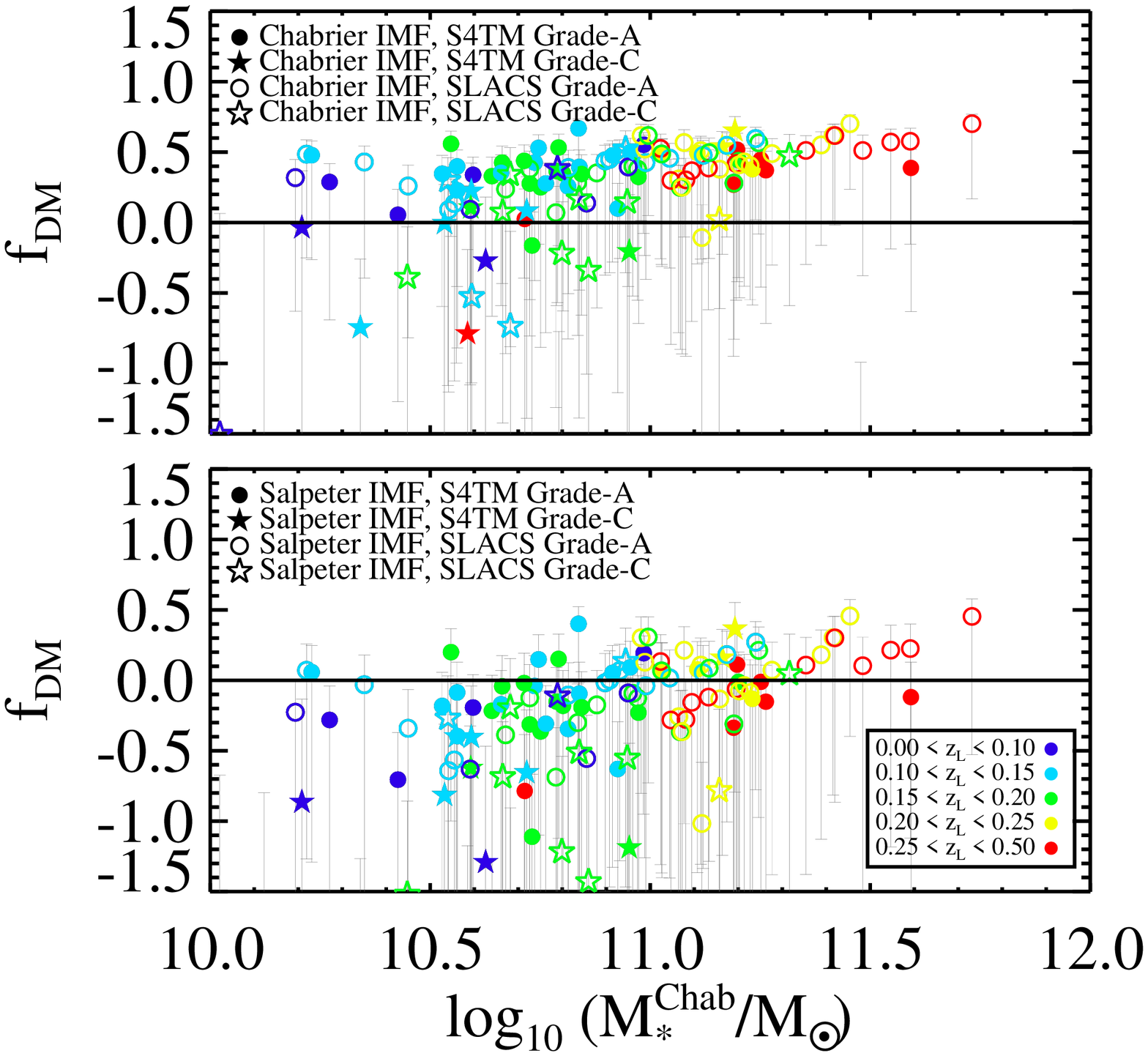} \label{fig:fig10a}}
\hfill
\subfigure[]{\includegraphics[width=0.49\textwidth]{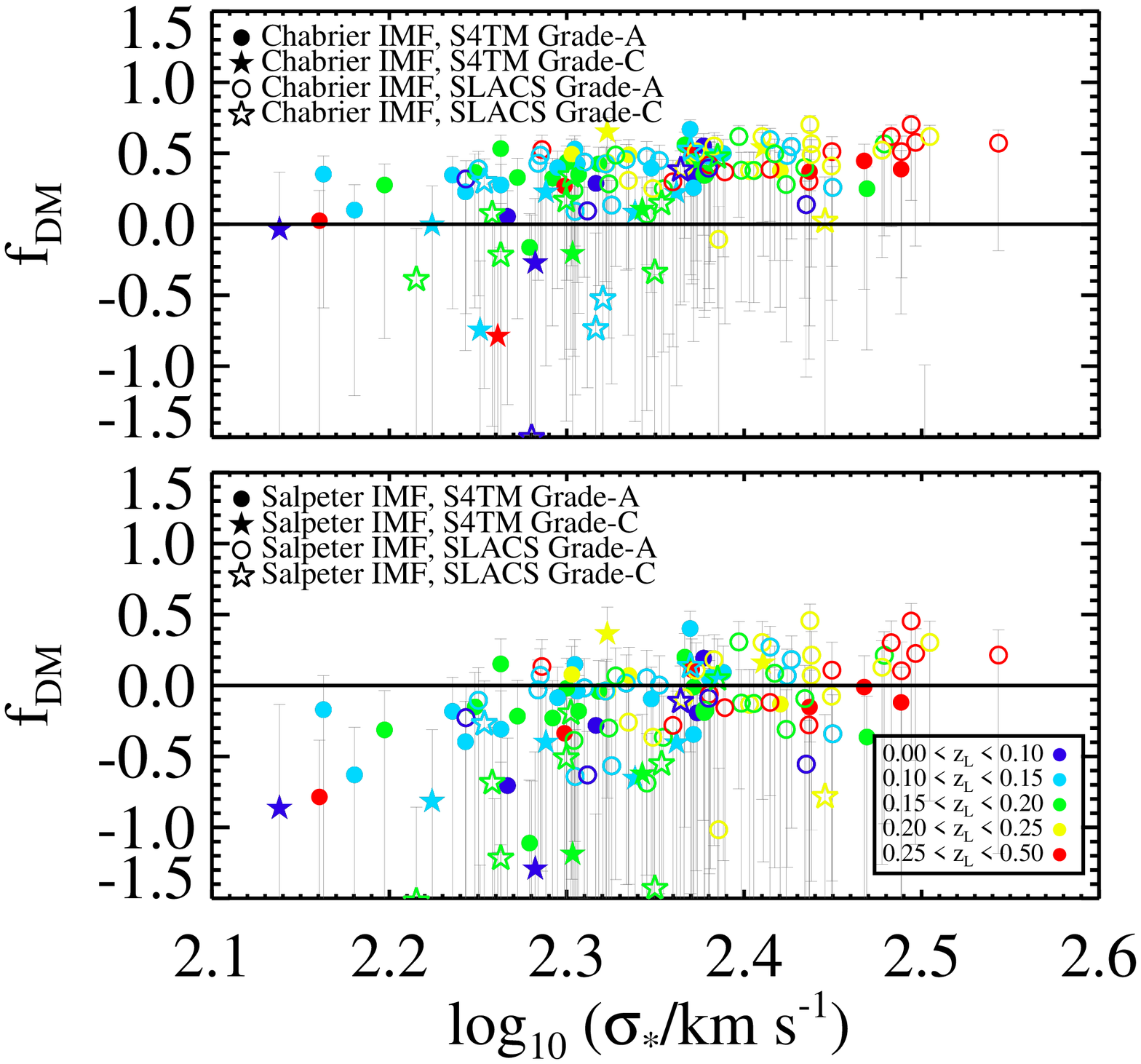} \label{fig:fig10b}}
\hfill
\subfigure[]{\includegraphics[width=0.49\textwidth]{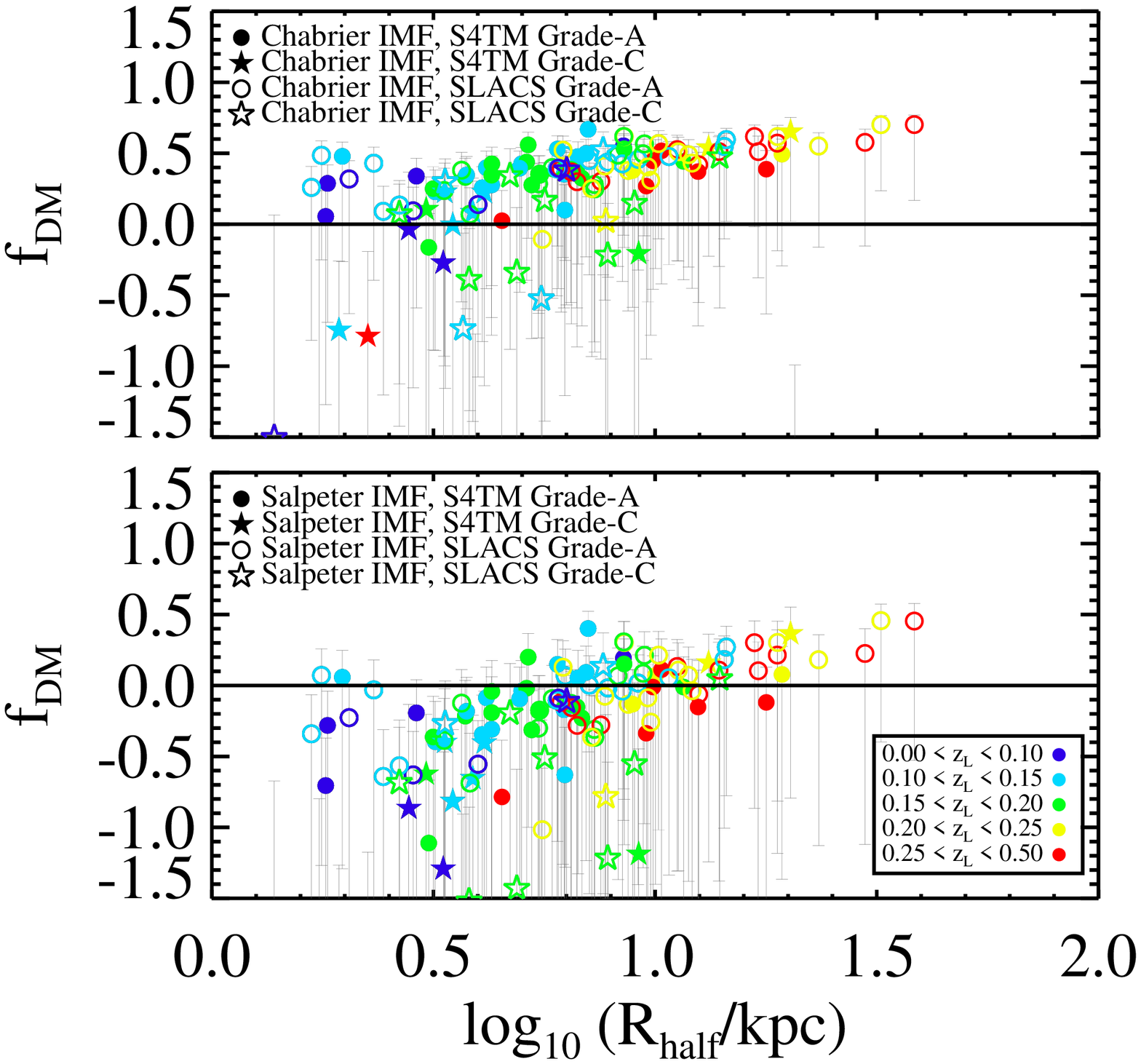} \label{fig:fig10c}}
\caption{\label{fig:f_dm}
Relations between the dark-matter fraction $f_{\rm dm}$ and $\log_{10} M_{*}^{\rm Chabrier}$, $\log_{10} \sigma_{*}$, and $\log_{10} \rm R_{half}$, respectively. Filled symbols represent S4TM lenses while open symbols are for SLACS lenses. Grade-C lenses are shown by stars and grade-A lenses are circles. Colors represent the redshifts of the lens galaxies. Gray error bars represent the systematic variations. Either a Chabrier or Salpeter IMF is considered.}
\end{figure*}

To translate photometry into stellar masses, we make use of SSP models obtained with the Flexible Stellar Population Synthesis \citep[FSPS:][]{Conroy09, Conroy10}.  Without colors or narrow-band indices, we must necessarily make assumptions about population parameters such as the formation time after the Big Bang $t_{\rm form}$, metallicity, and dust in the FSPS code.  Since all the lens galaxies are by selection ETGs at relatively low redshifts, we adopt a reference model with typical values of $t_{\rm form} = 4$ Gyrs, solar metallicity, and no dust \citep{Gallazzi06, Carson10}. We cross-check our stellar mass estimations with the values obtained by \citet{SLACSIX} from multi-band \textsl{HST} photometric data and a Bayesian stellar population analysis approach for 52 confirmed SLACS lenses in common and find good agreement with no bias observed.
To quantify the systematic uncertainty of the simple treatment, we also consider lower- and upper-bound models.  Our lower-bound model is dust-free and metal-poor ($\log_{10} Z/Z_{\sun} = -0.30$), while our upper-bound model is dusty and metal-rich (the optical depth for the dust attenuation $\tau=0.95$, $\log_{10} Z/Z_{\sun} = 0.20$)\footnote{The definition and physical meaning of the dust parameter can be found in \citet{Charlot00, Conroy09}.}. The level of the resulting systematic variation in estimated stellar mass is around 0.5 dex. Figure~\ref{fig:M_*} shows the stellar masses of the 130 lenses from both the S4TM survey (filled symbols) and the SLACS survey (open symbols) for the reference model as a function of the stellar velocity dispersion. In accordance with the well-known Faber-Jackson relation \citep[FJR,][]{FJR}, galaxies with higher velocity dispersions also have higher stellar masses on average.

We next examine the relationships between the projected dark-matter fraction within one half of the half-light radius $f_{\rm dm}$ defined as $f_{\rm dm} \equiv 1-M_* (<\rm R_{half}/2)/M_{\rm R_{half}/2}$ and the stellar mass, the stellar velocity dispersion, and the half-light radius, respectively. The half-light radius $\rm R_{half}$ is determined by the core-S\'{e}rsic fit as the radius within which the enclosed light is one half of the total profile light. The stellar mass within one half of the half-light radius $M_* (<\rm R_{half}/2)$ is interpolated according to the underlying core-S\'{e}rsic profile under the assumption of a constant stellar-mass-to-light ratio for each galaxy. In Figure~\ref{fig:f_dm}, symbols correspond to predictions by the model of $t_{\rm form} = 4$ Gyrs, solar metallicity and dust-free for the two IMFs, respectively. In particular, stars represent grade-C lenses, and circles represent grade-A lenses. Filled symbols are S4TM lenses, while open symbols are SLACS lenses. The colors encode the redshifts of lens galaxies. Gray error bars show the systematic variations in $f_{\rm dm}$ throughout the parameter space as explained above. 
From Figure~\ref{fig:f_dm}, we see general trends toward higher projected dark-matter fractions in galaxies with higher masses, larger velocity dispersions, and bigger sizes, consistent with detections by \citet{SLACSX} using confirmed SLACS lenses. In all cases, the intrinsic scatter in $f_{\rm dm}$ is appreciable. 
More specifically, Panel (a) shows a strong correlation between $f_{\rm dm}$ and the stellar mass $M_*^{\rm Chabrier} (<\rm R_{half}/2)$ for either IMF, which we will thoroughly discuss in Section~\ref{sect:discussion}. The same behaviour can also be seen between Panels (b), which displays the relations between $f_{\rm dm}$ and the stellar velocity dispersion $\sigma_{*}$. Panel (c) suggests that bigger lens galaxies (larger $R_{\rm half}$ values) have larger matter-to-luminousity ratios in their central regions. These can be explained by the idea that the surface density of dark matter varies less slowly (as compared to the luminous matter) as the galaxy becomes more massive. It is also in line with previously detected correlations between $\gamma^{\prime}$ and stellar mass density \citep{Sonnenfeld13, Dutton14}.
Of particular note, by implying a \textit{negative} dark-matter fraction, the data strongly disfavor a Salpeter IMF for large fraction of the lenses, especially for those with stellar velocity dispersions $\sigma_{*}$ smaller than approximately $180$ km/s. This confirms a similar finding from \citet{Brewer12} based on a much smaller number of spiral lens galaxies.

We have several avenues to improve the stellar-mass estimation. First of all, the significant scatter and systematic variation in $f_{\rm dm}$ and the IMF that we haven't taken into account here will weaken any observed trends. A proper way to handle them is heavily required. On the lensing side, a simple SIE model for the total mass distribution as considered in this work, although is a good approximation, is not able to distinguish the contributions from dark and baryonic matter. 
Further physical assumptions need to be made to break the mass-density-profile-IMF degeneracy \citep{Treu10, Oguri13}.
On the SPS side, age, metallicity, dust and other parameters in the SPS models need to be better constrained. Also it has been studied that the SPS technique highly relies on several IMF-sensitive spectral features such as NaI, CaII, FeH, TiO, and CaH1 \citep{Conroy12, Spiniello14}. Improper interpretations in the SPS model, or lack of coverage of these features in the observed spectra can lead to significant systematics in constraining the IMF \citep{Conroy12, Spiniello14}. 

\section{Discussion and Conclusion}

In this paper, we report the discovery of 40 strong gravitational lenses with clear and definite multiple images (classified as ``grade-A'') and another 18 single-image lenses (classified as ``grade-C'') from the S4TM survey \citep[\textsl{HST} Program ID 12210][Brownstein et al. \ 2014 (in preparation)]{}, which by design selects lens galaxies with lower masses and smaller Einstein radii compared to the previous SLACS survey. Along with findings in the SLACS survey, we construct a statistically significant and more complete ensemble of over 100 gravitational lenses, including 98 grade-As and 33 grade-Cs. This combined sample probes ETGs with a mean lens redshift of $\langle z_L \rangle = 0.18$. All lenses have been modeled individually and measurements/posterior PDFs of the Einstein radii have been obtained for grade-A/C lens galaxies appropriately. 

\subsection{Discussion}
\label{sect:discussion}

We have discovered clear evidence for the dependence of the total mass-density profile on galaxy velocity dispersion, in the sense that less massive (lower velocity-dispersion) lens galaxies have more centrally concentrated (super-isothermal) profiles.  We have obtained this result by performing a hierarchical Bayesian analysis of the relation between $\log_{10} \sigma_{*}$ and $\log_{10} \sigma_{\rm SIE}$ for the combined lens ensemble. The inclusion of grade-C lenses is essential to this discovery: the significance of the trend is about 6-$\sigma$ when including grade-A and grade-C lenses together in the analysis, but only about 1-$\sigma$ (i.e., consistent with no trend) when analyzing grade-A lenses alone. This can be attributed both to the fact that the grade-C lenses extend the mass baseline of the measurement to lower masses, and to the fact that excluding grade-C lenses will bias the sample towards higher values of $\sigma_{\rm SIE}$ at fixed $\sigma_{*}$. Note that this trend is in consistent with the overall trend seen going to even higher masses. In fact, in galaxy clusters, the slope is far from isothermal and closer to $\gamma^{\prime}=1$ \citep{Newman13a, Newman13b}.

A trend of mass-density profile $\gamma^\prime$ upon the surface stellar mass density $\Sigma_{*} = M_{*} / (2 \pi R_{\rm eff}^2)$ has been found in \citet{SLACSX}, \citet{Sonnenfeld13}, and \citet{Dutton14} using either SLACS lenses or lenses from the Strong Lensing Legacy Survey \citep[SL2S,][]{Gavazzi12, Gavazzi14}. In particular, they found that galaxies with denser stellar mass densities have steeper profiles. Since the stellar velocity dispersion can be approximated as $\sigma_{*}^2 \propto M_{*}/R_{\rm eff} \propto \Sigma_{*} R_{\rm eff}$, the $\gamma^\prime$-$\sigma_{*}$ relation and $\gamma^\prime$-$\Sigma_{*}$ relation suggest that more massive galaxies are spatially less concentrated and have higher velocity dispersion at the present time. 
This can be understood in terms of the expectations from baryonic physics within DM halos. Dissipative gas cooling processes lead to higher baryon densities in the center of DM halos, and also steepen the DM halo profile through the effect of adiabatic contraction (AC: e.g., \citealt{Blumenthal86})\@. Energetic feedback processes from SN and AGN tend to heat gas and counteract central condensation. These processes compete against each other in their effect on the mass distribution in DM halos. More importantly, the efficiencies of both cooling and feedback processes depend on galaxy mass/velocity dispersion. Our findings suggest that for less massive galaxies, the impact of feedback is less significant as compared to AC/cooling and hence leads to a more centrally concentrated halo. The importance of feedback increases as galaxies become more massive, resulting in shallower density profiles. Metallicity, environment and other processes are also responsible for this competition. In order to determine whether the effect is regulated primarily by velocity dispersion or by stellar density, multi-band data sufficient for detailed stellar-population analysis will be required for the full S4TM sample.

We have applied a simplified SPS analysis to the \textsl{HST} I-band photometric data to estimate the stellar masses of the lens galaxies assuming either a Chabrier or a Salpeter IMF. Age, metallicity and dust have been chosen to match the typical values for passively evolving ETGs at low redshifts. A clear correlation between the projected dark matter fraction and the total mass is observed for both IMFs, consistent with previous findings \citep[e.g.][]{Tortora09, SLACSX, Cappellari12, Conroy13, Brewer14}. There are two possible interpretations of this result: as a true increase in dark-matter fraction with velocity dispersion, or as a trend in the stellar IMF with velocity dispersion.  The first interpretation aligns with the overall expectation of decreased star-formation efficiency with increasing halo mass for halos above $\sim 10^8$\,$M_{\odot}$\citep[e.g.][]{Behroozi10}.  This interpretation could also explain our observed trend in dark-matter profile slope: for a fixed stellar profile shape, an increased fractional mass contribution from a more spatially extended DM halo will result in a shallower total-mass density profile.  Alternatively, a trend towards a more bottom-heavy IMF in more massive galaxies can cause the apparent effect of an increased DM fraction when a single IMF is assumed across all masses \citep[e.g.,][]{Treu10, vanDokkum10, Strader11, Sonnenfeld12, Spiniello12, Ferreras13, LaBarbera13, Geha13, Conroy13, Tortora13, Spiniello14}. Given the limited data set used in this paper, we are not able to distinguish between the two interpretations. Additional assumptions or data are necessary to break this degeneracy \citep[e.g.,][]{Sonnenfeld12, Sonnenfeld14}. However, the Salpeter IMF is in any event disfavored at the low-mass end, since it results in unphysical negative DM fractions.

\subsection{Conclusion and Future Work}

To conclude, in this paper, 
\begin{enumerate}
\item We report the discovery of 40 new grade-A and 33 new grade-C ETG lenses from the S4TM and SLACS surveys. Besides the measurements for grade-A lenses, posterior PDFs of the Einstein radii for grade-C lenses are determined for the first time using the \textsc{MultiNest} tool. Combining with 58 grade-A ETG lenses from the SLACS survey, we construct an ETG lens ensemble with wider mass coverage than previous strong-lens samples;
\item Applying a hierarchical Bayesian method which utilizes the posterior PDFs of the Einstein radii, 
we study the correlation between $\log_{10} {\sigma}_*$ and $\log_{10} {\sigma}_{\rm SIE}$, and find a less-than-unity slope of $0.64\pm0.06$ which corresponds to a significant ($\approx$6$\sigma$) dependence of total mass-density profile on the lens mass in the sense that more massive ETGs possess shallower profiles (as quantified by the ratio of $\sigma_{*}$ to $\sigma_{\rm SIE}$ which serves as a proxy for the logarithmic mass-density profile slope $\gamma^{\prime}$). We have shown that this trend is only significant when grade-C lenses are included (the slope is $0.93\pm0.06$ for the grade-A only subsample), which highlights the importance of grade-C lenses to enabling a wider coverage of lens masses;
\item Stellar masses of lens galaxies are estimated based on their \textsl{HST} I-band photometry and SPS models assuming either a Chabrier or Salpeter IMF. The resulting DM fractions within one half of the half-light radius $f_{\rm DM}$ for each IMF model are found to be strongly correlated with the lens mass/velocity dispersion in the sense that more massive ETGs have larger DM fractions, or alternatively mass-dependent IMFs (or a combination of both effects). A Salpeter IMF is ruled out for ETGs with velocity dispersion smaller than $180$ km/s by implying negative $f_{\rm DM}$.
\end{enumerate}

The analysis of our new S4TM lens sample in combination with other lens samples can be improved with spatially resolved long-slit or integral field spectroscopy in order to determine the two-dimensional stellar kinematics of the lenses, which can in turn enable detailed lensing-plus-dynamical modeling to better constrain the mass-density profile for individual galaxies and eventually break the mass-sheet degeneracy \citep{Barnabe09, McKean10, Vandeven10, Newman11, Barnabe11, Dutton11, Sonnenfeld12}. Multi-band photometry or spectroscopy covering a wide wavelength range from near ultra-violet (NUV) to near infrared (NIR) for the lens galaxies would similarly yield better constraints on the age, metallicity, dust, and other parameters in the lens-galaxy SPS models. Finally, more sophisticated lens models with separate components for dark matter and stars would also improve upon our current single-component total-mass models \citep[e.g.][]{SLACSIV, Dutton11, Barnabe12, Vegetti10, Vegetti12}. With uniform data on a comprehensive lens sample from the SLACS, S4TM, and BELLS surveys covering a wide range in lens redshift ($0.1 < z < 0.7$) and total enclosed mass ($\rm 10^{10} M_\sun < M_{Ein} < 10^{12}M_\sun$), we can fully explore the variation of ETG mass structure across galaxy mass and cosmic time through a joint analysis of strong lensing, stellar dynamics, and stellar populations.

\acknowledgments

The authors thank the anonymous referee for insightful comments and suggestions that substantially improved this paper. The support and resources from the Center for High Performance Computing at the University of Utah is gratefully acknowledged. TT acknowledges support from the Packard Foundation through a Packard Research Fellowship. RG acknowledges support for the Centre National des Etudes Spatiales. The work of LAM was carried out at Jet Propulsion Laboratory, California Institute of Technology, under a contract with NASA. Support for program \#12210 was provided by NASA through a grant from the Space Telescope Science Institute, which is operated by the Association of Universities for Research in Astronomy, Inc., under NASA contract NAS 5-26555.

\bibliographystyle{apj}

\begin{thebibliography}{150}
\expandafter\ifx\csname natexlab\endcsname\relax\def\natexlab#1{#1}\fi

\bibitem[{{Abadi} {et~al.}(2010){Abadi}, {Navarro}, {Fardal}, {Babul}, \&
  {Steinmetz}}]{Abadi10}
{Abadi}, M.~G., {Navarro}, J.~F., {Fardal}, M., {Babul}, A., \& {Steinmetz}, M.
  2010, \mnras, 407, 435

\bibitem[{{Abazajian} {et~al.}(2009){Abazajian}, {Adelman-McCarthy},
  {Ag{\"u}eros}, {Allam}, {Allende Prieto}, {An}, {Anderson}, {Anderson},
  {Annis}, {Bahcall}, \& et~al.}]{DR7}
{Abazajian}, K.~N., {et~al.} 2009, \apjs, 182, 543

\bibitem[{{Anderson} {et~al.}(2012){Anderson}, {Aubourg}, {Bailey}, {Bizyaev},
  {Blanton}, {Bolton}, {Brinkmann}, {Brownstein}, {Burden}, {Cuesta}, {da
  Costa}, {Dawson}, {de Putter}, {Eisenstein}, {Gunn}, {Guo}, {Hamilton},
  {Harding}, {Ho}, {Honscheid}, {Kazin}, {Kirkby}, {Kneib}, {Labatie},
  {Loomis}, {Lupton}, {Malanushenko}, {Malanushenko}, {Mandelbaum}, {Manera},
  {Maraston}, {McBride}, {Mehta}, {Mena}, {Montesano}, {Muna}, {Nichol},
  {Nuza}, {Olmstead}, {Oravetz}, {Padmanabhan}, {Palanque-Delabrouille}, {Pan},
  {Parejko}, {P{\^a}ris}, {Percival}, {Petitjean}, {Prada}, {Reid}, {Roe},
  {Ross}, {Ross}, {Samushia}, {S{\'a}nchez}, {Schlegel}, {Schneider},
  {Sc{\'o}ccola}, {Seo}, {Sheldon}, {Simmons}, {Skibba}, {Strauss}, {Swanson},
  {Thomas}, {Tinker}, {Tojeiro}, {Maga{\~n}a}, {Verde}, {Wagner}, {Wake},
  {Weaver}, {Weinberg}, {White}, {Xu}, {Y{\`e}che}, {Zehavi}, \&
  {Zhao}}]{Anderson12}
{Anderson}, L., {et~al.} 2012, \mnras, 427, 3435

\bibitem[{{Arneson} {et~al.}(2012){Arneson}, {Brownstein}, \&
  {Bolton}}]{Arneson12}
{Arneson}, R.~A., {Brownstein}, J.~R., \& {Bolton}, A.~S. 2012, \apj, 753, 4

\bibitem[{{Auger} {et~al.}(2009){Auger}, {Treu}, {Bolton}, {Gavazzi},
  {Koopmans}, {Marshall}, {Bundy}, \& {Moustakas}}]{SLACSIX}
{Auger}, M.~W., {Treu}, T., {Bolton}, A.~S., {Gavazzi}, R., {Koopmans},
  L.~V.~E., {Marshall}, P.~J., {Bundy}, K., \& {Moustakas}, L.~A. 2009, \apj,
  705, 1099

\bibitem[{{Auger} {et~al.}(2010{\natexlab{a}}){Auger}, {Treu}, {Bolton},
  {Gavazzi}, {Koopmans}, {Marshall}, {Moustakas}, \& {Burles}}]{SLACSX}
{Auger}, M.~W., {Treu}, T., {Bolton}, A.~S., {Gavazzi}, R., {Koopmans},
  L.~V.~E., {Marshall}, P.~J., {Moustakas}, L.~A., \& {Burles}, S.
  2010{\natexlab{a}}, \apj, 724, 511

\bibitem[{{Auger} {et~al.}(2010{\natexlab{b}}){Auger}, {Treu}, {Gavazzi},
  {Bolton}, {Koopmans}, \& {Marshall}}]{Auger10}
{Auger}, M.~W., {Treu}, T., {Gavazzi}, R., {Bolton}, A.~S., {Koopmans},
  L.~V.~E., \& {Marshall}, P.~J. 2010{\natexlab{b}}, \apjl, 721, L163

\bibitem[{{Barnab{\`e}} {et~al.}(2011){Barnab{\`e}}, {Czoske}, {Koopmans},
  {Treu}, \& {Bolton}}]{Barnabe11}
{Barnab{\`e}}, M., {Czoske}, O., {Koopmans}, L.~V.~E., {Treu}, T., \& {Bolton},
  A.~S. 2011, \mnras, 415, 2215

\bibitem[{{Barnab{\`e}} {et~al.}(2009){Barnab{\`e}}, {Czoske}, {Koopmans},
  {Treu}, {Bolton}, \& {Gavazzi}}]{Barnabe09}
{Barnab{\`e}}, M., {Czoske}, O., {Koopmans}, L.~V.~E., {Treu}, T., {Bolton},
  A.~S., \& {Gavazzi}, R. 2009, \mnras, 399, 21

\bibitem[{{Barnab{\`e}} {et~al.}(2012){Barnab{\`e}}, {Dutton}, {Marshall},
  {Auger}, {Brewer}, {Treu}, {Bolton}, {Koo}, \& {Koopmans}}]{Barnabe12}
{Barnab{\`e}}, M., {et~al.} 2012, \mnras, 423, 1073

\bibitem[{{Behroozi} {et~al.}(2010){Behroozi}, {Conroy}, \&
  {Wechsler}}]{Behroozi10}
{Behroozi}, P.~S., {Conroy}, C., \& {Wechsler}, R.~H. 2010, \apj, 717, 379

\bibitem[{{Blumenthal} {et~al.}(1986){Blumenthal}, {Faber}, {Flores}, \&
  {Primack}}]{Blumenthal86}
{Blumenthal}, G.~R., {Faber}, S.~M., {Flores}, R., \& {Primack}, J.~R. 1986,
  \apj, 301, 27

\bibitem[{{Bolton} {et~al.}(2008{\natexlab{a}}){Bolton}, {Burles}, {Koopmans},
  {Treu}, {Gavazzi}, {Moustakas}, {Wayth}, \& {Schlegel}}]{SLACSV}
{Bolton}, A.~S., {Burles}, S., {Koopmans}, L.~V.~E., {Treu}, T., {Gavazzi}, R.,
  {Moustakas}, L.~A., {Wayth}, R., \& {Schlegel}, D.~J. 2008{\natexlab{a}},
  \apj, 682, 964

\bibitem[{{Bolton} {et~al.}(2006){Bolton}, {Burles}, {Koopmans}, {Treu}, \&
  {Moustakas}}]{SLACSI}
{Bolton}, A.~S., {Burles}, S., {Koopmans}, L.~V.~E., {Treu}, T., \&
  {Moustakas}, L.~A. 2006, \apj, 638, 703

\bibitem[{{Bolton} {et~al.}(2008{\natexlab{b}}){Bolton}, {Treu}, {Koopmans},
  {Gavazzi}, {Moustakas}, {Burles}, {Schlegel}, \& {Wayth}}]{SLACSVII}
{Bolton}, A.~S., {Treu}, T., {Koopmans}, L.~V.~E., {Gavazzi}, R., {Moustakas},
  L.~A., {Burles}, S., {Schlegel}, D.~J., \& {Wayth}, R. 2008{\natexlab{b}},
  \apj, 684, 248

\bibitem[{{Bolton} {et~al.}(2012{\natexlab{a}}){Bolton}, {Schlegel}, {Aubourg},
  {Bailey}, {Bhardwaj}, {Brownstein}, {Burles}, {Chen}, {Dawson}, {Eisenstein},
  {Gunn}, {Knapp}, {Loomis}, {Lupton}, {Maraston}, {Muna}, {Myers}, {Olmstead},
  {Padmanabhan}, {P{\^a}ris}, {Percival}, {Petitjean}, {Rockosi}, {Ross},
  {Schneider}, {Shu}, {Strauss}, {Thomas}, {Tremonti}, {Wake}, {Weaver}, \&
  {Wood-Vasey}}]{Bolton12}
{Bolton}, A.~S., {et~al.} 2012{\natexlab{a}}, \aj, 144, 144

\bibitem[{{Bolton} {et~al.}(2012{\natexlab{b}}){Bolton}, {Brownstein},
  {Kochanek}, {Shu}, {Schlegel}, {Eisenstein}, {Wake}, {Connolly}, {Maraston},
  {Arneson}, \& {Weaver}}]{BELLSII}
---. 2012{\natexlab{b}}, \apj, 757, 82

\bibitem[Brewer et al.(2013)]{Brewer13} Brewer, B.~J., 
Foreman-Mackey, D., \& Hogg, D.~W.\ 2013, \aj, 146, 7 

\bibitem[{{Brewer} {et~al.}(2014){Brewer}, {Marshall}, {Auger}, {Treu},
  {Dutton}, \& {Barnab{\`e}}}]{Brewer14}
{Brewer}, B.~J., {Marshall}, P.~J., {Auger}, M.~W., {Treu}, T., {Dutton},
  A.~A., \& {Barnab{\`e}}, M. 2014, \mnras, 437, 1950

\bibitem[{{Brewer} {et~al.}(2012){Brewer}, {Dutton}, {Treu}, {Auger},
  {Marshall}, {Barnab{\`e}}, {Bolton}, {Koo}, \& {Koopmans}}]{Brewer12}
{Brewer}, B.~J., {et~al.} 2012, \mnras, 422, 3574

\bibitem[{{Brownstein} {et~al.}(2012){Brownstein}, {Bolton}, {Schlegel},
  {Eisenstein}, {Kochanek}, {Connolly}, {Maraston}, {Pandey}, {Seitz}, {Wake},
  {Wood-Vasey}, {Brinkmann}, {Schneider}, \& {Weaver}}]{Brownstein2012}
{Brownstein}, J.~R., {et~al.} 2012, \apj, 744, 41

\bibitem[{{Cappellari} {et~al.}(2006){Cappellari}, {Bacon}, {Bureau}, {Damen},
  {Davies}, {de Zeeuw}, {Emsellem}, {Falc{\'o}n-Barroso}, {Krajnovi{\'c}},
  {Kuntschner}, {McDermid}, {Peletier}, {Sarzi}, {van den Bosch}, \& {van de
  Ven}}]{Cappellari06}
{Cappellari}, M., {et~al.} 2006, \mnras, 366, 1126

\bibitem[{{Cappellari} {et~al.}(2012){Cappellari}, {McDermid}, {Alatalo},
  {Blitz}, {Bois}, {Bournaud}, {Bureau}, {Crocker}, {Davies}, {Davis}, {de
  Zeeuw}, {Duc}, {Emsellem}, {Khochfar}, {Krajnovi{\'c}}, {Kuntschner},
  {Lablanche}, {Morganti}, {Naab}, {Oosterloo}, {Sarzi}, {Scott}, {Serra},
  {Weijmans}, \& {Young}}]{Cappellari12}
---. 2012, \nat, 484, 485

\bibitem[{{Carson} \& {Nichol}(2010)}]{Carson10}
{Carson}, D.~P., \& {Nichol}, R.~C. 2010, \mnras, 408, 213

\bibitem[{{Chabrier}(2003)}]{Chabrier03}
{Chabrier}, G. 2003, \pasp, 115, 763

\bibitem[{{Charlot} \& {Fall}(2000)}]{Charlot00}
{Charlot}, S., \& {Fall}, S.~M. 2000, \apj, 539, 718

\bibitem[{{Cole} {et~al.}(2000){Cole}, {Lacey}, {Baugh}, \& {Frenk}}]{Cole2000}
{Cole}, S., {Lacey}, C.~G., {Baugh}, C.~M., \& {Frenk}, C.~S. 2000, \mnras,
  319, 168

\bibitem[{{Conroy} {et~al.}(2013){Conroy}, {Dutton}, {Graves}, {Mendel}, \&
  {van Dokkum}}]{Conroy13}
{Conroy}, C., {Dutton}, A.~A., {Graves}, G.~J., {Mendel}, J.~T., \& {van
  Dokkum}, P.~G. 2013, \apjl, 776, L26

\bibitem[{{Conroy} \& {Gunn}(2010)}]{Conroy10}
{Conroy}, C., \& {Gunn}, J.~E. 2010, \apj, 712, 833

\bibitem[{{Conroy} {et~al.}(2009){Conroy}, {Gunn}, \& {White}}]{Conroy09}
{Conroy}, C., {Gunn}, J.~E., \& {White}, M. 2009, \apj, 699, 486

\bibitem[{{Conroy} \& {van Dokkum}(2012)}]{Conroy12}
{Conroy}, C., \& {van Dokkum}, P.~G. 2012, \apj, 760, 71

\bibitem[{{Dawson} {et~al.}(2013){Dawson}, {Schlegel}, {Ahn}, {Anderson},
  {Aubourg}, {Bailey}, {Barkhouser}, {Bautista}, {Beifiori}, {Berlind},
  {Bhardwaj}, {Bizyaev}, {Blake}, {Blanton}, {Blomqvist}, {Bolton}, {Borde},
  {Bovy}, {Brandt}, {Brewington}, {Brinkmann}, {Brown}, {Brownstein}, {Bundy},
  {Busca}, {Carithers}, {Carnero}, {Carr}, {Chen}, {Comparat}, {Connolly},
  {Cope}, {Croft}, {Cuesta}, {da Costa}, {Davenport}, {Delubac}, {de Putter},
  {Dhital}, {Ealet}, {Ebelke}, {Eisenstein}, {Escoffier}, {Fan}, {Filiz Ak},
  {Finley}, {Font-Ribera}, {G{\'e}nova-Santos}, {Gunn}, {Guo}, {Haggard},
  {Hall}, {Hamilton}, {Harris}, {Harris}, {Ho}, {Hogg}, {Holder}, {Honscheid},
  {Huehnerhoff}, {Jordan}, {Jordan}, {Kauffmann}, {Kazin}, {Kirkby}, {Klaene},
  {Kneib}, {Le Goff}, {Lee}, {Long}, {Loomis}, {Lundgren}, {Lupton}, {Maia},
  {Makler}, {Malanushenko}, {Malanushenko}, {Mandelbaum}, {Manera}, {Maraston},
  {Margala}, {Masters}, {McBride}, {McDonald}, {McGreer}, {McMahon}, {Mena},
  {Miralda-Escud{\'e}}, {Montero-Dorta}, {Montesano}, {Muna}, {Myers},
  {Naugle}, {Nichol}, {Noterdaeme}, {Nuza}, {Olmstead}, {Oravetz}, {Oravetz},
  {Owen}, {Padmanabhan}, {Palanque-Delabrouille}, {Pan}, {Parejko},
  {P{\^a}ris}, {Percival}, {P{\'e}rez-Fournon}, {P{\'e}rez-R{\`a}fols},
  {Petitjean}, {Pfaffenberger}, {Pforr}, {Pieri}, {Prada}, {Price-Whelan},
  {Raddick}, {Rebolo}, {Rich}, {Richards}, {Rockosi}, {Roe}, {Ross}, {Ross},
  {Rossi}, {Rubi{\~n}o-Martin}, {Samushia}, {S{\'a}nchez}, {Sayres}, {Schmidt},
  {Schneider}, {Sc{\'o}ccola}, {Seo}, {Shelden}, {Sheldon}, {Shen}, {Shu},
  {Slosar}, {Smee}, {Snedden}, {Stauffer}, {Steele}, {Strauss}, {Streblyanska},
  {Suzuki}, {Swanson}, {Tal}, {Tanaka}, {Thomas}, {Tinker}, {Tojeiro},
  {Tremonti}, {Vargas Maga{\~n}a}, {Verde}, {Viel}, {Wake}, {Watson}, {Weaver},
  {Weinberg}, {Weiner}, {West}, {White}, {Wood-Vasey}, {Yeche}, {Zehavi},
  {Zhao}, \& {Zheng}}]{Dawson2013}
{Dawson}, K.~S., {et~al.} 2013, \aj, 145, 10

\bibitem[Dubois et al.(2013)]{Dubois13} Dubois, Y., Gavazzi, R., 
Peirani, S., \& Silk, J.\ 2013, \mnras, 433, 3297 

\bibitem[{{Duffy} {et~al.}(2010){Duffy}, {Schaye}, {Kay}, {Dalla Vecchia},
  {Battye}, \& {Booth}}]{Duffy10}
{Duffy}, A.~R., {Schaye}, J., {Kay}, S.~T., {Dalla Vecchia}, C., {Battye},
  R.~A., \& {Booth}, C.~M. 2010, \mnras, 405, 2161

\bibitem[{{Dutton} {et~al.}(2011){Dutton}, {Brewer}, {Marshall}, {Auger},
  {Treu}, {Koo}, {Bolton}, {Holden}, \& {Koopmans}}]{Dutton11}
{Dutton}, A.~A., {et~al.} 2011, \mnras, 417, 1621

\bibitem[Dutton 
\& Treu(2014)]{Dutton14} Dutton, A.~A., \& Treu, T.\ 2014, \mnras, 438, 3594 

\bibitem[{{Eisenstein} {et~al.}(2001){Eisenstein}, {Annis}, {Gunn}, {Szalay},
  {Connolly}, {Nichol}, {Bahcall}, {Bernardi}, {Burles}, {Castander},
  {Fukugita}, {Hogg}, {Ivezi{\'c}}, {Knapp}, {Lupton}, {Narayanan}, {Postman},
  {Reichart}, {Richmond}, {Schneider}, {Schlegel}, {Strauss}, {SubbaRao},
  {Tucker}, {Vanden Berk}, {Vogeley}, {Weinberg}, \& {Yanny}}]{Eisenstein2001}
{Eisenstein}, D.~J., {et~al.} 2001, \aj, 122, 2267

\bibitem[{{Eisenstein} {et~al.}(2005){Eisenstein}, {Zehavi}, {Hogg},
  {Scoccimarro}, {Blanton}, {Nichol}, {Scranton}, {Seo}, {Tegmark}, {Zheng},
  {Anderson}, {Annis}, {Bahcall}, {Brinkmann}, {Burles}, {Castander},
  {Connolly}, {Csabai}, {Doi}, {Fukugita}, {Frieman}, {Glazebrook}, {Gunn},
  {Hendry}, {Hennessy}, {Ivezi{\'c}}, {Kent}, {Knapp}, {Lin}, {Loh}, {Lupton},
  {Margon}, {McKay}, {Meiksin}, {Munn}, {Pope}, {Richmond}, {Schlegel},
  {Schneider}, {Shimasaku}, {Stoughton}, {Strauss}, {SubbaRao}, {Szalay},
  {Szapudi}, {Tucker}, {Yanny}, \& {York}}]{Eisenstein05}
---. 2005, \apj, 633, 560

\bibitem[{{Eisenstein} {et~al.}(2011){Eisenstein}, {Weinberg}, {Agol},
  {Aihara}, {Allende Prieto}, {Anderson}, {Arns}, {Aubourg}, {Bailey},
  {Balbinot}, \& et~al.}]{SDSSIII}
---. 2011, \aj, 142, 72

\bibitem[{{Faber} \& {Jackson}(1976)}]{FJR}
{Faber}, S.~M., \& {Jackson}, R.~E. 1976, \apj, 204, 668

\bibitem[{{Falco} {et~al.}(1985){Falco}, {Gorenstein}, \& {Shapiro}}]{Falco85}
{Falco}, E.~E., {Gorenstein}, M.~V., \& {Shapiro}, I.~I. 1985, \apjl, 289, L1

\bibitem[{{Ferreras} {et~al.}(2013){Ferreras}, {La Barbera}, {de la Rosa},
  {Vazdekis}, {de Carvalho}, {Falc{\'o}n-Barroso}, \&
  {Ricciardelli}}]{Ferreras13}
{Ferreras}, I., {La Barbera}, F., {de la Rosa}, I.~G., {Vazdekis}, A., {de
  Carvalho}, R.~R., {Falc{\'o}n-Barroso}, J., \& {Ricciardelli}, E. 2013,
  \mnras, 429, L15

\bibitem[{{Ferreras} {et~al.}(2010){Ferreras}, {Saha}, {Leier}, {Courbin}, \&
  {Falco}}]{Ferreras10}
{Ferreras}, I., {Saha}, P., {Leier}, D., {Courbin}, F., \& {Falco}, E.~E. 2010,
  \mnras, 409, L30

\bibitem[{{Gallazzi} {et~al.}(2006){Gallazzi}, {Charlot}, {Brinchmann}, \&
  {White}}]{Gallazzi06}
{Gallazzi}, A., {Charlot}, S., {Brinchmann}, J., \& {White}, S.~D.~M. 2006,
  \mnras, 370, 1106

\bibitem[{{Gavazzi} {et~al.}(2008){Gavazzi}, {Treu}, {Koopmans}, {Bolton},
  {Moustakas}, {Burles}, \& {Marshall}}]{SLACSVI}
{Gavazzi}, R., {Treu}, T., {Koopmans}, L.~V.~E., {Bolton}, A.~S., {Moustakas},
  L.~A., {Burles}, S., \& {Marshall}, P.~J. 2008, \apj, 677, 1046

\bibitem[{{Gavazzi} {et~al.}(2012){Gavazzi}, {Treu}, {Marshall}, {Brault}, \&
  {Ruff}}]{Gavazzi12}
{Gavazzi}, R., {Treu}, T., {Marshall}, P.~J., {Brault}, F., \& {Ruff}, A. 2012,
  \apj, 761, 170

\bibitem[{{Gavazzi} {et~al.}(2007){Gavazzi}, {Treu}, {Rhodes}, {Koopmans},
  {Bolton}, {Burles}, {Massey}, \& {Moustakas}}]{SLACSIV}
{Gavazzi}, R., {Treu}, T., {Rhodes}, J.~D., {Koopmans}, L.~V.~E., {Bolton},
  A.~S., {Burles}, S., {Massey}, R.~J., \& {Moustakas}, L.~A. 2007, \apj, 667,
  176

\bibitem[Gavazzi et al.(2014)]{Gavazzi14} Gavazzi, R., Marshall, 
P.~J., Treu, T., \& Sonnenfeld, A.\ 2014, \apj, 785, 144 


\bibitem[{{Geha} {et~al.}(2013){Geha}, {Brown}, {Tumlinson}, {Kalirai},
  {Simon}, {Kirby}, {VandenBerg}, {Mu{\~n}oz}, {Avila}, {Guhathakurta}, \&
  {Ferguson}}]{Geha13}
{Geha}, M., {et~al.} 2013, \apj, 771, 29

\bibitem[{{Gnedin} {et~al.}(2004){Gnedin}, {Kravtsov}, {Klypin}, \&
  {Nagai}}]{Gnedin04}
{Gnedin}, O.~Y., {Kravtsov}, A.~V., {Klypin}, A.~A., \& {Nagai}, D. 2004, \apj,
  616, 16

\bibitem[{{Goldberg} \& {Leonard}(2007)}]{Goldberg07}
{Goldberg}, D.~M., \& {Leonard}, A. 2007, \apj, 660, 1003

\bibitem[{{Gorenstein} {et~al.}(1988){Gorenstein}, {Shapiro}, \&
  {Falco}}]{Gorenstein88}
{Gorenstein}, M.~V., {Shapiro}, I.~I., \& {Falco}, E.~E. 1988, \apj, 327, 693

\bibitem[{{Governato} {et~al.}(2010){Governato}, {Brook}, {Mayer}, {Brooks},
  {Rhee}, {Wadsley}, {Jonsson}, {Willman}, {Stinson}, {Quinn}, \&
  {Madau}}]{Governato10}
{Governato}, F., {et~al.} 2010, \nat, 463, 203

\bibitem[{{Graham} {et~al.}(2003){Graham}, {Erwin}, {Trujillo}, \& {Asensio
  Ramos}}]{Graham03}
{Graham}, A.~W., {Erwin}, P., {Trujillo}, I., \& {Asensio Ramos}, A. 2003, \aj,
  125, 2951

\bibitem[{{Graham} {et~al.}(2006){Graham}, {Merritt}, {Moore}, {Diemand}, \&
  {Terzi{\'c}}}]{Graham06}
{Graham}, A.~W., {Merritt}, D., {Moore}, B., {Diemand}, J., \& {Terzi{\'c}}, B.
  2006, \aj, 132, 2701

\bibitem[{{Grillo} \& {Christensen}(2011)}]{Grillo11}
{Grillo}, C., \& {Christensen}, L. 2011, \mnras, 418, 929

\bibitem[{{Gustafsson} {et~al.}(2006){Gustafsson}, {Fairbairn}, \&
  {Sommer-Larsen}}]{Gustafsson06}
{Gustafsson}, M., {Fairbairn}, M., \& {Sommer-Larsen}, J. 2006, \prd, 74,
  123522

\bibitem[{{Hubble}(1926)}]{Hubble26}
{Hubble}, E. 1926, Contributions from the Mount Wilson Observatory / Carnegie
  Institution of Washington, 324, 1

\bibitem[{{Hubble}(1936)}]{Hubble36}
{Hubble}, E.~P. 1936, {Realm of the Nebulae}

\bibitem[{{Huterer} {et~al.}(2005){Huterer}, {Keeton}, \& {Ma}}]{Huterer05}
{Huterer}, D., {Keeton}, C.~R., \& {Ma}, C.-P. 2005, \apj, 624, 34

\bibitem[{{Jorgensen} {et~al.}(1995){Jorgensen}, {Franx}, \&
  {Kjaergaard}}]{Jorgensen95}
{Jorgensen}, I., {Franx}, M., \& {Kjaergaard}, P. 1995, \mnras, 276, 1341

\bibitem[{{Kassiola} \& {Kovner}(1993)}]{Kassiola93}
{Kassiola}, A., \& {Kovner}, I. 1993, \apj, 417, 450

\bibitem[{{Kauffmann} {et~al.}(1993){Kauffmann}, {White}, \&
  {Guiderdoni}}]{Kauffmann93}
{Kauffmann}, G., {White}, S.~D.~M., \& {Guiderdoni}, B. 1993, \mnras, 264, 201

\bibitem[{{Keeton} \& {Kochanek}(1998)}]{Keeton98}
{Keeton}, C.~R., \& {Kochanek}, C.~S. 1998, \apj, 495, 157

\bibitem[{{Kochanek}(1994)}]{Kochanek94}
{Kochanek}, C.~S. 1994, \apj, 436, 56

\bibitem[{{Kochanek} {et~al.}(2000){Kochanek}, {Falco}, {Impey}, {Leh{\'a}r},
  {McLeod}, {Rix}, {Keeton}, {Mu{\~n}oz}, \& {Peng}}]{Kochanek2000}
{Kochanek}, C.~S., {et~al.} 2000, \apj, 543, 131

\bibitem[{{Komatsu} {et~al.}(2011){Komatsu}, {Smith}, {Dunkley}, {Bennett},
  {Gold}, {Hinshaw}, {Jarosik}, {Larson}, {Nolta}, {Page}, {Spergel},
  {Halpern}, {Hill}, {Kogut}, {Limon}, {Meyer}, {Odegard}, {Tucker}, {Weiland},
  {Wollack}, \& {Wright}}]{WMAP7}
{Komatsu}, E., {et~al.} 2011, \apjs, 192, 18

\bibitem[{{Koopmans} \& {Treu}(2002)}]{KT02}
{Koopmans}, L.~V.~E., \& {Treu}, T. 2002, \apjl, 568, L5

\bibitem[{{Koopmans} \& {Treu}(2003)}]{KT03}
---. 2003, \apj, 583, 606

\bibitem[{{Koopmans} {et~al.}(2006){Koopmans}, {Treu}, {Bolton}, {Burles}, \&
  {Moustakas}}]{SLACSIII}
{Koopmans}, L.~V.~E., {Treu}, T., {Bolton}, A.~S., {Burles}, S., \&
  {Moustakas}, L.~A. 2006, \apj, 649, 599

\bibitem[Koopmans et al.(2009)]{Koopmans09} Koopmans, L.~V.~E., 
Bolton, A., Treu, T., et al.\ 2009, \apjl, 703, L51 

\bibitem[{{Kormann} {et~al.}(1994){Kormann}, {Schneider}, \&
  {Bartelmann}}]{Kormann94}
{Kormann}, R., {Schneider}, P., \& {Bartelmann}, M. 1994, \aap, 284, 285

\bibitem[{{Kroupa}(2001)}]{Kroupa01}
{Kroupa}, P. 2001, \mnras, 322, 231

\bibitem[{{La Barbera} {et~al.}(2013){La Barbera}, {Ferreras}, {Vazdekis}, {de
  la Rosa}, {de Carvalho}, {Trevisan}, {Falc{\'o}n-Barroso}, \&
  {Ricciardelli}}]{LaBarbera13}
{La Barbera}, F., {Ferreras}, I., {Vazdekis}, A., {de la Rosa}, I.~G., {de
  Carvalho}, R.~R., {Trevisan}, M., {Falc{\'o}n-Barroso}, J., \&
  {Ricciardelli}, E. 2013, \mnras, 433, 3017

\bibitem[{{Markwardt}(2009)}]{MPFIT}
{Markwardt}, C.~B. 2009, in Astronomical Society of the Pacific Conference
  Series, Vol. 411, Astronomical Data Analysis Software and Systems XVIII, ed.
  D.~A. {Bohlender}, D.~{Durand}, \& P.~{Dowler}, 251

\bibitem[{{Martizzi} {et~al.}(2012){Martizzi}, {Teyssier}, {Moore}, \&
  {Wentz}}]{Martizzi12}
{Martizzi}, D., {Teyssier}, R., {Moore}, B., \& {Wentz}, T. 2012, \mnras, 422,
  3081

\bibitem[{{McKean} {et~al.}(2010){McKean}, {Auger}, {Koopmans}, {Vegetti},
  {Czoske}, {Fassnacht}, {Treu}, {More}, \& {Kocevski}}]{McKean10}
{McKean}, J.~P., {et~al.} 2010, \mnras, 404, 749

\bibitem[{{Mehlert} {et~al.}(2003){Mehlert}, {Thomas}, {Saglia}, {Bender}, \&
  {Wegner}}]{Mehlert03}
{Mehlert}, D., {Thomas}, D., {Saglia}, R.~P., {Bender}, R., \& {Wegner}, G.
  2003, \aap, 407, 423

\bibitem[{{Moore} {et~al.}(1999){Moore}, {Quinn}, {Governato}, {Stadel}, \&
  {Lake}}]{Moore99}
{Moore}, B., {Quinn}, T., {Governato}, F., {Stadel}, J., \& {Lake}, G. 1999,
  \mnras, 310, 1147

\bibitem[{Mor{\'e}(1978)}]{More78}
Mor{\'e}, J. 1978, in Lecture Notes in Mathematics, Vol. 630, Numerical
  Analysis, ed. G.~Watson (Springer Berlin Heidelberg), 105--116

\bibitem[{{Narayan} \& {Bartelmann}(1996)}]{Narayan96}
{Narayan}, R., \& {Bartelmann}, M. 1996, ArXiv Astrophysics e-prints

\bibitem[{{Navarro} {et~al.}(1996){Navarro}, {Frenk}, \& {White}}]{NFW96}
{Navarro}, J.~F., {Frenk}, C.~S., \& {White}, S.~D.~M. 1996, \apj, 462, 563

\bibitem[{{Navarro} {et~al.}(1997){Navarro}, {Frenk}, \& {White}}]{NFW97}
---. 1997, \apj, 490, 493

\bibitem[{{Navarro} {et~al.}(2010){Navarro}, {Ludlow}, {Springel}, {Wang},
  {Vogelsberger}, {White}, {Jenkins}, {Frenk}, \& {Helmi}}]{Navarro10}
{Navarro}, J.~F., {et~al.} 2010, \mnras, 402, 21

\bibitem[{{Newman} {et~al.}(2011){Newman}, {Treu}, {Ellis}, \&
  {Sand}}]{Newman11}
{Newman}, A.~B., {Treu}, T., {Ellis}, R.~S., \& {Sand}, D.~J. 2011, \apjl, 728,
  L39

\bibitem[Newman et al.(2013{\natexlab{a}})]{Newman13a} Newman, A.~B., Treu, T., 
Ellis, R.~S., et al.\ 2013, \apj, 765, 24   

\bibitem[Newman et al.(2013{\natexlab{b}})]{Newman13b} Newman, A.~B., Treu, T., 
Ellis, R.~S., \& Sand, D.~J.\ 2013, \apj, 765, 25 

\bibitem[{{Newton} {et~al.}(2011){Newton}, {Marshall}, {Treu}, {Auger},
  {Gavazzi}, {Bolton}, {Koopmans}, \& {Moustakas}}]{SLACSXI}
{Newton}, E.~R., {Marshall}, P.~J., {Treu}, T., {Auger}, M.~W., {Gavazzi}, R.,
  {Bolton}, A.~S., {Koopmans}, L.~V.~E., \& {Moustakas}, L.~A. 2011, \apj, 734,
  104

\bibitem[{{Nipoti} {et~al.}(2004){Nipoti}, {Treu}, {Ciotti}, \&
  {Stiavelli}}]{Nipoti04}
{Nipoti}, C., {Treu}, T., {Ciotti}, L., \& {Stiavelli}, M. 2004, \mnras, 355,
  1119

\bibitem[{{Oguri} {et~al.}(2013){Oguri}, {Rusu}, \& {Falco}}]{Oguri13}
{Oguri}, M., {Rusu}, C.~E., \& {Falco}, E.~E. 2013, ArXiv e-prints

\bibitem[{{Palacios} {et~al.}(2010){Palacios}, {Gebran}, {Josselin}, {Martins},
  {Plez}, {Belmas}, \& {L{\`e}bre}}]{POLLUX}
{Palacios}, A., {Gebran}, M., {Josselin}, E., {Martins}, F., {Plez}, B.,
  {Belmas}, M., \& {L{\`e}bre}, A. 2010, \aap, 516, A13

\bibitem[{{Percival} {et~al.}(2007){Percival}, {Cole}, {Eisenstein}, {Nichol},
  {Peacock}, {Pope}, \& {Szalay}}]{Percival07}
{Percival}, W.~J., {Cole}, S., {Eisenstein}, D.~J., {Nichol}, R.~C., {Peacock},
  J.~A., {Pope}, A.~C., \& {Szalay}, A.~S. 2007, \mnras, 381, 1053

\bibitem[{{Prugniel} \& {Soubiran}(2001)}]{ELODIE}
{Prugniel}, P., \& {Soubiran}, C. 2001, \aap, 369, 1048

\bibitem[{{Romano-D{\'{\i}}az} {et~al.}(2008){Romano-D{\'{\i}}az}, {Shlosman},
  {Hoffman}, \& {Heller}}]{Romano-Diaz08}
{Romano-D{\'{\i}}az}, E., {Shlosman}, I., {Hoffman}, Y., \& {Heller}, C. 2008,
  \apjl, 685, L105

\bibitem[{{Salpeter}(1955)}]{Salpeter55}
{Salpeter}, E.~E. 1955, \apj, 121, 161

\bibitem[{{Schneider}(2014)}]{Schneider14}
{Schneider}, P. 2014, ArXiv e-prints

\bibitem[{{Schneider} \& {Sluse}(2013)}]{Schneider13}
{Schneider}, P., \& {Sluse}, D. 2013, \aap, 559, A37

\bibitem[{{Shu} {et~al.}(2012){Shu}, {Bolton}, {Schlegel}, {Dawson}, {Wake},
  {Brownstein}, {Brinkmann}, \& {Weaver}}]{Shu12}
{Shu}, Y., {Bolton}, A.~S., {Schlegel}, D.~J., {Dawson}, K.~S., {Wake}, D.~A.,
  {Brownstein}, J.~R., {Brinkmann}, J., \& {Weaver}, B.~A. 2012, \aj, 143, 90

\bibitem[{{Sonnenfeld} {et~al.}(2012){Sonnenfeld}, {Treu}, {Gavazzi},
  {Marshall}, {Auger}, {Suyu}, {Koopmans}, \& {Bolton}}]{Sonnenfeld12}
{Sonnenfeld}, A., {Treu}, T., {Gavazzi}, R., {Marshall}, P.~J., {Auger}, M.~W.,
  {Suyu}, S.~H., {Koopmans}, L.~V.~E., \& {Bolton}, A.~S. 2012, \apj, 752, 163

\bibitem[{{Sonnenfeld} {et~al.}(2013){Sonnenfeld}, {Treu}, {Gavazzi}, {Suyu},
  {Marshall}, {Auger}, \& {Nipoti}}]{Sonnenfeld13}
{Sonnenfeld}, A., {Treu}, T., {Gavazzi}, R., {Suyu}, S.~H., {Marshall}, P.~J.,
  {Auger}, M.~W., \& {Nipoti}, C. 2013, \apj, 777, 98

\bibitem[Sonnenfeld et al.(2014)]{Sonnenfeld14} Sonnenfeld, A., 
Treu, T., Marshall, P.~J., et al.\ 2014, arXiv:1410.1881 

\bibitem[{{Spiniello} {et~al.}(2011){Spiniello}, {Koopmans}, {Trager},
  {Czoske}, \& {Treu}}]{Spiniello11}
{Spiniello}, C., {Koopmans}, L.~V.~E., {Trager}, S.~C., {Czoske}, O., \&
  {Treu}, T. 2011, \mnras, 417, 3000

\bibitem[{{Spiniello} {et~al.}(2014){Spiniello}, {Trager}, {Koopmans}, \&
  {Conroy}}]{Spiniello14}
{Spiniello}, C., {Trager}, S., {Koopmans}, L.~V.~E., \& {Conroy}, C. 2014,
  \mnras, 438, 1483

\bibitem[{{Spiniello} {et~al.}(2012){Spiniello}, {Trager}, {Koopmans}, \&
  {Chen}}]{Spiniello12}
{Spiniello}, C., {Trager}, S.~C., {Koopmans}, L.~V.~E., \& {Chen}, Y.~P. 2012,
  \apjl, 753, L32

\bibitem[{{Strader} {et~al.}(2011){Strader}, {Caldwell}, \& {Seth}}]{Strader11}
{Strader}, J., {Caldwell}, N., \& {Seth}, A.~C. 2011, \aj, 142, 8

\bibitem[{{Strauss} {et~al.}(2002){Strauss}, {Weinberg}, {Lupton}, {Narayanan},
  {Annis}, {Bernardi}, {Blanton}, {Burles}, {Connolly}, {Dalcanton}, {Doi},
  {Eisenstein}, {Frieman}, {Fukugita}, {Gunn}, {Ivezi{\'c}}, {Kent}, {Kim},
  {Knapp}, {Kron}, {Munn}, {Newberg}, {Nichol}, {Okamura}, {Quinn}, {Richmond},
  {Schlegel}, {Shimasaku}, {SubbaRao}, {Szalay}, {Vanden Berk}, {Vogeley},
  {Yanny}, {Yasuda}, {York}, \& {Zehavi}}]{Strauss2002}
{Strauss}, M.~A., {et~al.} 2002, \aj, 124, 1810

\bibitem[{{Toomre} \& {Toomre}(1972)}]{TT72}
{Toomre}, A., \& {Toomre}, J. 1972, \apj, 178, 623

\bibitem[{{Tortora} {et~al.}(2009){Tortora}, {Napolitano}, {Romanowsky},
  {Capaccioli}, \& {Covone}}]{Tortora09}
{Tortora}, C., {Napolitano}, N.~R., {Romanowsky}, A.~J., {Capaccioli}, M., \&
  {Covone}, G. 2009, \mnras, 396, 1132

\bibitem[{{Tortora} {et~al.}(2013){Tortora}, {Romanowsky}, \&
  {Napolitano}}]{Tortora13}
{Tortora}, C., {Romanowsky}, A.~J., \& {Napolitano}, N.~R. 2013, \apj, 765, 8

\bibitem[Treu et al.(2011)]{SWELLSI} Treu, T., Dutton, A.~A., 
Auger, M.~W., et al.\ 2011, \mnras, 417, 1601 

\bibitem[{{Treu} {et~al.}(2010){Treu}, {Auger}, {Koopmans}, {Gavazzi},
  {Marshall}, \& {Bolton}}]{Treu10}
{Treu}, T., {Auger}, M.~W., {Koopmans}, L.~V.~E., {Gavazzi}, R., {Marshall},
  P.~J., \& {Bolton}, A.~S. 2010, \apj, 709, 1195

\bibitem[{{Treu} {et~al.}(2009){Treu}, {Gavazzi}, {Gorecki}, {Marshall},
  {Koopmans}, {Bolton}, {Moustakas}, \& {Burles}}]{SLACSVIII}
{Treu}, T., {Gavazzi}, R., {Gorecki}, A., {Marshall}, P.~J., {Koopmans},
  L.~V.~E., {Bolton}, A.~S., {Moustakas}, L.~A., \& {Burles}, S. 2009, \apj,
  690, 670

\bibitem[{{Treu} {et~al.}(2006){Treu}, {Koopmans}, {Bolton}, {Burles}, \&
  {Moustakas}}]{SLACSII}
{Treu}, T., {Koopmans}, L.~V., {Bolton}, A.~S., {Burles}, S., \& {Moustakas},
  L.~A. 2006, \apj, 640, 662

\bibitem[{{Treu} \& {Koopmans}(2002)}]{TK02a}
{Treu}, T., \& {Koopmans}, L.~V.~E. 2002, \apj, 575, 87

\bibitem[{{Treu} \& {Koopmans}(2004)}]{TK04}
---. 2004, \apj, 611, 739

\bibitem[{{Valdes} {et~al.}(2004){Valdes}, {Gupta}, {Rose}, {Singh}, \&
  {Bell}}]{Indo-US}
{Valdes}, F., {Gupta}, R., {Rose}, J.~A., {Singh}, H.~P., \& {Bell}, D.~J.
  2004, \apjs, 152, 251

\bibitem[{{van de Ven} {et~al.}(2010){van de Ven}, {Falc{\'o}n-Barroso},
  {McDermid}, {Cappellari}, {Miller}, \& {de Zeeuw}}]{Vandeven10}
{van de Ven}, G., {Falc{\'o}n-Barroso}, J., {McDermid}, R.~M., {Cappellari},
  M., {Miller}, B.~W., \& {de Zeeuw}, P.~T. 2010, \apj, 719, 1481

\bibitem[{{van Dokkum} \& {Conroy}(2010)}]{vanDokkum10}
{van Dokkum}, P.~G., \& {Conroy}, C. 2010, \nat, 468, 940

\bibitem[{{Vegetti} {et~al.}(2010){Vegetti}, {Koopmans}, {Bolton}, {Treu}, \&
  {Gavazzi}}]{Vegetti10}
{Vegetti}, S., {Koopmans}, L.~V.~E., {Bolton}, A., {Treu}, T., \& {Gavazzi}, R.
  2010, \mnras, 408, 1969

\bibitem[{{Vegetti} {et~al.}(2012){Vegetti}, {Lagattuta}, {McKean}, {Auger},
  {Fassnacht}, \& {Koopmans}}]{Vegetti12}
{Vegetti}, S., {Lagattuta}, D.~J., {McKean}, J.~P., {Auger}, M.~W.,
  {Fassnacht}, C.~D., \& {Koopmans}, L.~V.~E. 2012, \nat, 481, 341

\bibitem[{{Velliscig} {et~al.}(2014){Velliscig}, {van Daalen}, {Schaye},
  {McCarthy}, {Cacciato}, {Le Brun}, \& {Dalla Vecchia}}]{Velliscig14}
{Velliscig}, M., {van Daalen}, M.~P., {Schaye}, J., {McCarthy}, I.~G.,
  {Cacciato}, M., {Le Brun}, A.~M.~C., \& {Dalla Vecchia}, C. 2014, ArXiv
  e-prints

\bibitem[{{White} \& {Frenk}(1991)}]{White91}
{White}, S.~D.~M., \& {Frenk}, C.~S. 1991, \apj, 379, 52

\bibitem[{{York} {et~al.}(2000){York}, {Adelman}, {Anderson}, {Anderson},
  {Annis}, {Bahcall}, {Bakken}, {Barkhouser}, {Bastian}, {Berman}, {Boroski},
  {Bracker}, {Briegel}, {Briggs}, {Brinkmann}, {Brunner}, {Burles}, {Carey},
  {Carr}, {Castander}, {Chen}, {Colestock}, {Connolly}, {Crocker}, {Csabai},
  {Czarapata}, {Davis}, {Doi}, {Dombeck}, {Eisenstein}, {Ellman}, {Elms},
  {Evans}, {Fan}, {Federwitz}, {Fiscelli}, {Friedman}, {Frieman}, {Fukugita},
  {Gillespie}, {Gunn}, {Gurbani}, {de Haas}, {Haldeman}, {Harris}, {Hayes},
  {Heckman}, {Hennessy}, {Hindsley}, {Holm}, {Holmgren}, {Huang}, {Hull},
  {Husby}, {Ichikawa}, {Ichikawa}, {Ivezi{\'c}}, {Kent}, {Kim}, {Kinney},
  {Klaene}, {Kleinman}, {Kleinman}, {Knapp}, {Korienek}, {Kron}, {Kunszt},
  {Lamb}, {Lee}, {Leger}, {Limmongkol}, {Lindenmeyer}, {Long}, {Loomis},
  {Loveday}, {Lucinio}, {Lupton}, {MacKinnon}, {Mannery}, {Mantsch}, {Margon},
  {McGehee}, {McKay}, {Meiksin}, {Merelli}, {Monet}, {Munn}, {Narayanan},
  {Nash}, {Neilsen}, {Neswold}, {Newberg}, {Nichol}, {Nicinski}, {Nonino},
  {Okada}, {Okamura}, {Ostriker}, {Owen}, {Pauls}, {Peoples}, {Peterson},
  {Petravick}, {Pier}, {Pope}, {Pordes}, {Prosapio}, {Rechenmacher}, {Quinn},
  {Richards}, {Richmond}, {Rivetta}, {Rockosi}, {Ruthmansdorfer}, {Sandford},
  {Schlegel}, {Schneider}, {Sekiguchi}, {Sergey}, {Shimasaku}, {Siegmund},
  {Smee}, {Smith}, {Snedden}, {Stone}, {Stoughton}, {Strauss}, {Stubbs},
  {SubbaRao}, {Szalay}, {Szapudi}, {Szokoly}, {Thakar}, {Tremonti}, {Tucker},
  {Uomoto}, {Vanden Berk}, {Vogeley}, {Waddell}, {Wang}, {Watanabe},
  {Weinberg}, {Yanny}, {Yasuda}, \& {SDSS Collaboration}}]{SDSSI}
{York}, D.~G., {et~al.} 2000, \aj, 120, 1579

\bibitem[Feroz 
\& Hobson(2008)]{Feroz08} Feroz, F., \& Hobson, M.~P.\ 2008, \mnras, 384, 449 

\bibitem[Feroz et al.(2009)]{Feroz09} Feroz, F., Hobson, M.~P., 
\& Bridges, M.\ 2009, \mnras, 398, 1601 

\bibitem[Feroz et al.(2013)]{Feroz13} Feroz, F., Hobson, M.~P., 
Cameron, E., \& Pettitt, A.~N.\ 2013, arXiv:1306.2144 

\bibitem[Buchner et 
al.(2014)]{Buchner14} Buchner, J., Georgakakis, A., Nandra, K., et al.\ 2014, \aap, 564, AA125 

\bibitem[Holder 
\& Schechter(2003)]{Holder03} Holder, G.~P., \& Schechter, P.~L.\ 2003, \apj, 589, 688 
\bibitem[Wong et al.(2011)]{Wong11} Wong, K.~C., Keeton, 
C.~R., Williams, K.~A., Momcheva, I.~G., 
\& Zabludoff, A.~I.\ 2011, \apj, 726, 84 

\end{thebibliography}

\clearpage

\appendix
\begin{figure*}[htbp]
\centering
\includegraphics[width=0.495\textwidth]{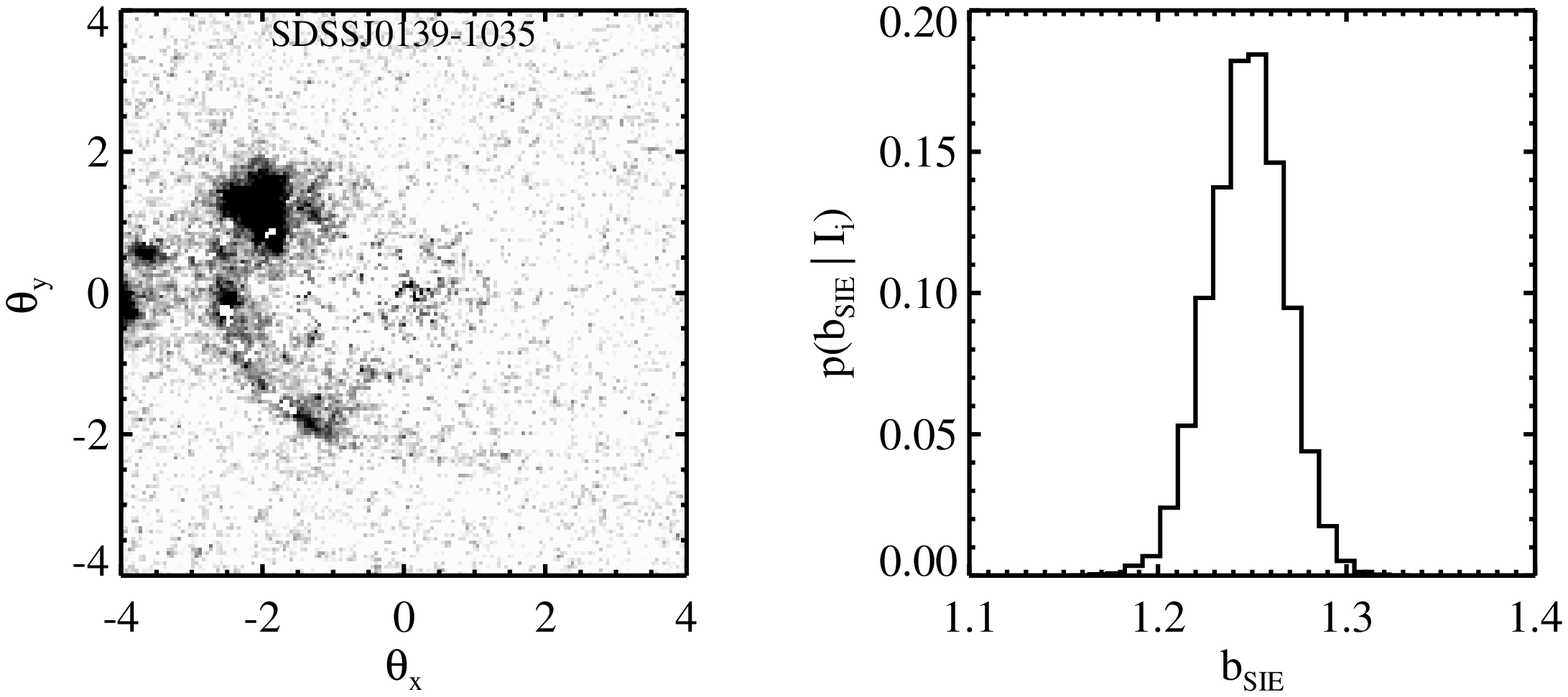}
\includegraphics[width=0.495\textwidth]{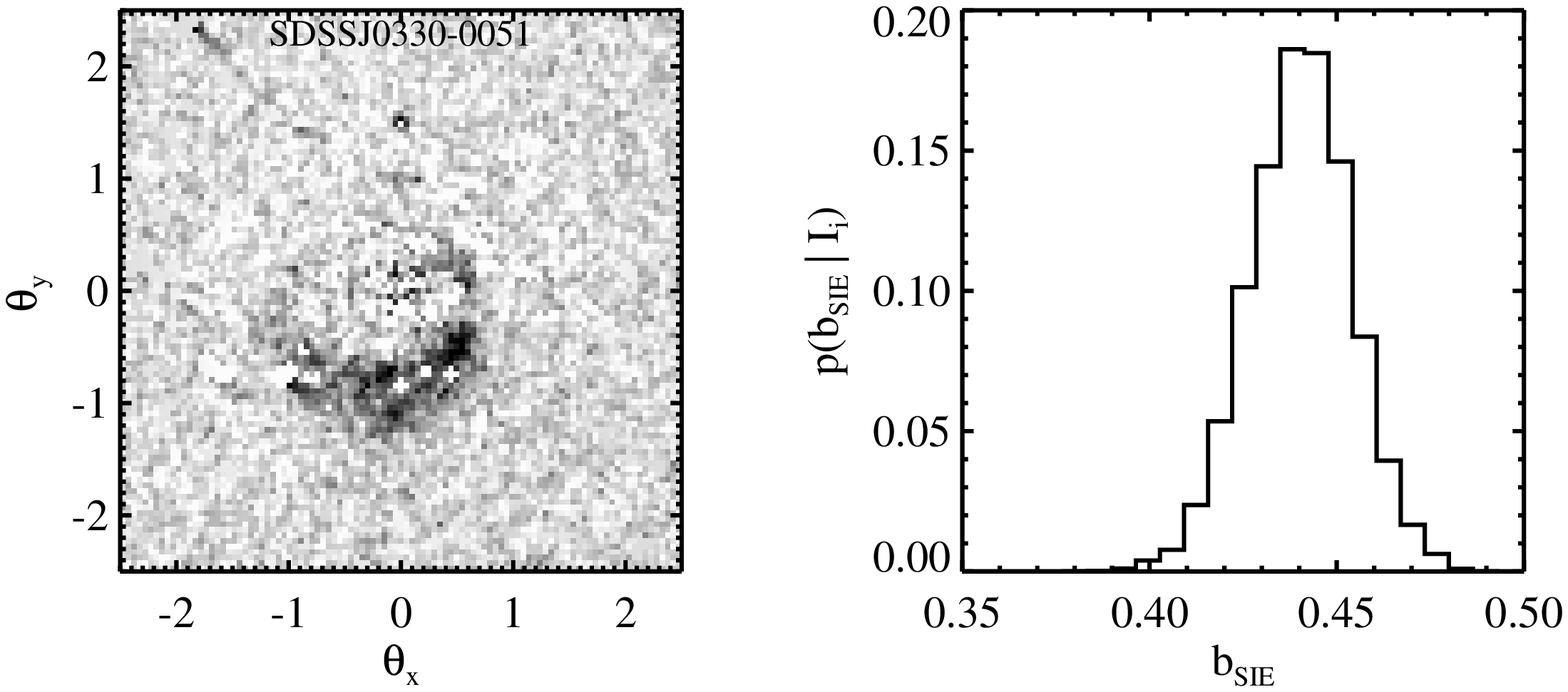}
\includegraphics[width=0.495\textwidth]{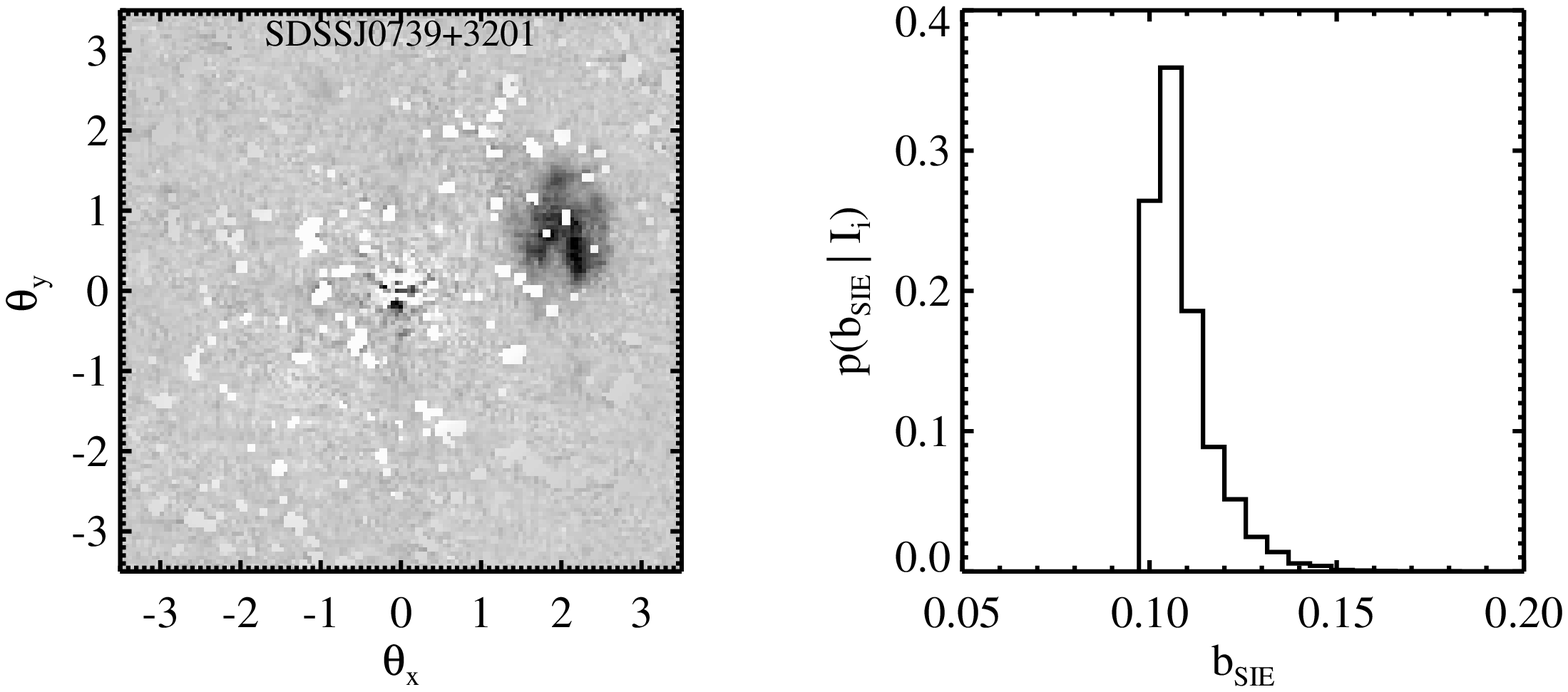}
\includegraphics[width=0.495\textwidth]{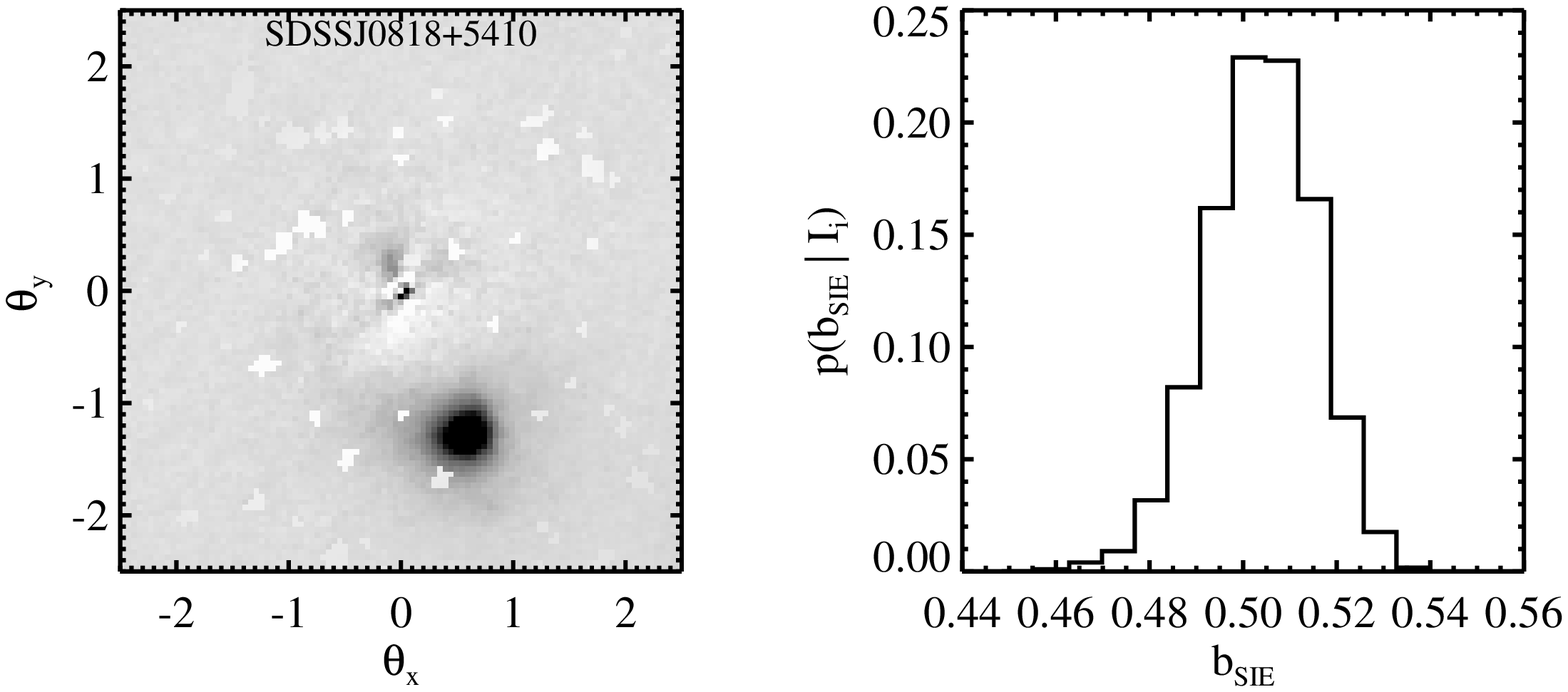}
\includegraphics[width=0.495\textwidth]{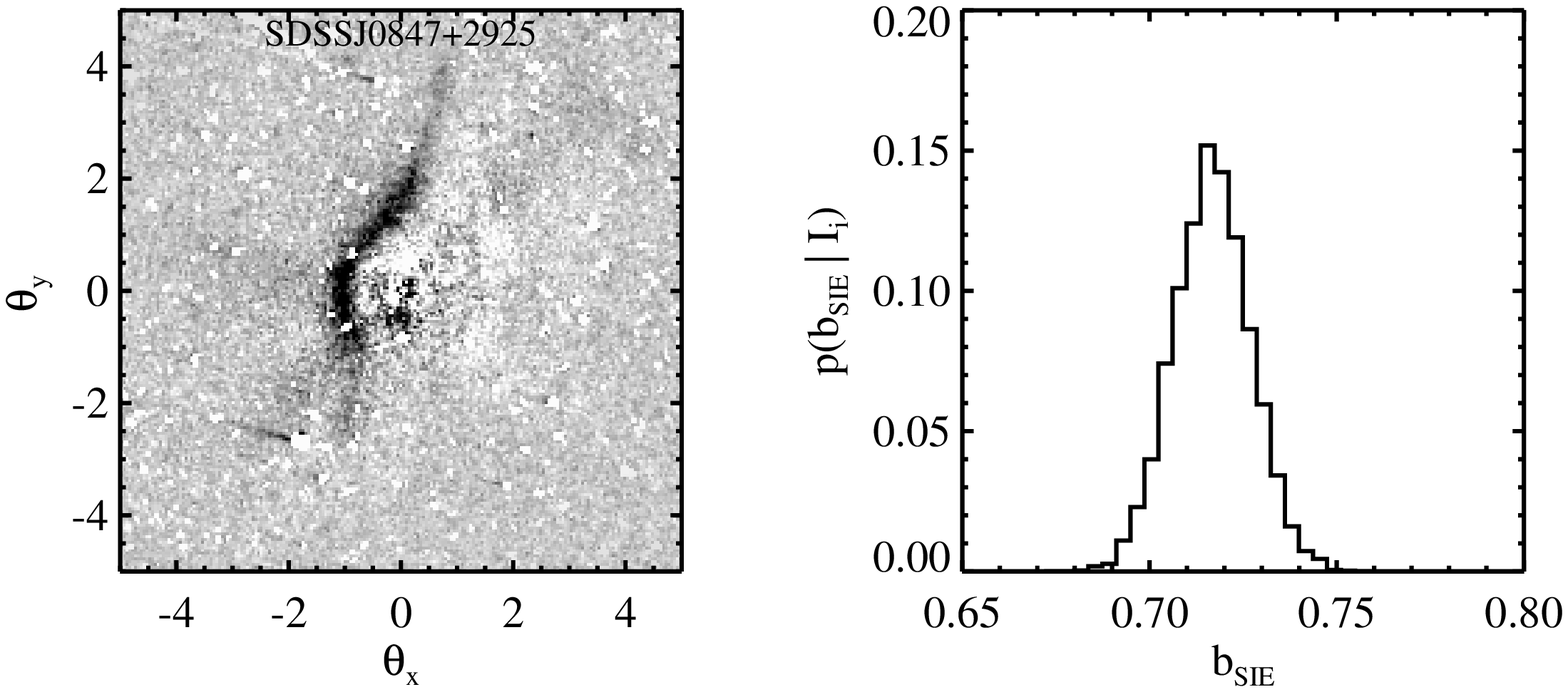}
\includegraphics[width=0.495\textwidth]{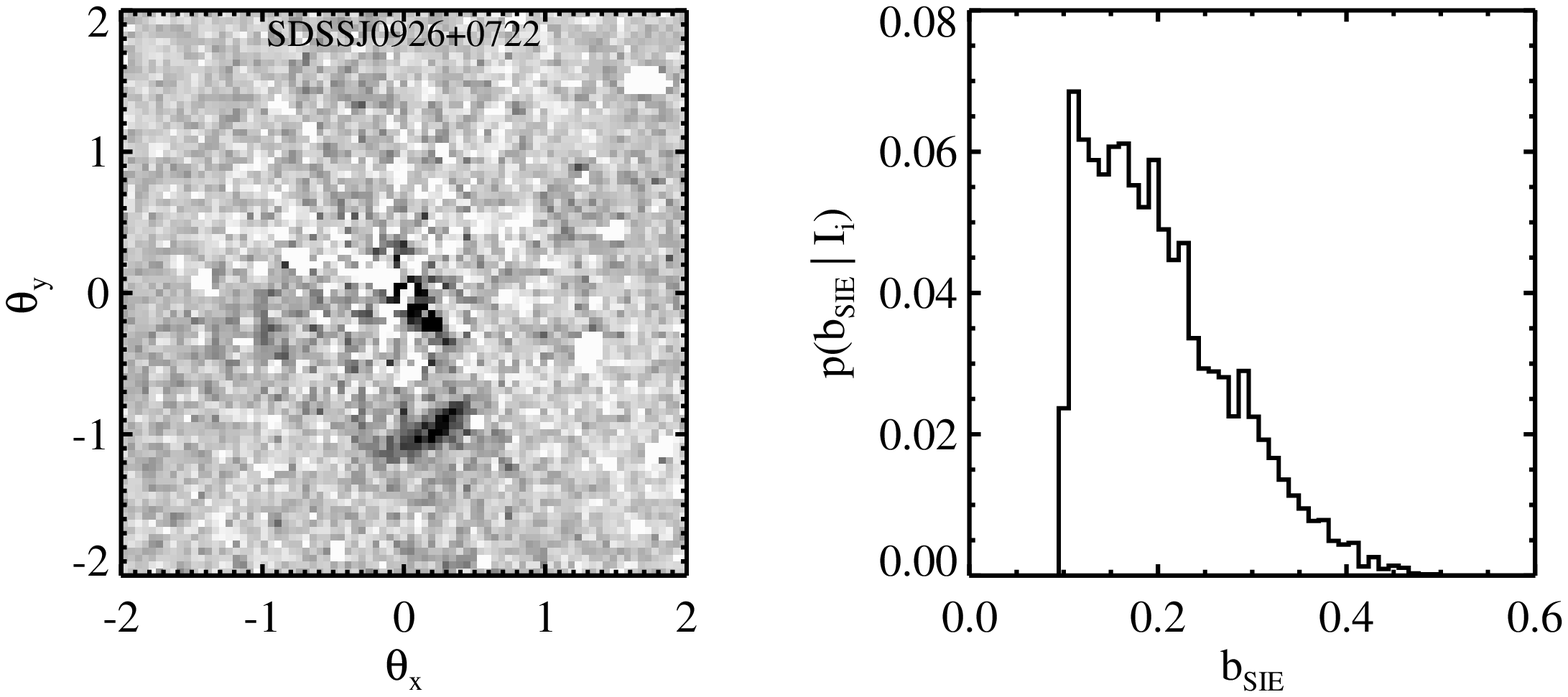}
\includegraphics[width=0.495\textwidth]{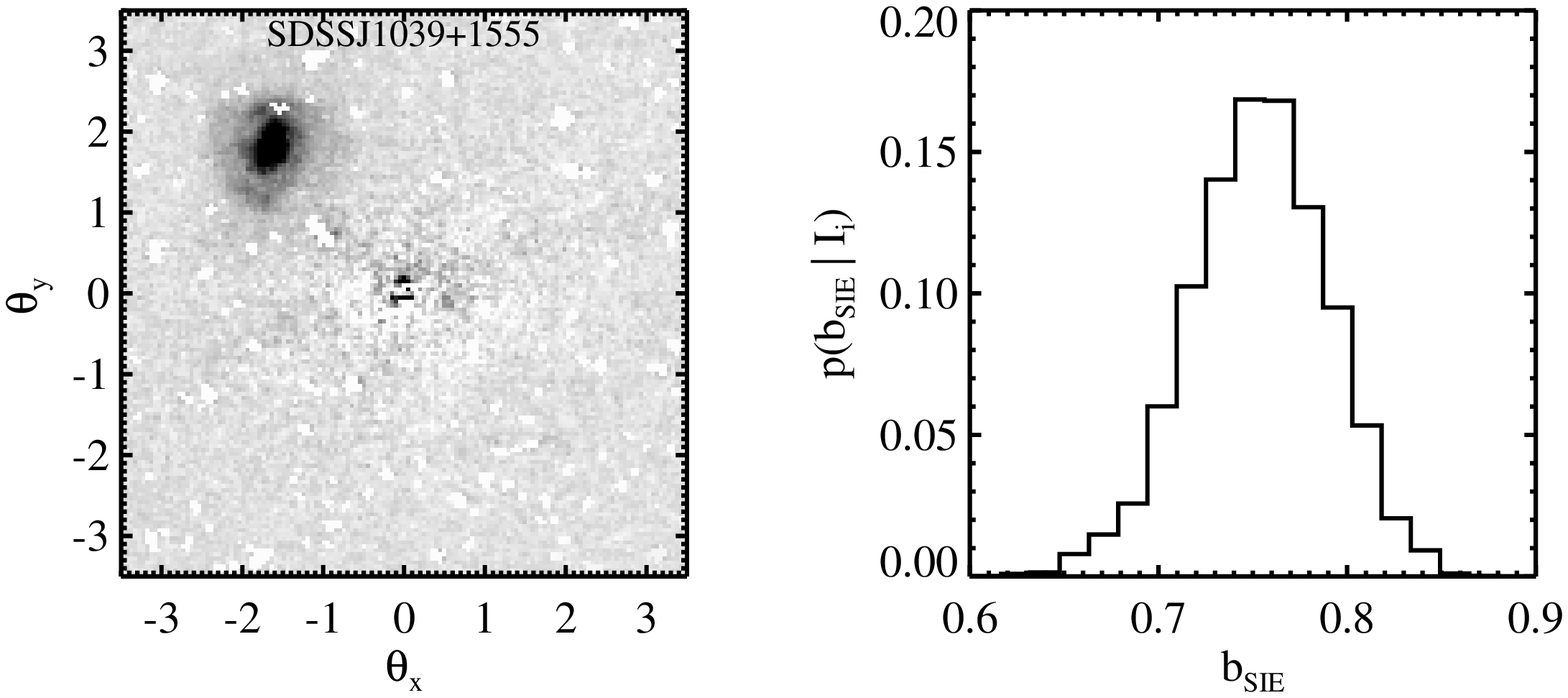}
\includegraphics[width=0.495\textwidth]{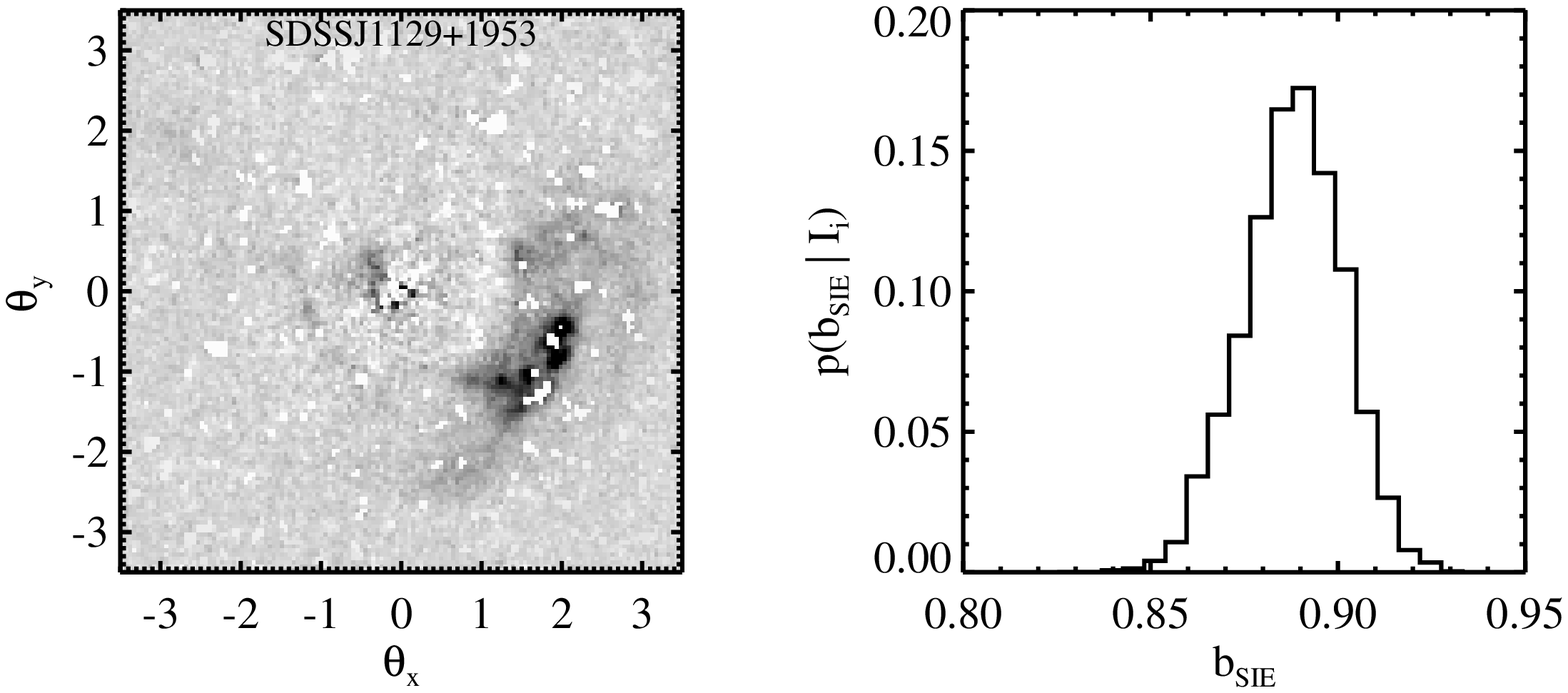}
\includegraphics[width=0.495\textwidth]{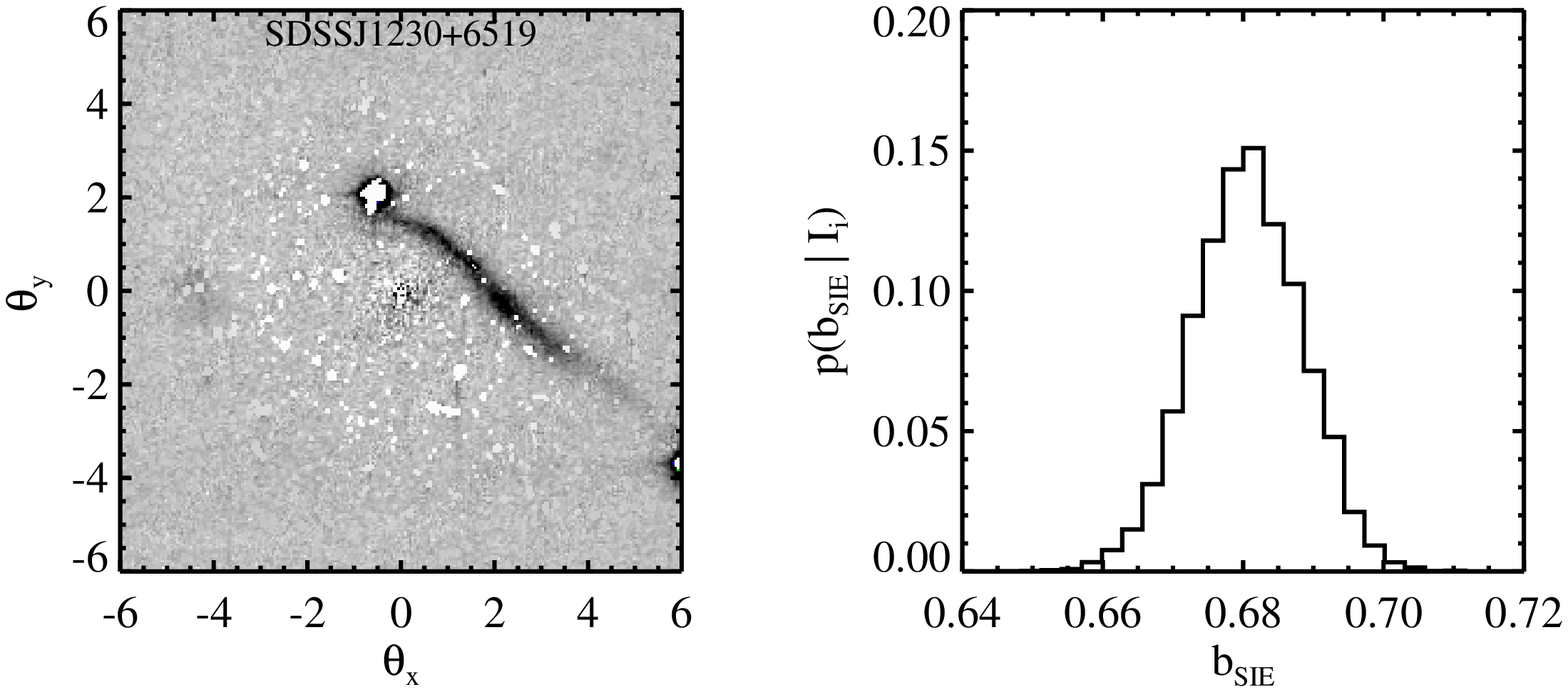}
\includegraphics[width=0.495\textwidth]{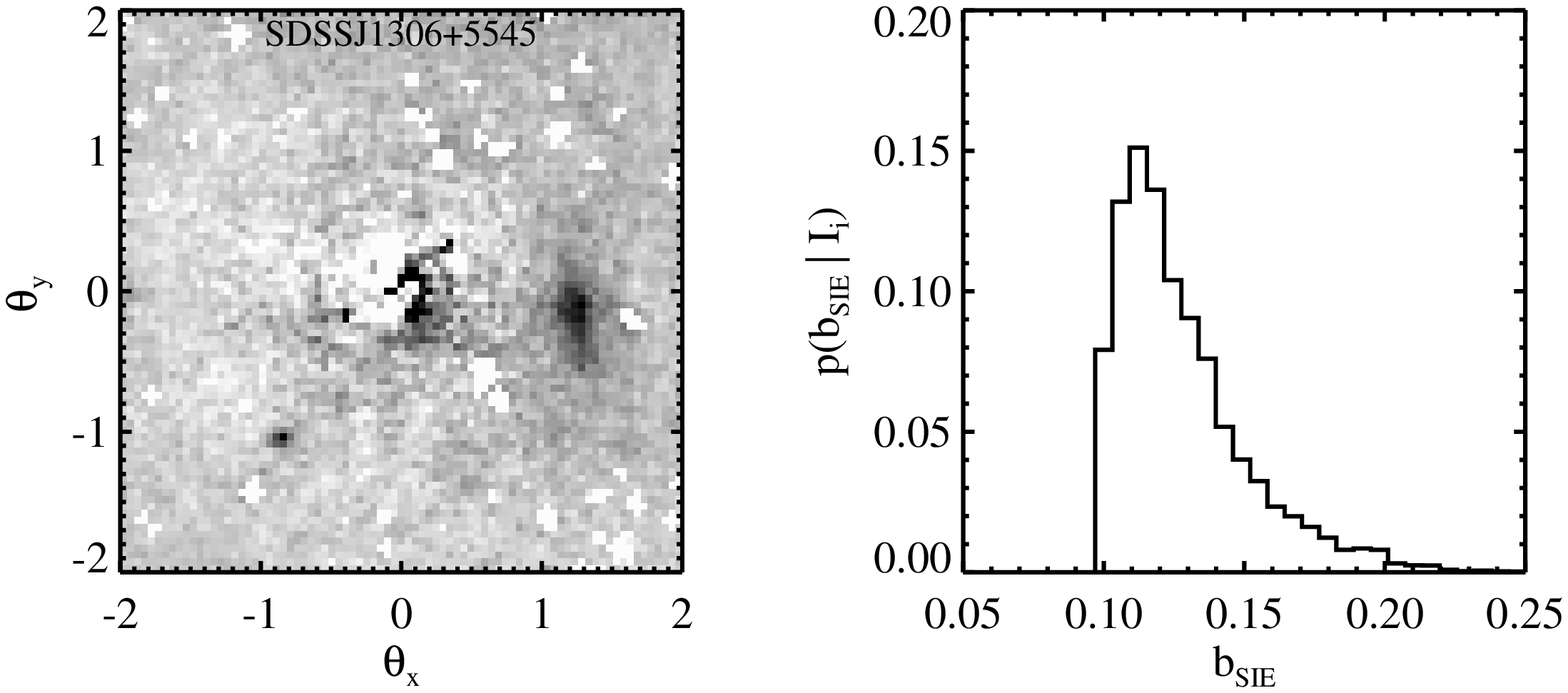}
\includegraphics[width=0.495\textwidth]{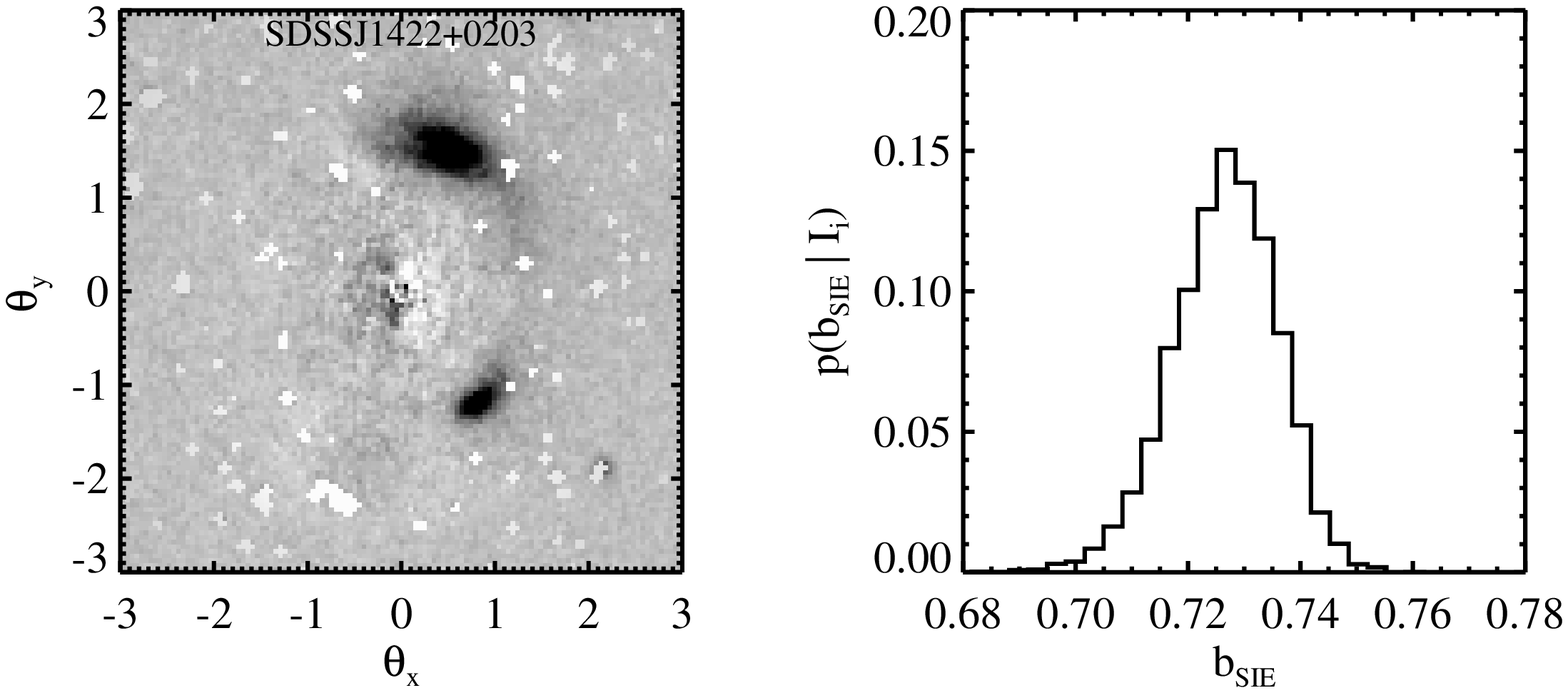}
\includegraphics[width=0.495\textwidth]{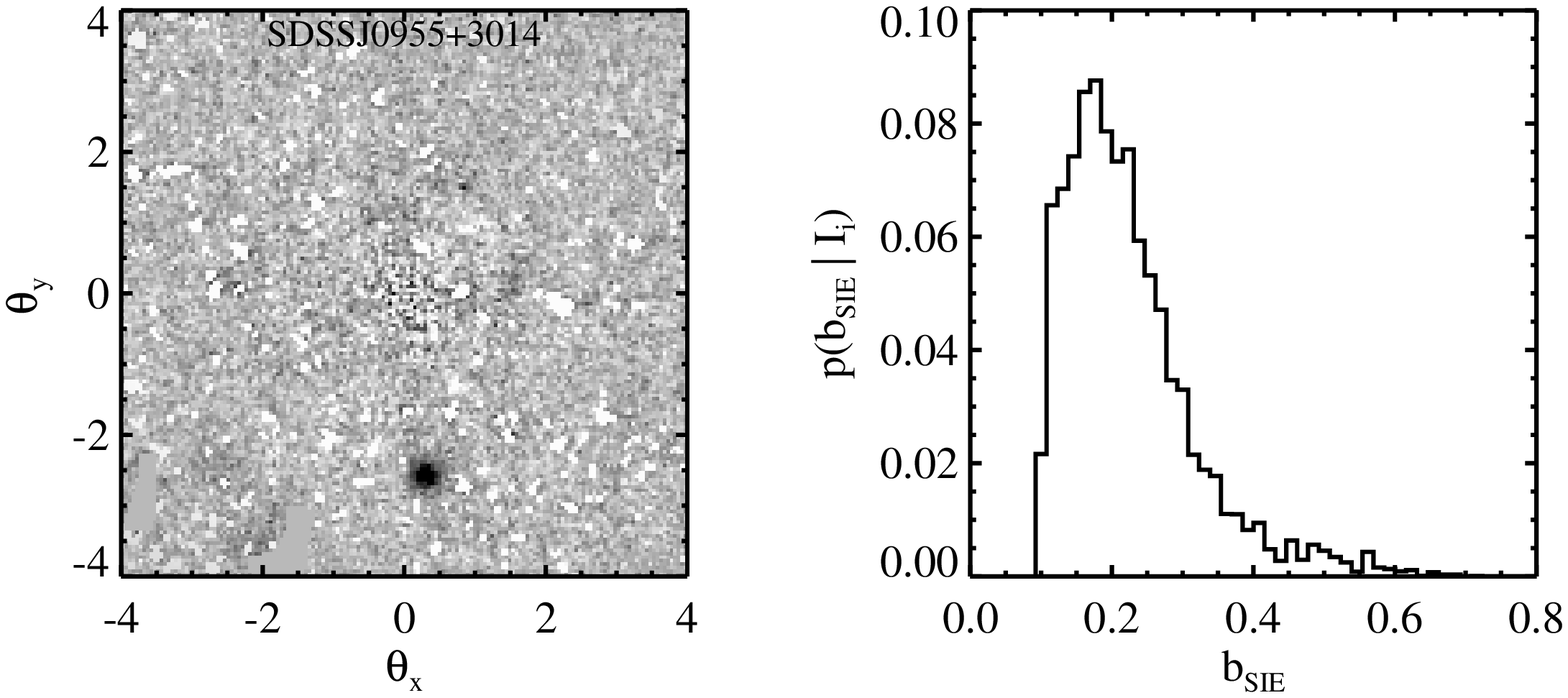}
\caption{{\label{fig:PPDF_12210_1}}
Foreground-subtracted images and corresponding posterior PDFs of $b_{\rm SIE}$ for all the 17 grade-C lenses in the S4TM survey.}
\end{figure*}
\addtocounter{figure}{-1}
\begin{figure*}[htbp]
\includegraphics[width=0.495\textwidth]{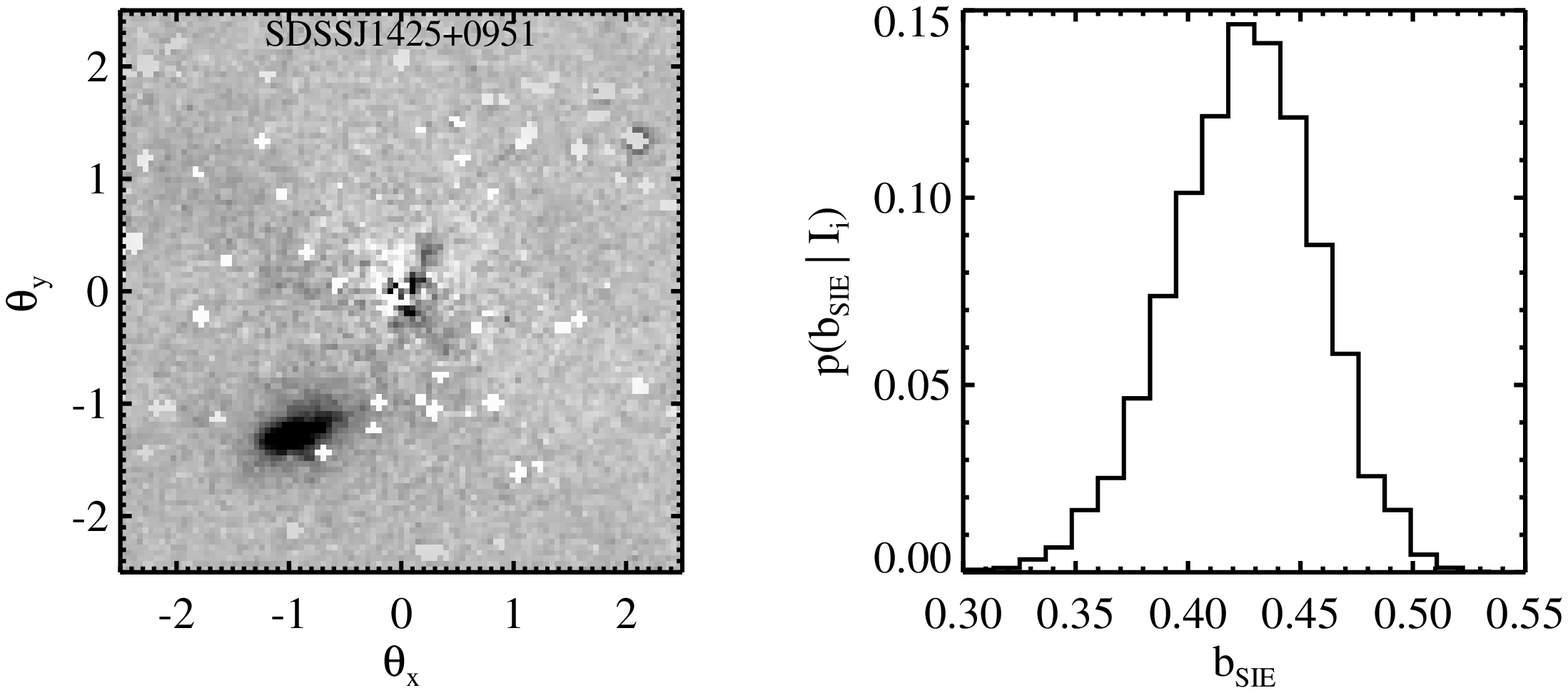}
\includegraphics[width=0.495\textwidth]{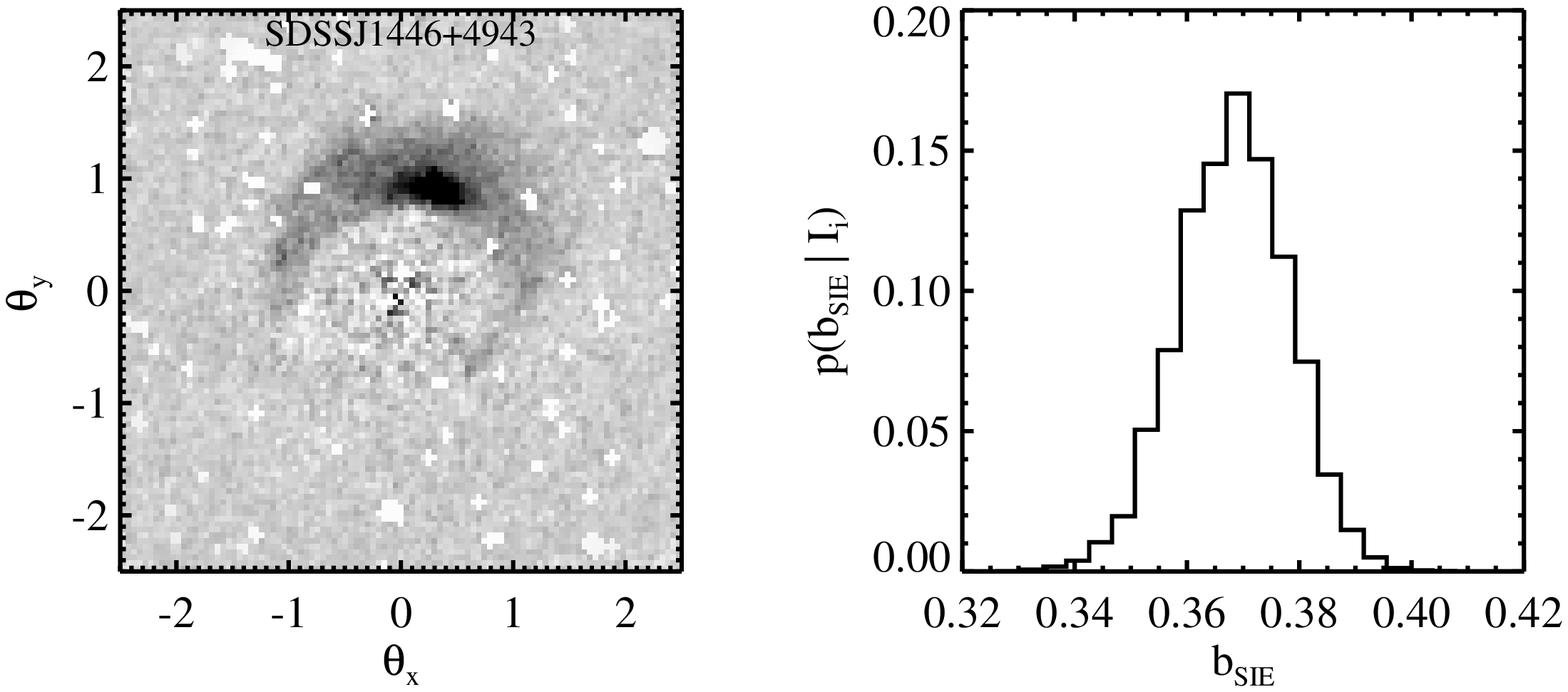}
\includegraphics[width=0.495\textwidth]{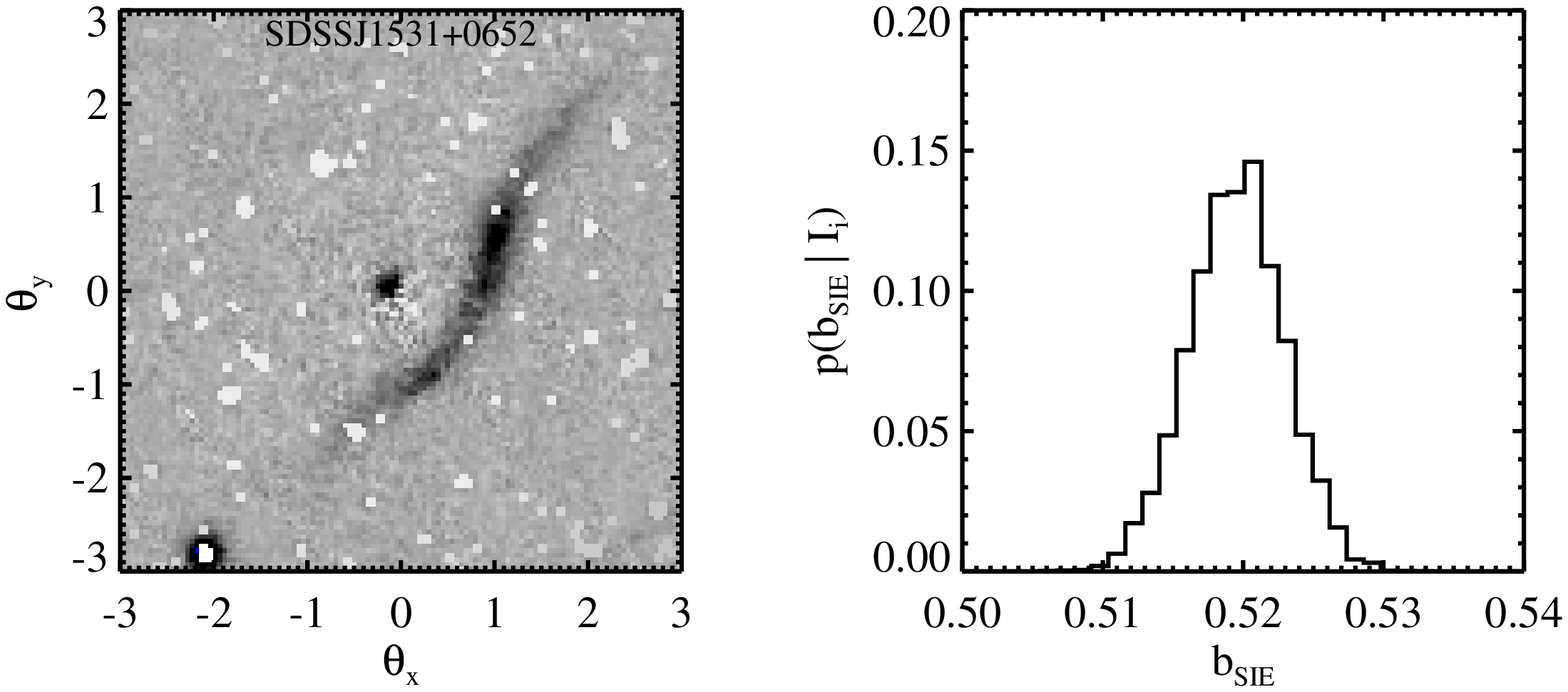}
\includegraphics[width=0.495\textwidth]{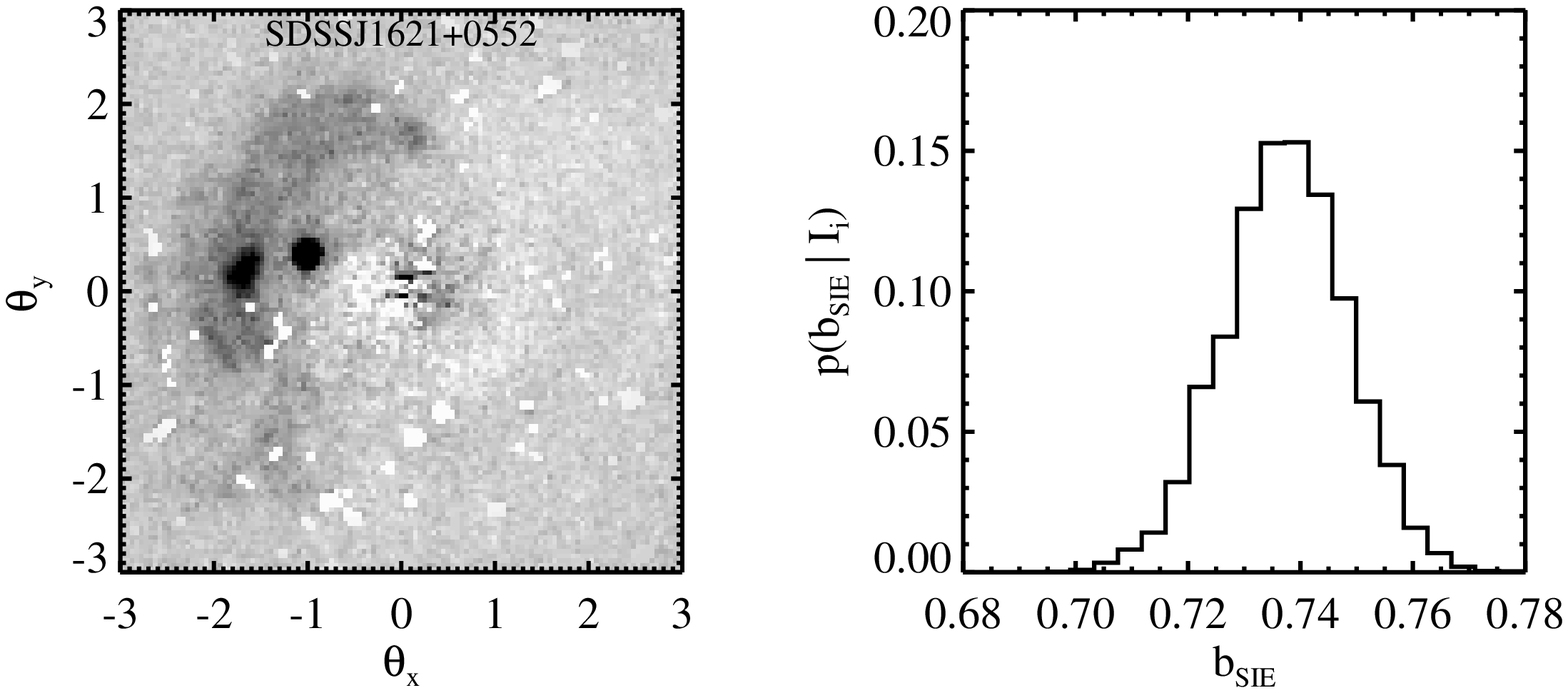}
\includegraphics[width=0.495\textwidth]{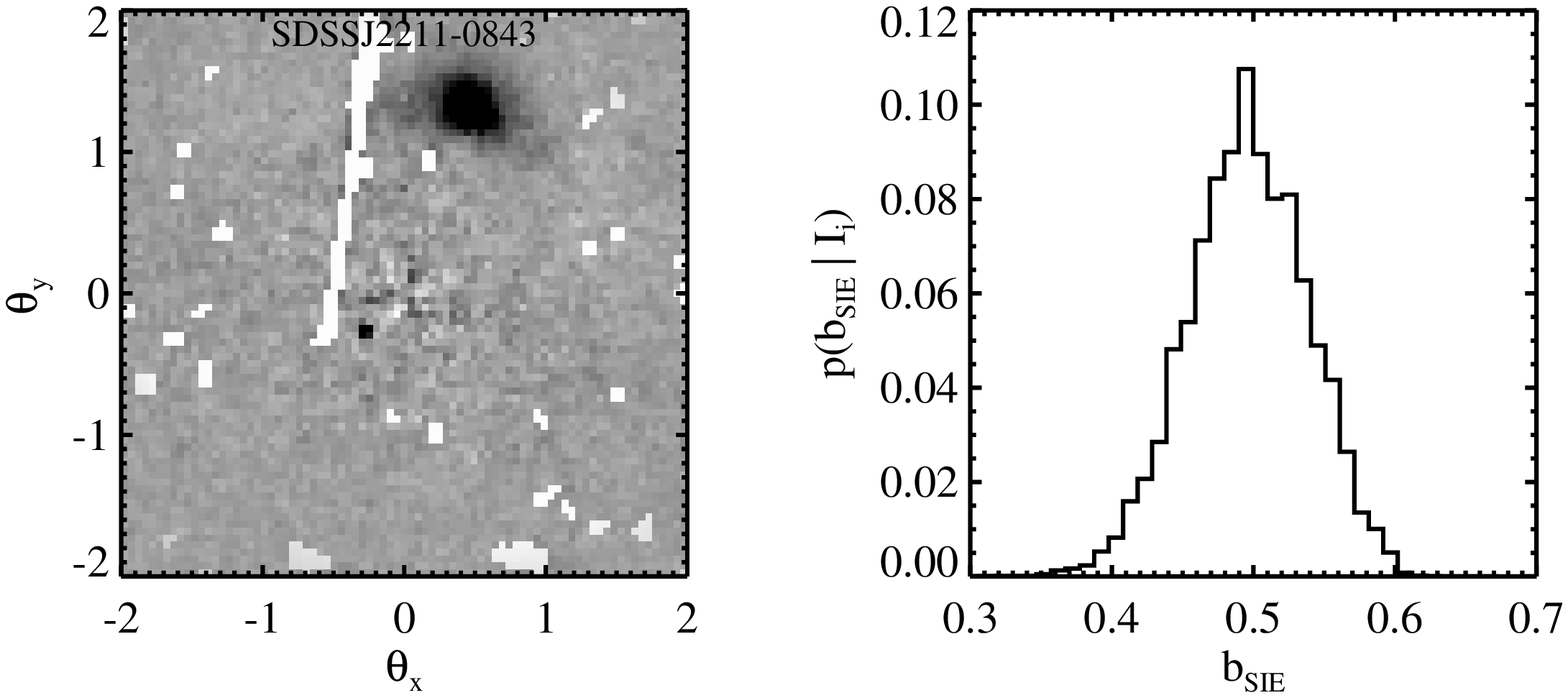}
\caption{{\label{fig:PPDF_12210_2}} \textit{Continued}}
\end{figure*}

\begin{figure*}[htbp]
\centering
\includegraphics[width=0.495\textwidth]{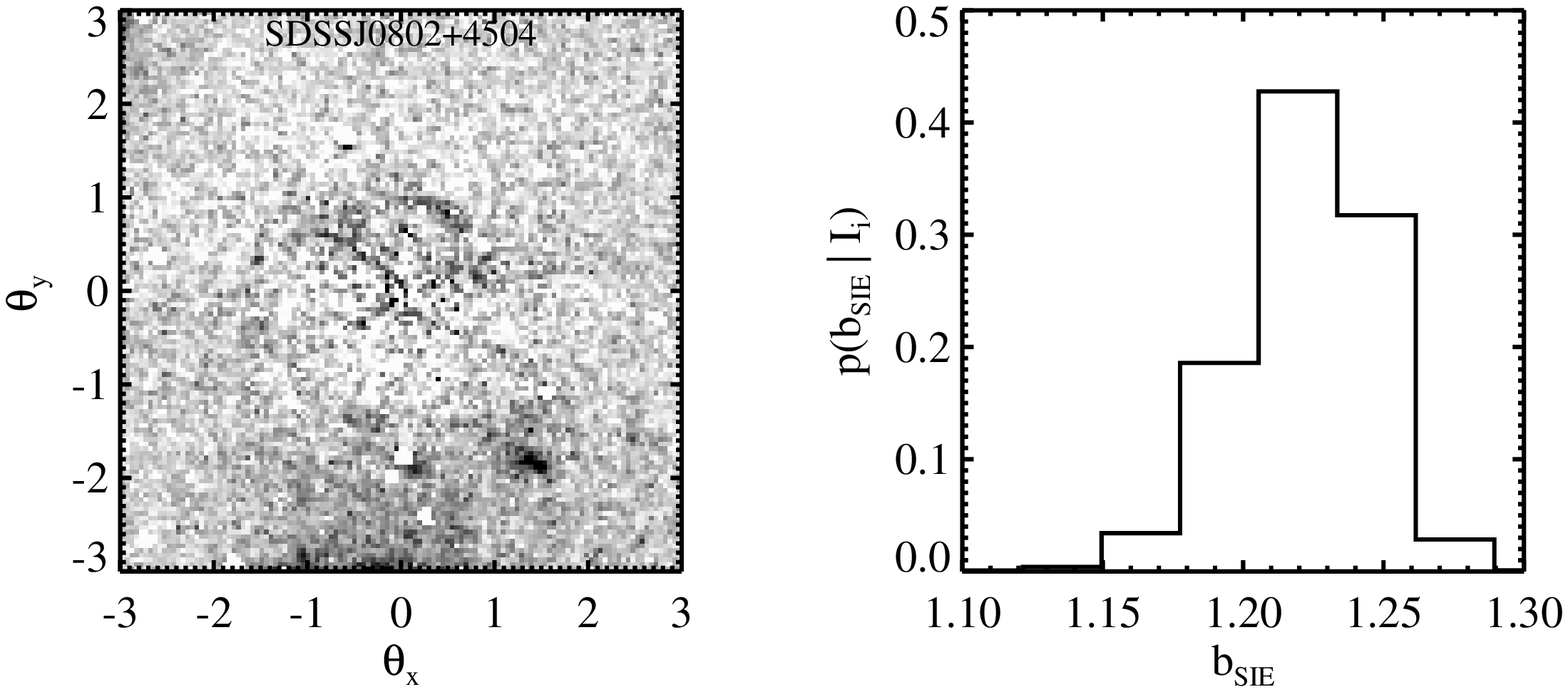}
\includegraphics[width=0.495\textwidth]{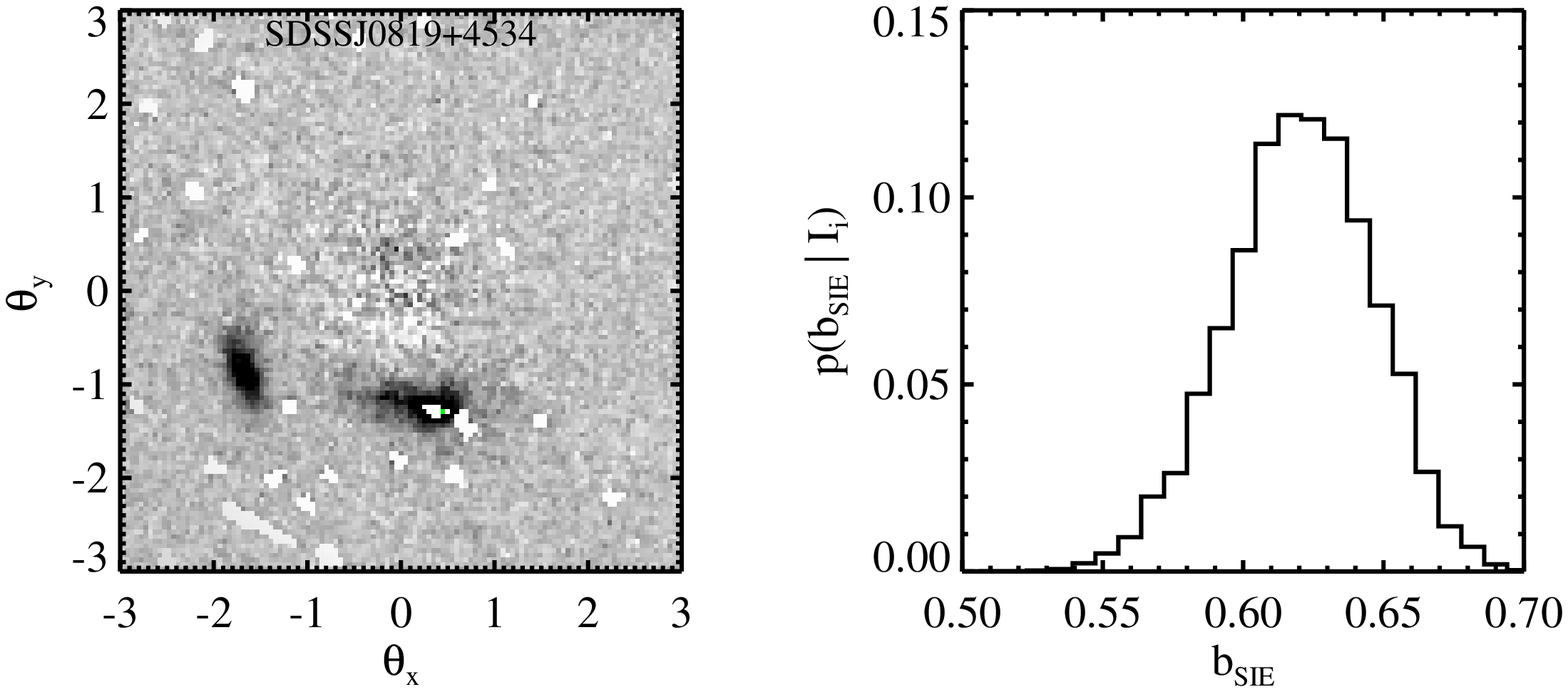}
\includegraphics[width=0.495\textwidth]{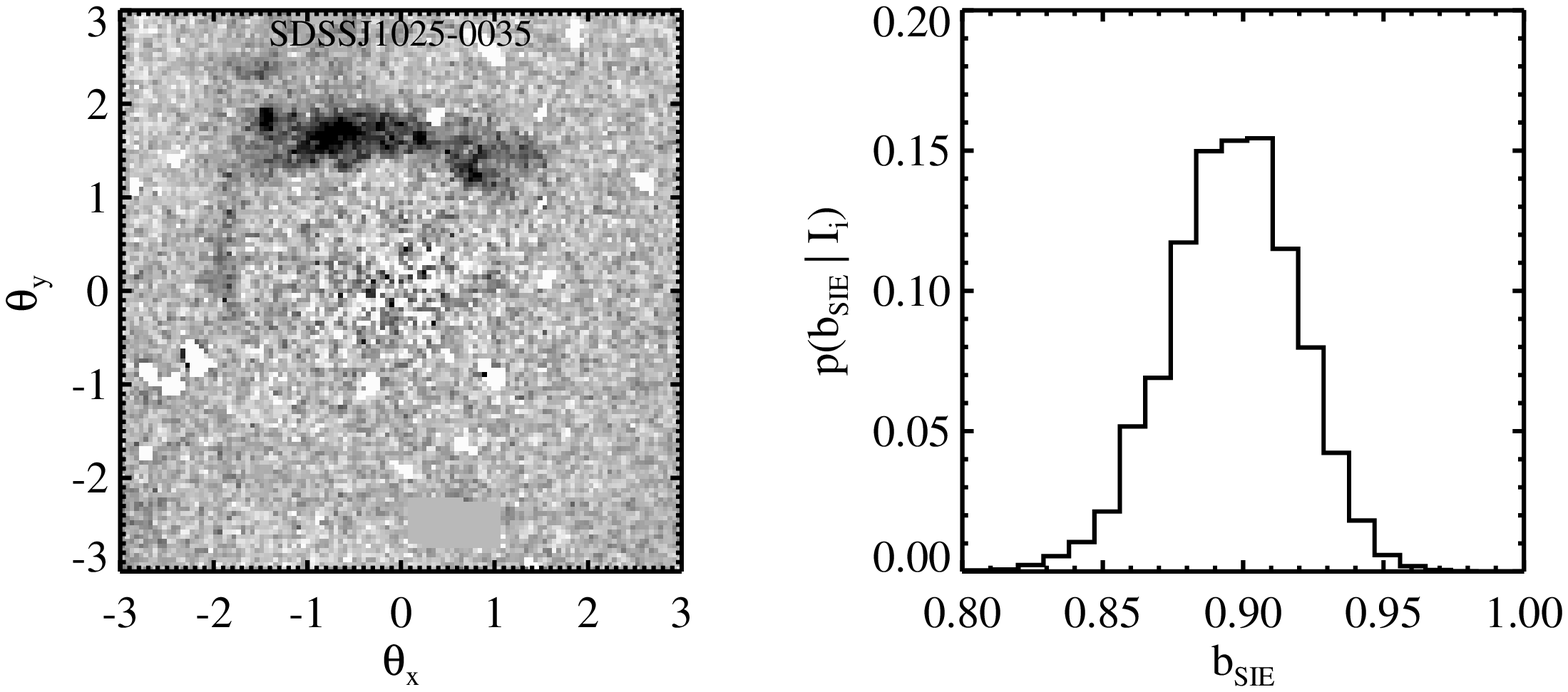}
\includegraphics[width=0.495\textwidth]{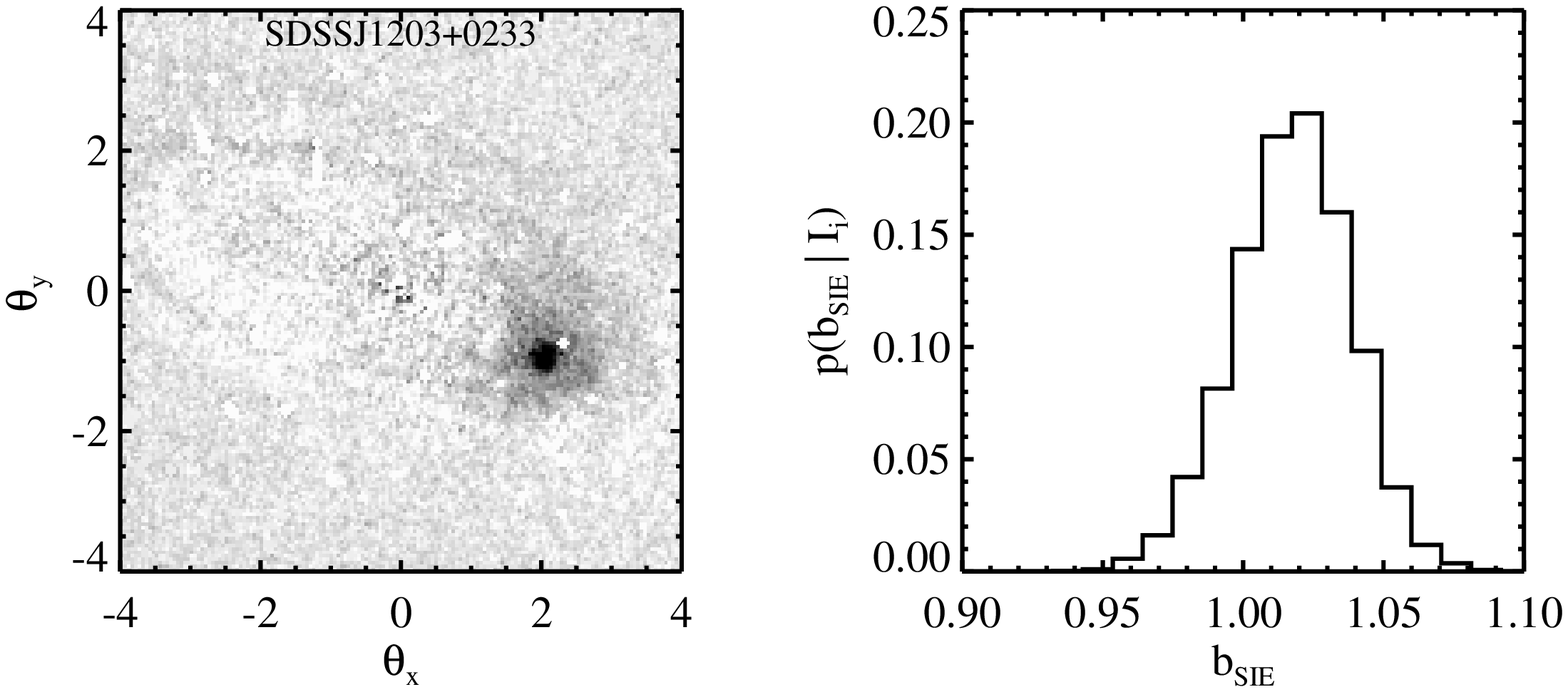}
\includegraphics[width=0.495\textwidth]{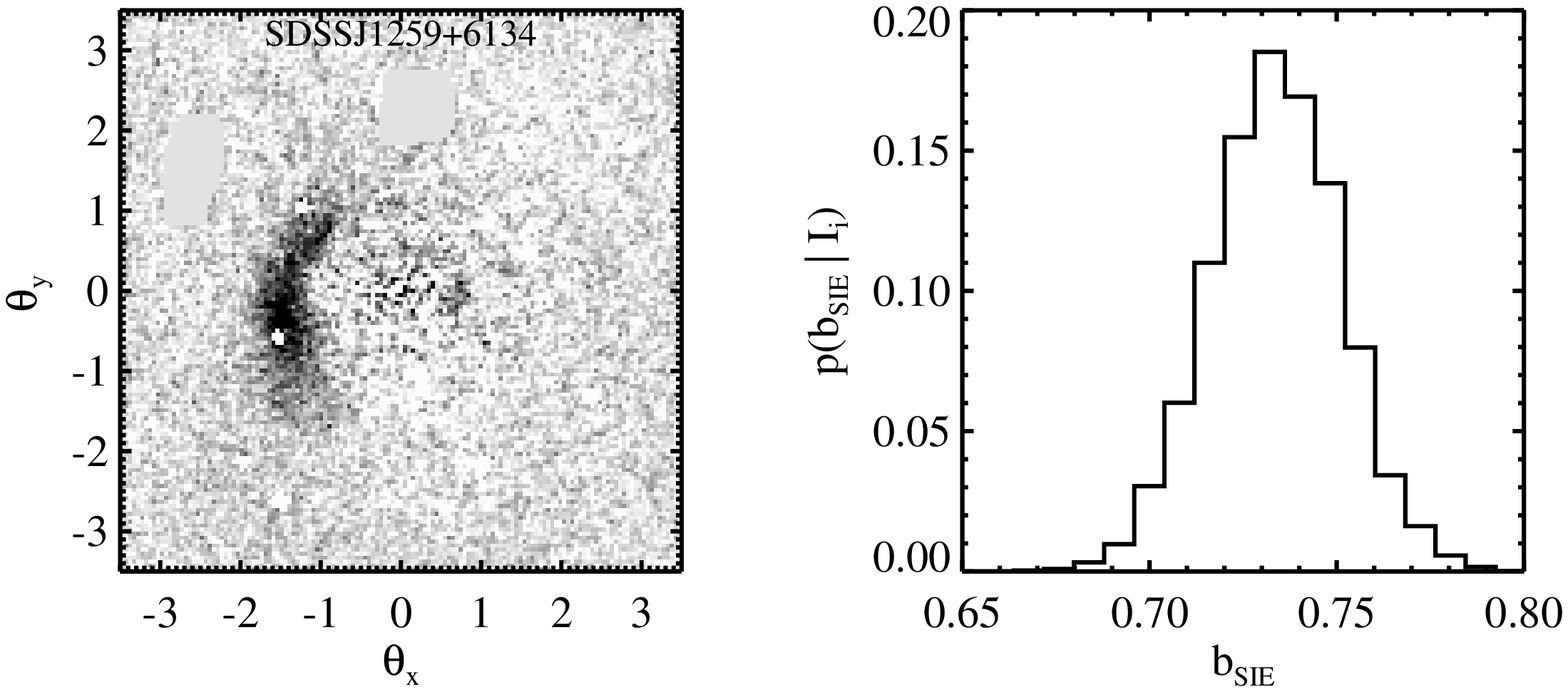}
\includegraphics[width=0.495\textwidth]{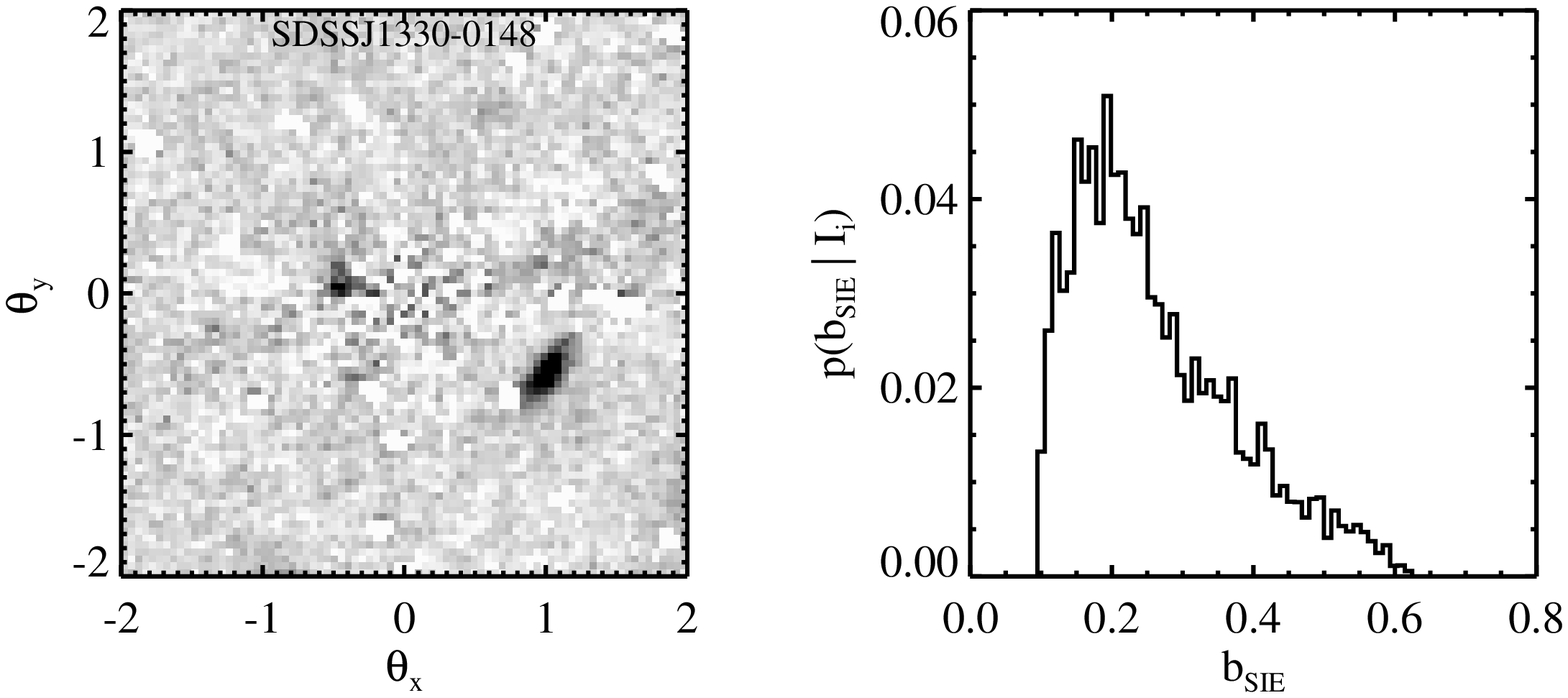}
\includegraphics[width=0.495\textwidth]{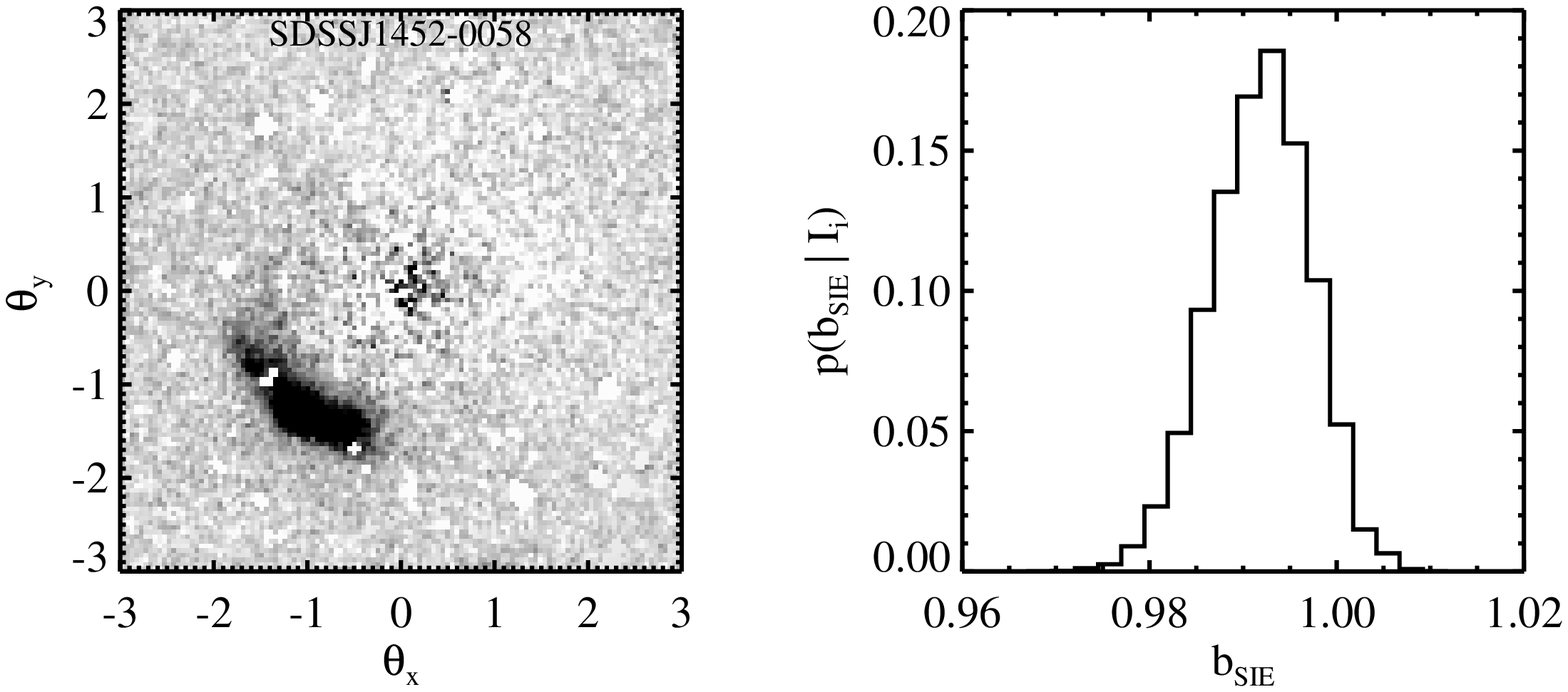}
\includegraphics[width=0.495\textwidth]{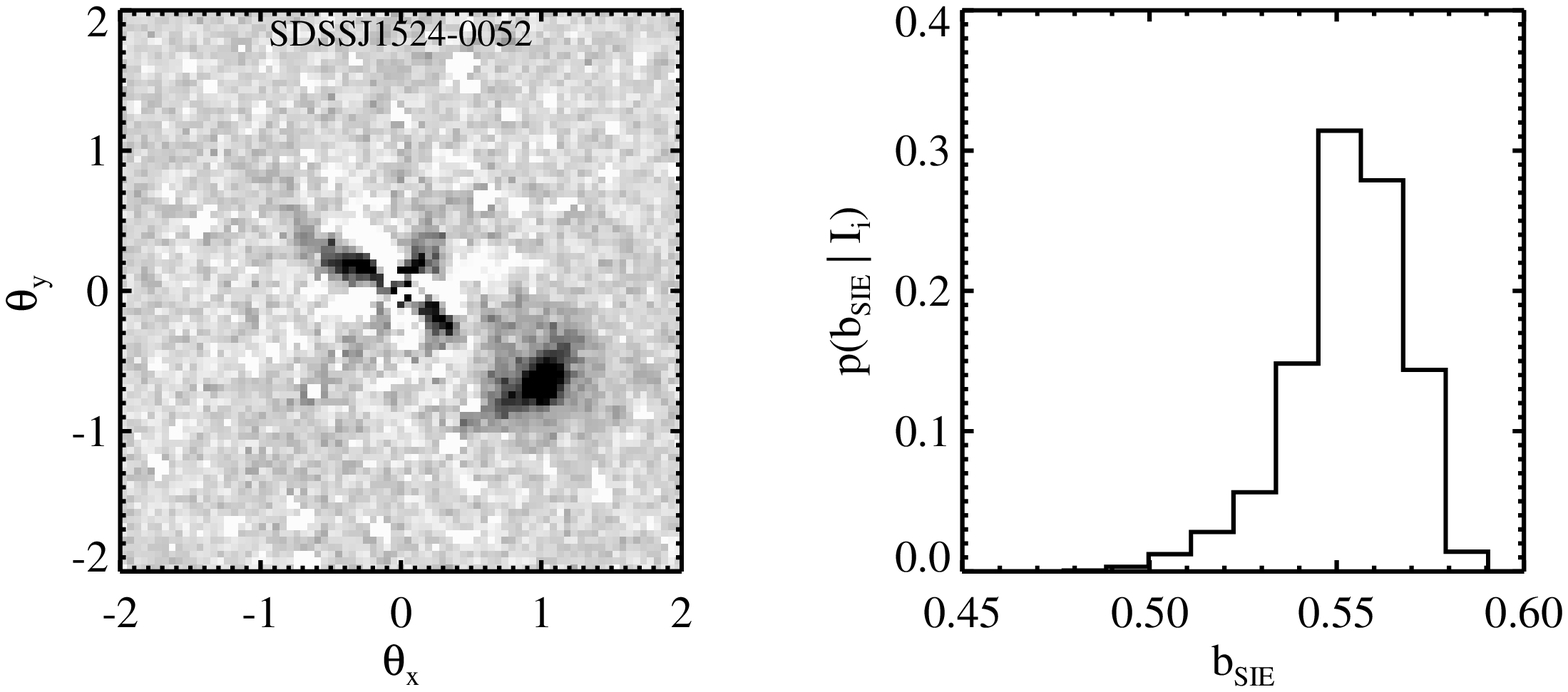}
\includegraphics[width=0.495\textwidth]{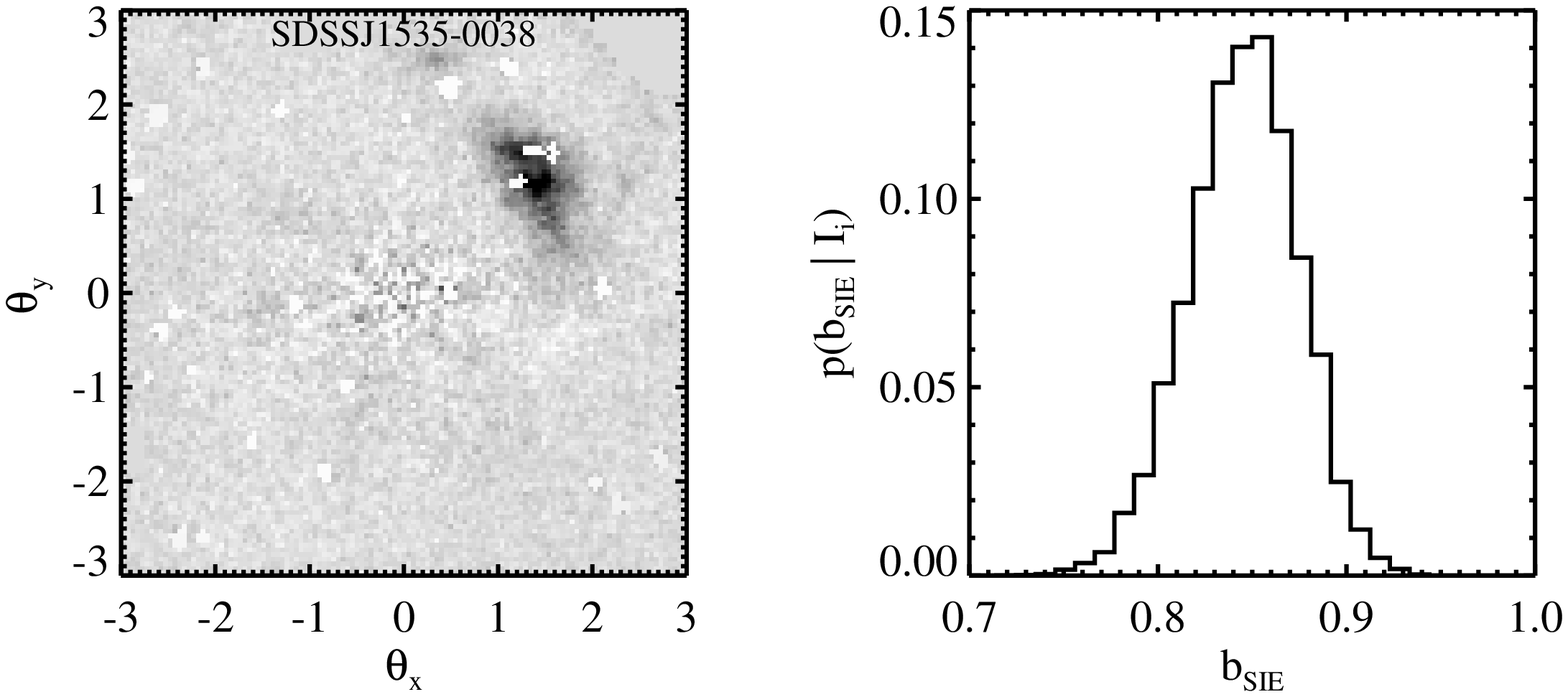}
\includegraphics[width=0.495\textwidth]{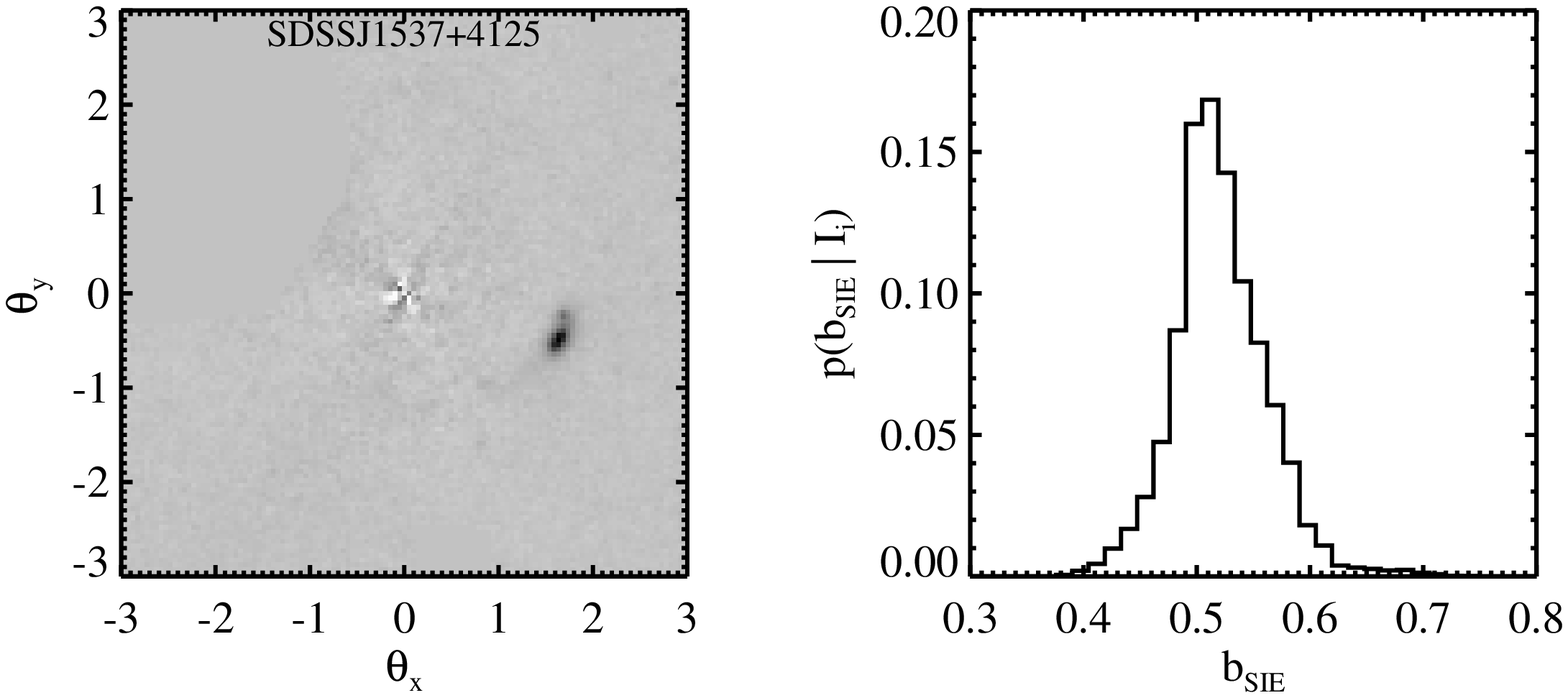}
\includegraphics[width=0.495\textwidth]{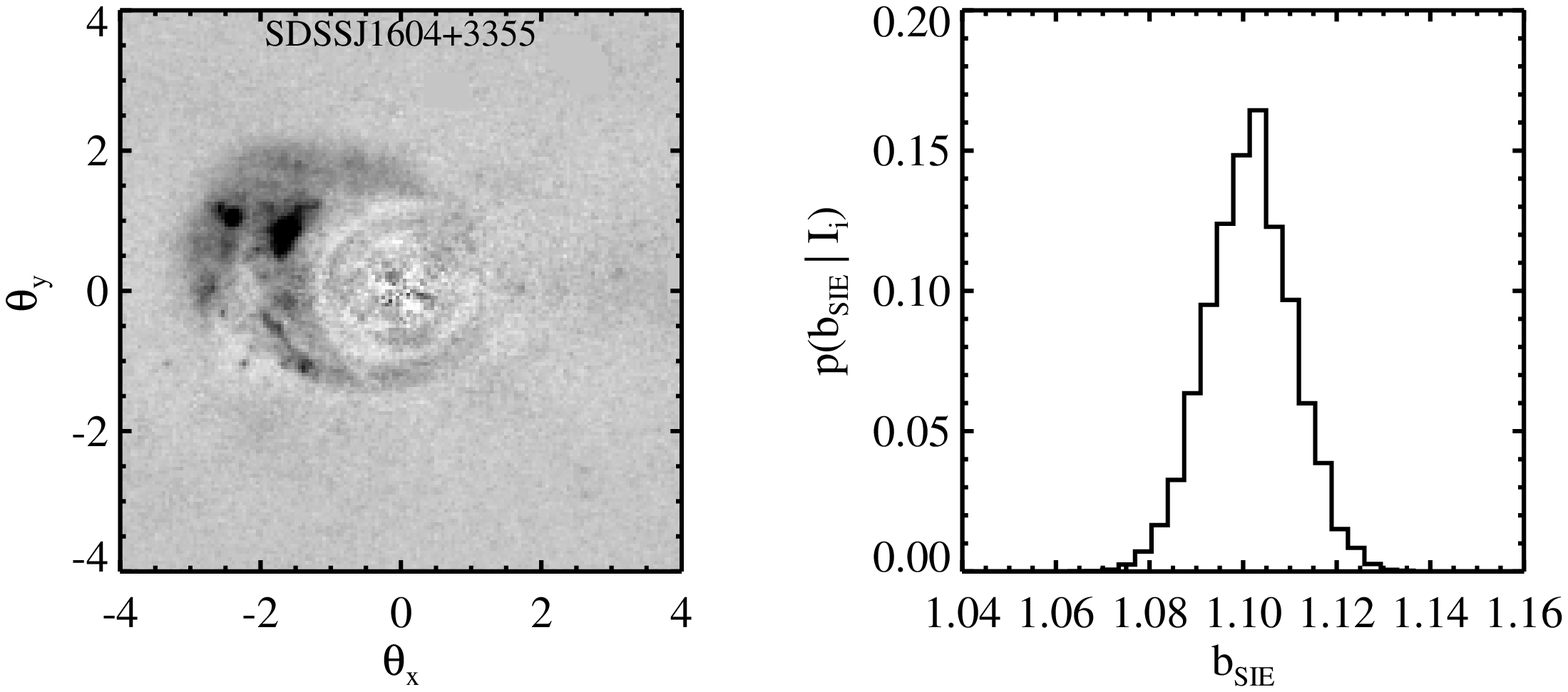}
\includegraphics[width=0.495\textwidth]{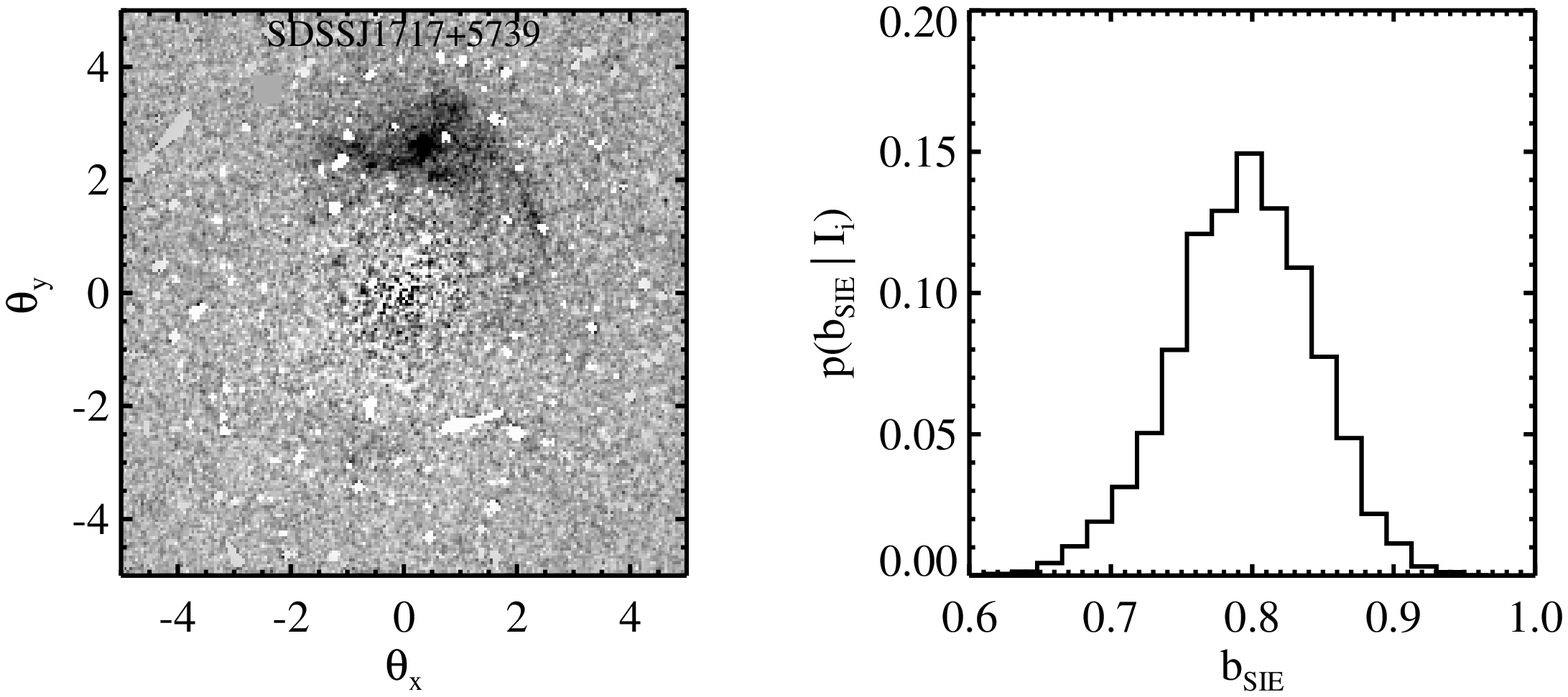}
\caption{{\label{fig:PPDF_slacs_1}}
Foreground-subtracted images and corresponding posterior PDFs of $b_{\rm SIE}$ for all the 15 grade-C lenses in the SLACS survey.}
\end{figure*}
\addtocounter{figure}{-1}
\begin{figure*}[htbp]
\includegraphics[width=0.495\textwidth]{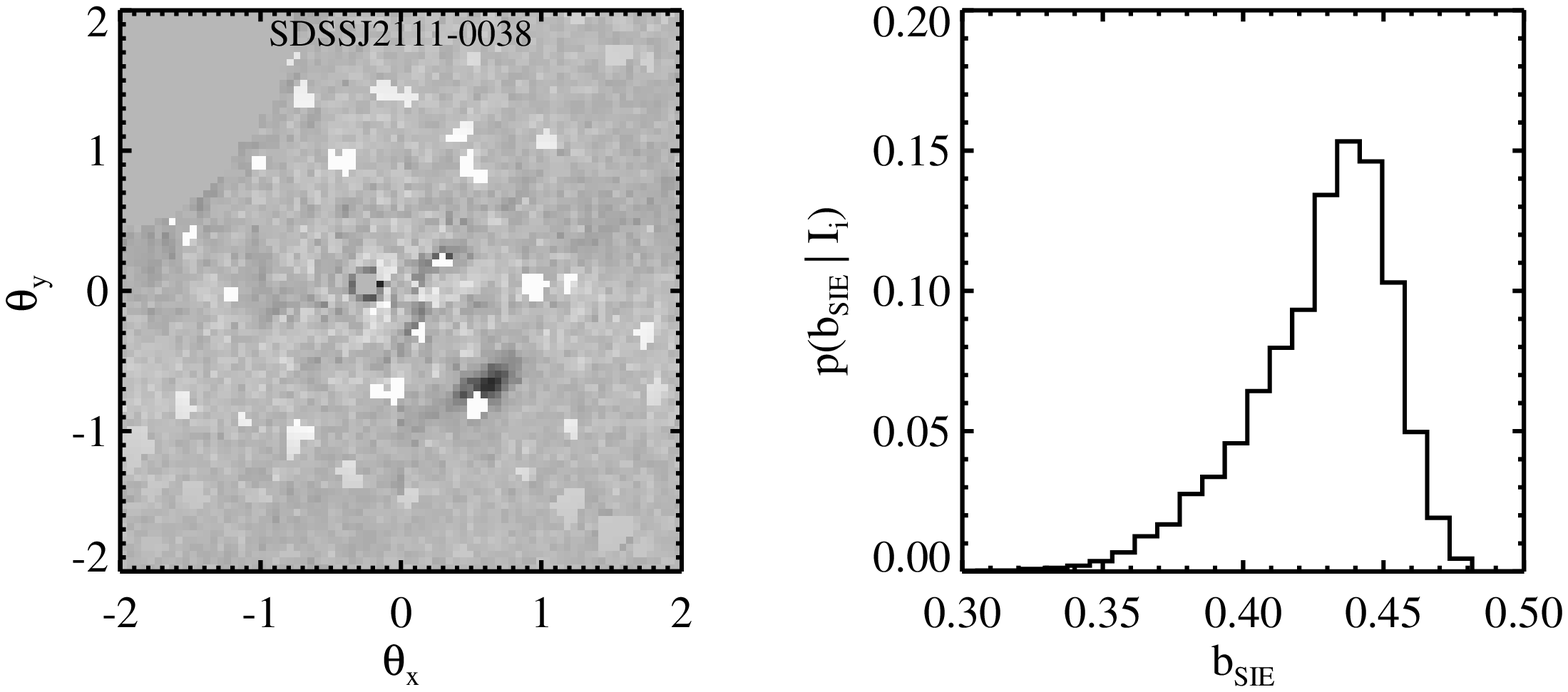}
\includegraphics[width=0.495\textwidth]{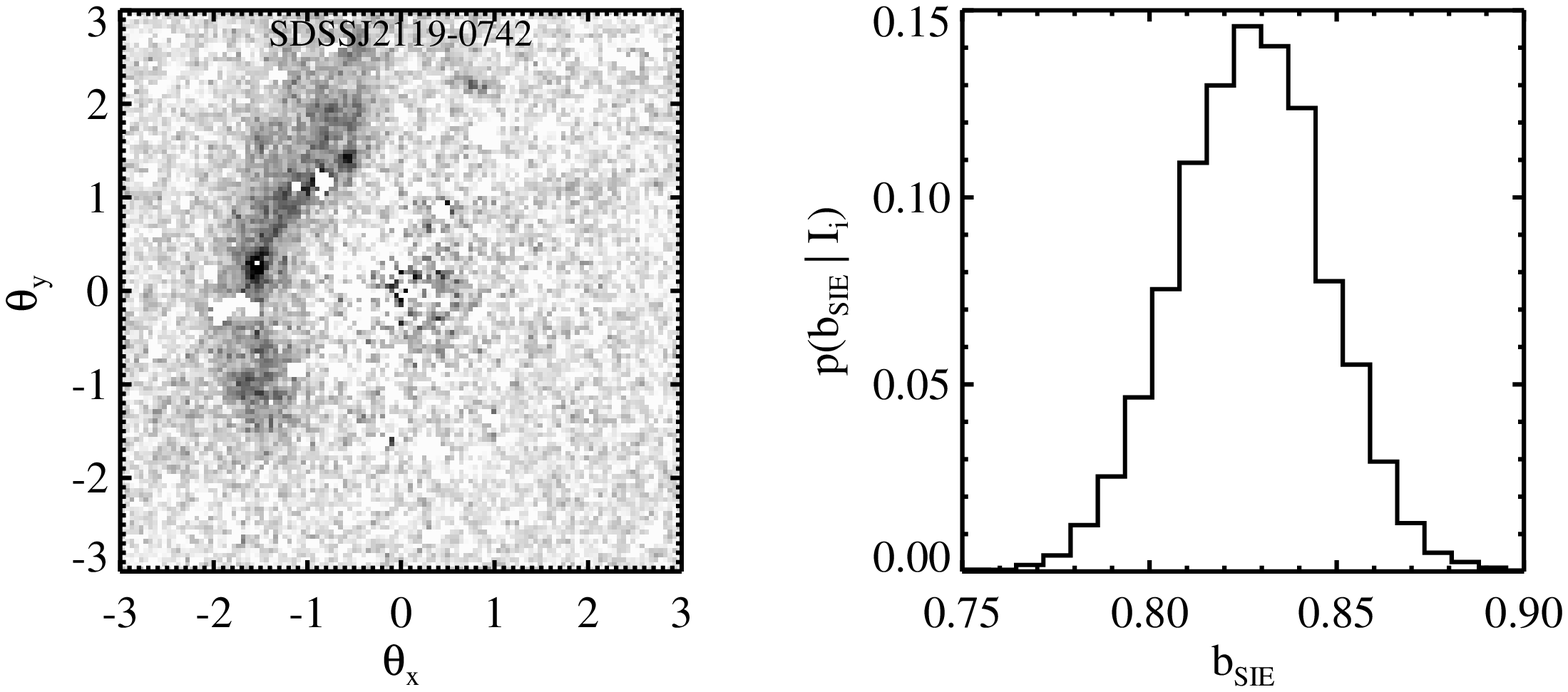}
\includegraphics[width=0.495\textwidth]{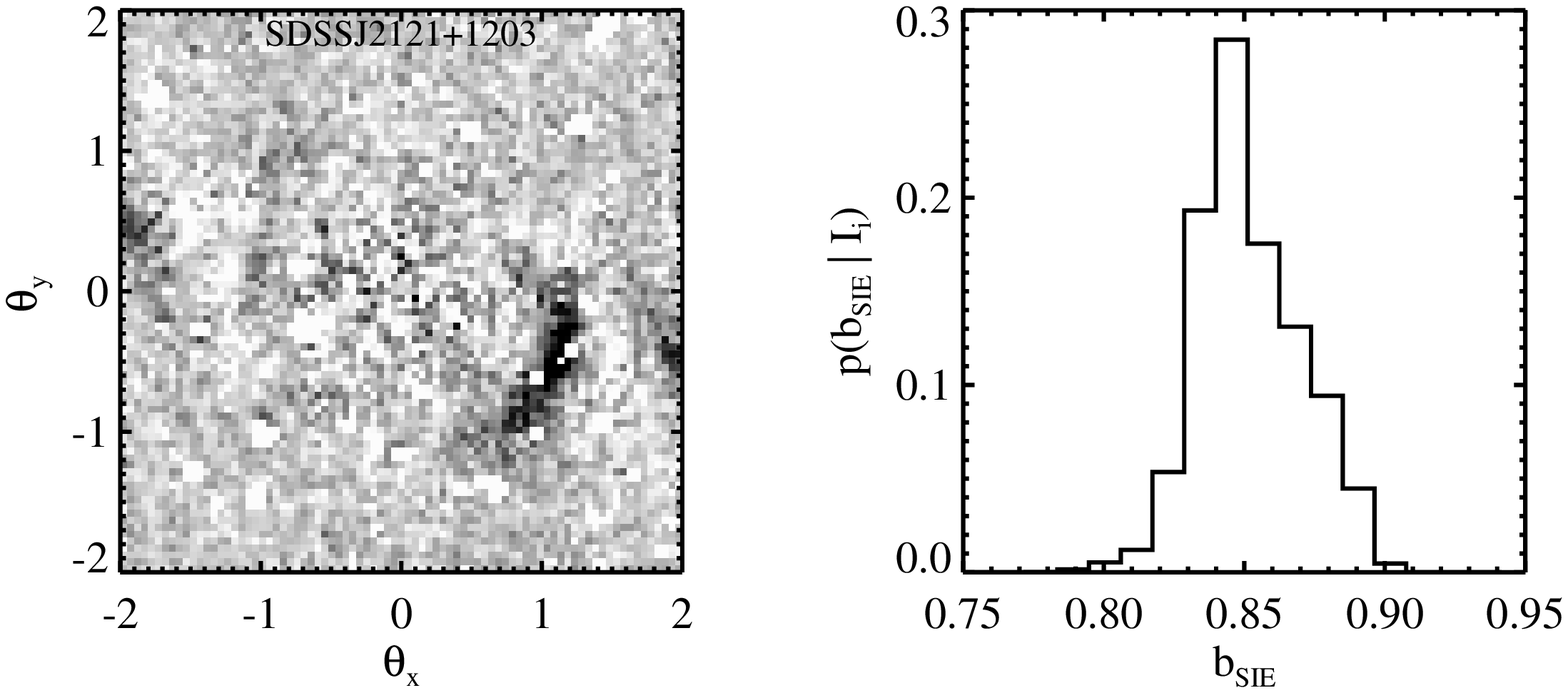}
\caption{{\label{fig:PPDF_slacs_2}}\textit{Continued}}
\end{figure*}

\end{document}